\documentclass[fleqn,usenatbib]{mnras}
\input{gridline_fig.sty}
\usepackage{newtxtext,newtxmath}

\usepackage[T1]{fontenc}

\DeclareRobustCommand{\VAN}[3]{#2}
\let\VANthebibliography\thebibliography
\def\thebibliography{\DeclareRobustCommand{\VAN}[3]{##3}\VANthebibliography}

\usepackage{graphicx}	
\usepackage{amsmath}	

\title[Reliability study of AGN torus morphology]{Inferring the morphology of AGN torus using X-ray spectra: A reliability study}

\author[T. Saha, A. G. Markowitz and J. Buchner]{
Tathagata Saha,$^{1}$\thanks{E-mail: tathagata@camk.edu.pl}
Alex G. Markowitz,$^{1,2}$
\& Johannes Buchner$^{3}$
\\
$^{1}$Nicolaus Copernicus Astronomical Center of the Polish Academy of Sciences, Bartycka 18, PL-00-716 Warszawa, Poland\\
$^{2}$Center for Astrophysics and Space Sciences, University of California, San Diego,  9500 Gilman Drive, La Jolla, CA 92093-0424, USA\\
$^{3}$Max Planck Institute for Extraterrestrial Physics, Giessenbachstrasse, D-85741 Garching, Germany\\
}

\date{Accepted 2021 November 4. Received BBB; in original form 2021 June 9}

\pubyear{20YY}

\begin{document}
\label{firstpage}
\pagerange{\pageref{firstpage}--\pageref{lastpage}}
\maketitle


\begin{abstract}
Numerous X-ray spectral models have been developed to model emission reprocessed by the torus of an active galactic nucleus (AGN), e.g., UXCLUMPY, CTORUS, and MYTORUS. They span a range of assumed torus geometries and morphologies --- some posit smooth gas distributions, and others posit distributions of clouds. It is suspected that given the quality of currently available data, certain model parameters, such as coronal power law photon index and parameters determining the morphology of the AGN torus, may be poorly constrained due to model degeneracies. In this work, we test the reliability of these models in terms of recovery of parameters and the ability to discern the morphology of the torus using \textit{XMM--Newton} and \textit{NuSTAR} spectral data.  We perform extensive simulations of X-ray spectra of Compton-thick AGNs under six X-ray spectral models of the torus.  We use Bayesian methods to investigate degeneracy between model parameters, distinguish models and determine the dependence of the parameter constraints on the instruments used. For typical exposure times and fluxes for nearby Compton-thick AGN, we find that several parameters across the models used here cannot be well constrained, e.g., the distribution of clouds, the number of clouds in the radial direction, even when the applied model is correct. We also find that Bayesian evidence values can robustly distinguish between a correct and a wrong model only if there is sufficient energy coverage and only if the intrinsic flux of the object is above a particular value determined by the instrument combination and the model considered.
\end{abstract}




\section{INTRODUCTION} \label{sec:intro}

The optical spectra of quasars and Seyfert galaxies are subject to broad classification: only Type 1 active galactic nuclei (AGN) display Doppler-broadened lines (Balmer lines, etc.), while both types exhibit narrow emission lines. A gaseous and dusty "torus" was hypothesized to explain this diversity: under an orientation-dependent unification scheme, it blocks the line of sight (LOS) to the central engine and the broad line emission region \citep{antonucci1993} in the type 2 objects. For example, one simple depiction of the morphology of the AGN torus was visualized in \cite{u95} where it was approximated to be a contiguous axis-symmetric dusty doughnut. In the X-rays, optical type 2s (and a few type 1s) exhibit strong line-of-sight absorption  \cite[e.g.][]{awaki91_1} which can be attributable to a circumnuclear obscuring torus. This finding implied that the X-ray spectral properties of the AGN were dependent on whether and at which angle the observer's line of sight intersected the torus. Physically such an obscuring torus can be a matter reservoir which feeds the AGN over its duty cycle and additionally can have links to radiatively driven outflows \citep{honig12}. The outflows driving matter away from the central black hole into the galactic environment \citep{honig19} thus are a potential source of AGN feedback, limiting the mass of the central black hole \citep{murray05}. Thus the different roles the torus plays in the AGN or galactic environment can be correlated with the complexities in its structure. Hence, studying the morphology and nature of the torus is important not just for understanding AGN/Seyfert orientation-dependent unification schemes, but also for luminosity-dependent unification schemes \citep{ricci13}, contributions of obscured AGN to the cosmic X-ray background (CXB) \citep{comastri95, risaliti99, gilli01, triester05, gilli07} and potential anisotropic radiative effects on host galaxy processes, such as star formation \citep{murray05,fabian06} or ionization of diffuse plasma \citep[e.g.][]{yang01}.

Several studies in IR and X-rays have shed light on the possible structure of the AGN torus. The X-ray study by \cite{awaki91} showed that the iron line Fe K edge could be reproduced sufficiently by a torus model where viewing angle is the only parameter. The sufficiency of a one parameter torus model in \citet{awaki91} could be the result of poor quality of data (in terms of signal-to-noise ratio and degrees of freedom available) analysed. A subsequent study by \cite{aherrero03} discovered that a simplistic torus model with a higher optical depth towards the equator was unable to explain the absence of predicted dichotomy in the steepness in the IR spectra for type 1 and type 2 AGNs. \cite{lutz04} and \cite{horst06} demonstrated that infrared (IR) emission from dust structures is relatively more isotropic than predicted from a simple continuous-doughnut shape; additionally the ratio of X-ray and mid-infrared (MIR) luminosities is similar for both type-1 and type-2 Seyferts. All these results provide observational evidence against the orientation-only classification of type-1 and type-2 AGN. Also, X-ray eclipse events \citep[e.g.][]{risality02,markowitz14} produced strong evidence that the obscurer is clumpy, thus challenging the pure orientation-dependent paradigm of AGN classification. Additional evidence for clumpiness in the circumnuclear material comes from MIR spectral fits \citep{almeida11}. It has thus been established that it is not just the orientation but also the nature of the obscurer that determines whether an AGN is seen as type-1 or 2 in optical or obscured or unobscured in X-rays \cite[e.g.][]{almeida2017}. \cite{elitzur08} summarized the observational signatures which indicated the presence of the clumpiness of the torus. An even further complication is a likely dependence of covering fraction on intrinsic luminosity \citep[e.g.][]{burlon11,ricci15}.

In this work, our attention is focussed towards Compton-thick X-ray obscured AGN and we investigate how accurately the properties of the torus and the central engine can be discerned from their X-ray spectra.  The X-ray spectrum of a Compton-thick obscured AGN is comprised of two components viz. the photoelectrically absorbed (zeroth-order continuum) and Compton scattered component (reflection component) of the direct power law originating in the AGN corona. In addition to the continuum, there are several distinct emission lines from the most abundant elements, notably the 6.4 keV iron line, accompanied by their Compton shoulders \citep[e.g.][]{ghisellini94,done96,matt02,yaqoob10_1}. The zeroth-order continuum and the scattered power law has information about the torus morphology and the obscured central engine. Previously in X-ray spectral data fits \citep[e.g.][]{bianchi05}, the assumed model for the zeroth-order continuum was an absorbed power law and the model for the scattered component was approximated by using simplistic models viz. reflection from infinite slab \citep{magdziarz95}. Reflection models incorporating a semi-infinite slab are more consistent with an accretion disc rather than a torus. Hence, no physically important parameters corresponding to the torus morphology can be derived when a geometrically inconsistent model \citep[e.g.][]{murphy09} is used for data analysis. Thus in an attempt to study the physics of the X-ray obscurer, numerous physically motivated models simulating the reprocessed X-ray spectral emission from the AGN torus have been developed over the past decade. These models use complex radiative transfer codes to calculate the X-ray spectrum of a Compton-thick AGN. Some assume a contiguous dust and gas structure; others assume clumpy distribution of dust and gas. Other than the differences in the morphology, the models also assume different aspects of radiative physics like different scattering cross-sections and consideration of some different radiative components. The differences in morphology and/or radiative physics introduce different features in the scattered spectra of the AGN torus e.g. like different nature of the Compton-Reflection Hump (hereafter ``CRH"), different shape of the iron lines, and Compton shoulders. Most of the simulated models are available as FITS tables and can be used for fitting to real data to estimate the model parameters via spectral analysis software like \textsc{xspec} and \textsc{isis}. 

Spectral model fits, including those involving torus models, can be potentially misleading in two different ways. Firstly, some parameters in the given model are degenerate with some other parameters and these degeneracies can affect the quality of simple spectral model fits. From the statistical point of view in complex parameter space, simple $\rm \chi^2$-fit algorithms can get stuck and falsely return wrong values of parameter and uncertainties. Secondly, it might be possible to fit a given data set with different models and obtain good measure of goodness of fit ($\chi^2/$dof), but with differing value of fit parameters \citep{guainazzi16} for different models assuming different geometry. \cite{ogawa19} has demonstrated similar cases of model degeneracy for un-obscured cases, where both relativistic reflection from the accretion disc, modelled using \texttt{xillver} and the scattered continuum from the AGN torus, modelled using \texttt{XCLUMPY} \citep{tanimoto19} could fit the same data set. This problem of model degeneracy will aggravate as we descend down to low data quality shorter exposures or lower fluxes where the signal-to-noise ratio is poor. The data-model residuals alone can become ineffective in determining the best model for a given data. This brings challenges in measurement of quantities derived from fit parameters; e.g. Eddington ratio ($\lambda_{\rm edd}$) or bolometric luminosity ($L_{\rm bol}$) calculated from $\Gamma$ (e.g. \cite{brightman13}). So from the viewpoint of data analysis, X-ray astronomers must be informed about which parameters from the models can be constrained correctly, which parameters have the potential to mislead and what are the signatures that can be helpful to realize that the fitting model is not appropriate.

In this paper, we investigate the problem from the viewpoint of an observational astronomer. The goal of our work partially mirrors that of \cite{gonzalez_martin19} in their testing reliability of AGN IR spectral model fits. We use data simulated using the torus models and the instrument functions of two X-ray satellites, \textit{XMM--Newton} and \textit{NuSTAR}, whose combined X-ray coverage spans 0.2 to 78 keV. We use Bayesian analysis to calculate posteriors and evidence value of a model given a simulated data set(s). The case where data is simulated under and analysed with the same torus model is referred to as the intramodel analysis. The case where data is analysed using a torus model different from the one that has been used to simulate it, is referred to as cross-model analysis. We analyse the behaviour of parameter posteriors and the values of Bayesian evidence for both intra- and cross-model analyses in different conditions to understand their implications on model distinction and parameter determination.

The remainder of this paper is organized as follows: In Section 2, we introduce the spectral models and each of their geometries and parameters, we introduce the Compton-thick model used to test them, and we describe aspects of data simulation and the Bayesian spectral fitting method used. In Sections 3 and 4, we present the results of our model fits, for intra- and cross-model analyses, respectively. In Section 5, we briefly discuss feasibility of detection of a broad diskline component in these spectra. In Section 6, we present considerations for exposure time requirements. In Section 7, we discuss our results, including guidelines to the X-ray community regarding making conclusions about the original obscurer morphology and the values of parameter that are estimated from the fits, and caveats associated with our analysis. We summarize our findings in section~\ref{summary}.

\section{METHODOLOGY} \label{sec:methods}
In this section we discuss the models we test, their morphology and the parameters. We also discuss the methodology of our data simulation using the models and our data analysis. This section also contains important abbreviations that we use throughout the work in the subsequent sections.
\subsection{Description of models used} \label{sub:model-desc}
 In this subsection we summarize the key features of the different models we tested here.
\begin{itemize}
\item[1.]MYTORUS: \cite{murphy09} assumes an axi-symmetric doughnut geometry. The model thus implements the classic orientation dependent AGN classification paradigm. The  size of the torus is determined by the ratio $c/a$ where $c$ is the distance from the point X-ray source and $a$ is the radius of the torus cross section. The ratio is $c/a = 0.5$ (fixed) in the published model. The gas density is uniform throughout the torus. The column density at the equator ($N_{\rm H,eq}$) is a variable parameter; the LOS column density ($N_{\rm H,los}$) is derived from the ratio $c/a$ and angle of inclination ($\theta_{\rm i}$). Here and throughout the paper, we refer $\theta_{i} = 0^{\circ}$ or $90^{\circ}$ as face-on or edge-on respectively. The mathematical relation between $N_{\rm H,eq}$ and $N_{\rm H,los}$ is given by:
\begin{equation}\label{eq-1}
N_{\rm H,los} = N_{\rm H,eq}\sqrt{1-\left( \frac{c}{a} \right) ^2 \rm cos^2\theta_{\rm i}}.
\end{equation}
The primary radiative processes that reprocesses the input radiation are photoelectric absorption, Compton scattering and fluorescent line emission. The published model in its most recent version has three separate FITS files: each for the zeroth-order continuum, the scattered continuum and the iron fluorescent lines with their Compton shoulders. The input to the torus is assumed to be a simple power law.

\item[2.]RXTORUS: This model is based on the radiative transfer code REFLEX \citep{paltani17}. The geometry assumed is an axi-symmetric doughnut. However, there are important additions to the radiative physics of the model, where photon scattering from electrons bound to the metallic atoms are also taken into account. The scattering cross-sections of bound electrons are modified by binding corrections, which decrease the Compton scattering cross section and the Rayleigh scattering dominates, mainly at the low energies. This is manifested as excess soft band emission, when compared to the corresponding spectra for MYTORUS. Additionally, $c/a$ is a variable parameter (unlike in MYTORUS). $N_{\rm H,los}$ is a derived parameter calculated from Equation \ref{eq-1}.

\item[3.]ETORUS: \cite{ikeda09} assume a continuous torus. This model has a spherical geometry with a biconical cutout at the poles; the cone vertices lie above/below the central point on the symmetry axis. The X-ray source is a point source at the sphere's center. The ratio of the inner to the outer radius of the torus ($r = r_{\rm in}/r_{\rm out}$) is kept fixed at 0.01 in the published model. The opening angle ($\theta_{\rm o}$) is variable and for obscured AGN, the angle of inclination ($\theta_{\rm i}$) has to be greater ($\theta_{\rm i}>\theta_{\rm o}$). The density of the torus is constant all throughout the volume. The model uses $N_{\rm H,eq}$ as a parameter in data-model fits and the line of sight absorption is a function of $r$, $N_{\rm H,eq}$, $\theta_{\rm i}$ and $\theta_{\rm o}$ and is given by:
\begin{equation}
N_{\rm H,los} = N_{\rm H,eq}\frac{r \rm (cos \theta_{\rm i} - cos \theta_{\rm o}) + sin(\theta_{\rm i} -\theta_{\rm o}) }{(1-r)(r \rm cos \theta_{\rm i} + sin(\theta_{\rm i} -\theta_{\rm o}))}.
\end{equation}
However for our analysis we only test the case where $\theta_{\rm i}$ and $\theta_{\rm o}$ are such that $N_{\rm H,los} \simeq N_{\rm H,eq}$. The model assumes a Thompson scattering cross-section to calculate the scattered continuum. The published model has only the reflected continuum. The zeroth-order continuum is modelled by a simple \texttt{zTBABS}$\times$\texttt{CABS}$\times$\texttt{CUTOFFPL} in \textsc{xspec} notation, where \texttt{zTBABS} models the direct photoelectric absorption and \texttt{CABS} models the Compton scattering losses. \texttt{ZGAUSS} is used to estimate emission lines.

\item[4.]BORUS: BORUS \citep{balokovic18} assumes a continuous torus with a biconically cutout spherical geometry. Qualitatively, similar to ETORUS but the with the cone vertices coincide with the central point X-ray source. The free parameters and model setup are similar to that of ETORUS. Similar to ETORUS the zeroth-order continuum is modelled with a power law attenuated \texttt{zTBABS} $\times$ \texttt{CABS}. BORUS allows for the iron abundance ($A_{\rm Fe}$) to be a free parameter in the torus, thus simulating the iron line consistently with the scattered continuum.

\item[5.]CTORUS:  The model \citep{liu14} is based on \texttt{Geant4} biconical cut-out  geometry except the gas distribution is clumpy. The clumps are distributed uniformly  between the inner radius $R_{\rm in}$ and the outer radius $R_{\rm out}$, effectively forming a thick shell filled with clouds. The angular limitation of the clump distribution is put by a conical surface for which $\sigma$ \citep{liu14} is 60$^{\rm \circ}$. $\sigma$ is not a variable parameter here.

\item[6.]UXCLUMPY: UXCLUMPY \citep{buchner19} uses the radiative transfer code \texttt{XARS} and cloud distribution used in \citep{nenkova08} given by the mathematical formula:
\begin{equation}
\mathcal{N}(\beta) = \mathcal{N}_0 e^{ -\left(\frac{\beta}{\sigma_{\rm o}} \right)^m},
\end{equation}
where $\beta$ is the latitude angle. The value of $m$ is set to 2. $\sigma_o$ sets the width of the torus cloud distribution about the equator. The radial distribution of the cloud is uniform \citep{nenkova08}. \texttt{UXCLUMPY} uses such a cloud distribution with an additional optional inner Compton-thick gaseous ring. The Compton-thick ring results in additional absorption in the 7~keV to $\sim$ 20~keV band, thus increasing the curvature of the CRH. UXCLUMPY uses $N_{\rm H,los}$ as a free parameter and it assumes that the line of sight always intersects with at least one clump irrespective of the inclination ($\theta_{\rm i}$). However, the distribution of the clouds imply that, as $\theta_{\rm i}$ tends towards the edge or equator ($90^{\rm \circ}$), the number of clumps and hence the probability of obscuration increases.
\end{itemize}
In Table \ref{tab:1}, we summarize the models and their most important features and parameters. 
\begin{table*}
\begin{tabular}{p{0.15\textwidth}|p{0.2\textwidth}|p{0.08\textwidth}|p{0.28\textwidth}|p{0.18\textwidth}}
\hline
\hline
Models  &Gas distribution \& morphology & Energy range (keV) & Free parameters and ranges                                                             & Parameter input\\
(1)     & (2) & (3) & (4) & (5)  \\
\hline
MYTORUS (MYT)   & Continuous :                  & 0.6-78 & $N_{\rm H,eq}(\rm 10^{22} cm^{-2}) = [1,10^3]$                                           &150(MCT), 680(HCT) \\        
\citep{murphy09}& Classic Doughnut              &        & $\Gamma = [1.6,2.5]$                                                             &1.9 \\
                &                               &        & $\theta_{\rm i}(^{\circ}) = [0,90]$($\theta_{i}<60^{\circ}$)                     &70 \\
                &                               &        & $C_{\rm sc-pl} = [10^{-7},1]$                                                    &$10^{-3}$\\
                &                               &        & $T/R = [0.5,5.0]$                                                                &1,1.8\\
\hline               
RXTORUS (RXT)  & Continuous :                   & 0.6-78 & $N_{\rm H,eq}(\rm 10^{22} cm^{-2}) = [1,10^3]$                                  &150\\
\citep{paltani17}&Classic Doughnut              &        & $\Gamma = [1.6,2.5]$                                                            &1.9\\
               &                                &        & $\theta_{\rm i}(^{\circ}) = [0,90]$($\theta_{i}<60^{\circ}$)                    &70\\
               &                                &        & $c/a = [0.1,1]$                                                                 &0.5\\
               &                                &        & $C_{\rm sc-pl} = [10^{-7},1]$                                                   &$10^{-3}$\\
               &                                &        & $T/R = [0.5,5.0]$                                                               &1,1.8\\
\hline         
ETORUS (ETOR)  & Continuous :                   & 1-78    & $N_{\rm H,eq} \simeq N_{\rm H,los} (10^{22}\rm cm^{-2}) = [1,10^3]$                   &150,500\\
\citep{ikeda09}&Sphere with biconical cutout    &         & $\Gamma = [1.6,2.5]$                                                           &1.9\\ 
               &cone vertex above/below the &         & $\theta_{\rm o}(^{\circ}) = [0,70]$                                            &45\\
               &axis X-ray source on the symmetry &         & $\theta_{\rm i}(^{\circ}) = [0,90]$($\theta_{i}>\theta_{\rm o}$)               &60\\
               &                                &         & $C_{\rm sc-pl} = [10^{-7},1]$                                                  &$1E-3$\\
               &                                &         & $T/R = [0.5,5.0]$                                                              &1\\
\hline         
BORUS (BOR)    & Continuous :                   & 1-78    & $N_{\rm H,los}(\rm 10^{22} cm^{-2}) = [1,10^3]$                                        &100(MCT), 500(HCT)\\
\citep{balokovic18}& Sphere with biconical cutout &       & $\Gamma = [1.6,2.5]$                                                           &1.9\\
               & cone vertex co-incident        &         & $E_{\rm cut}(\rm keV) = [60,500]$                                              &400\\
               & with the X-ray source          &         & $\rm C_{\rm frac,tor} = cos \theta_{\rm o} = [0,1]$                            &0.58\\
               &                                &         & $\rm cos \theta_{\rm i} = [0,1]$ ($\rm cos \theta_{i}<\rm cos \theta_{\rm o}$ ) &0.48\\
               &                                &         & $A_{\rm Fe} = [0.1,10]$                                                        &1.0\\
               &                                &         & $C_{\rm sc-pl} = [10^{-7},1]$                                                  &1e-3\\
               &                                &         & $T/R = [0.5,5.0]$                                                              &1\\
\hline         
CTORUS (CTOR)  & Clumpy :                       & 1.2-78  & $N_{\rm H,los}(\rm 10^{22} cm^{-2}) = [1,10^3]$                                     &100(MCT), 500(HCT)\\
\citep{liu14}  &  Uniformly distributed         &         & $\Gamma = [1.4,2.5]$                                                           &1.9\\
               &  clouds in a thick spherical   &         & $N_{\rm cloud} = [2,10]$                                                       &4\\
               &  shell                         &         & cos$\theta_{\rm i} = [0.05,0.95]$($\theta_{i}>30^{\circ}$)                     &0.35\\
               &                                &         & $C_{\rm sc-pl} = [10^{-7},1]$                                                  &$10^{-3}$\\
               &                                &         & $T/R = [0.5,5.0]$                                                              &1,1.8\\
\hline  
UXCLUMPY (UXCL)& Clumpy :                       & 0.3-78  & $N_{\rm H,los}(\rm 10^{22} cm^{-2}) = [1,10^4]$                                        &100(MCT), 500(HCT)\\
\citep{buchner19}&                              &         & $\Gamma = [1.4,2.5]$                                                           &1.9\\
               &  Distribution proposed in      &         & $E_{\rm cut}(\rm keV) = [60,500]$                                              &200,400\\
               &  \citep{nenkova08}             &         & $\sigma_{\rm o}(^{\circ}) = [6,84]$                                            &45\\
               &                                &         & $C_{\rm frac}=[0.0,0.6]$(inner ring)                                                         &0.4\\
               &                                &         & $C_{\rm sc-pl} = [10^{-7},1]$                                                  &$10^{-2}$(MCT)\\
               &                                &         &                                                                                &$3.5E-3$(HCT)\\
               &                                &         & $\theta_{\rm i}(^{\circ}) = [0,90]$                                            &60\\
         
\hline
\hline
\end{tabular}
\caption{In this table we summarize: the models names, abbreviations that we occasionally use in the parentheses and the corresponding papers (Col. 1), the type of torus and the morphology of gas or clump distribution (Col. 2), energy range set by the upper limit of \textit{NuSTAR} (Col. 3), their corresponding parameter and the prior range we use in our work for Bayesian analysis (Col. 4), the input values of these parameters we used in our data simulations (Col. 5) \label{tab:1}} 
\end{table*}

\subsection{The implemented model} \label{sub:model-implement}
\begin{figure}
	\includegraphics[scale=0.6]{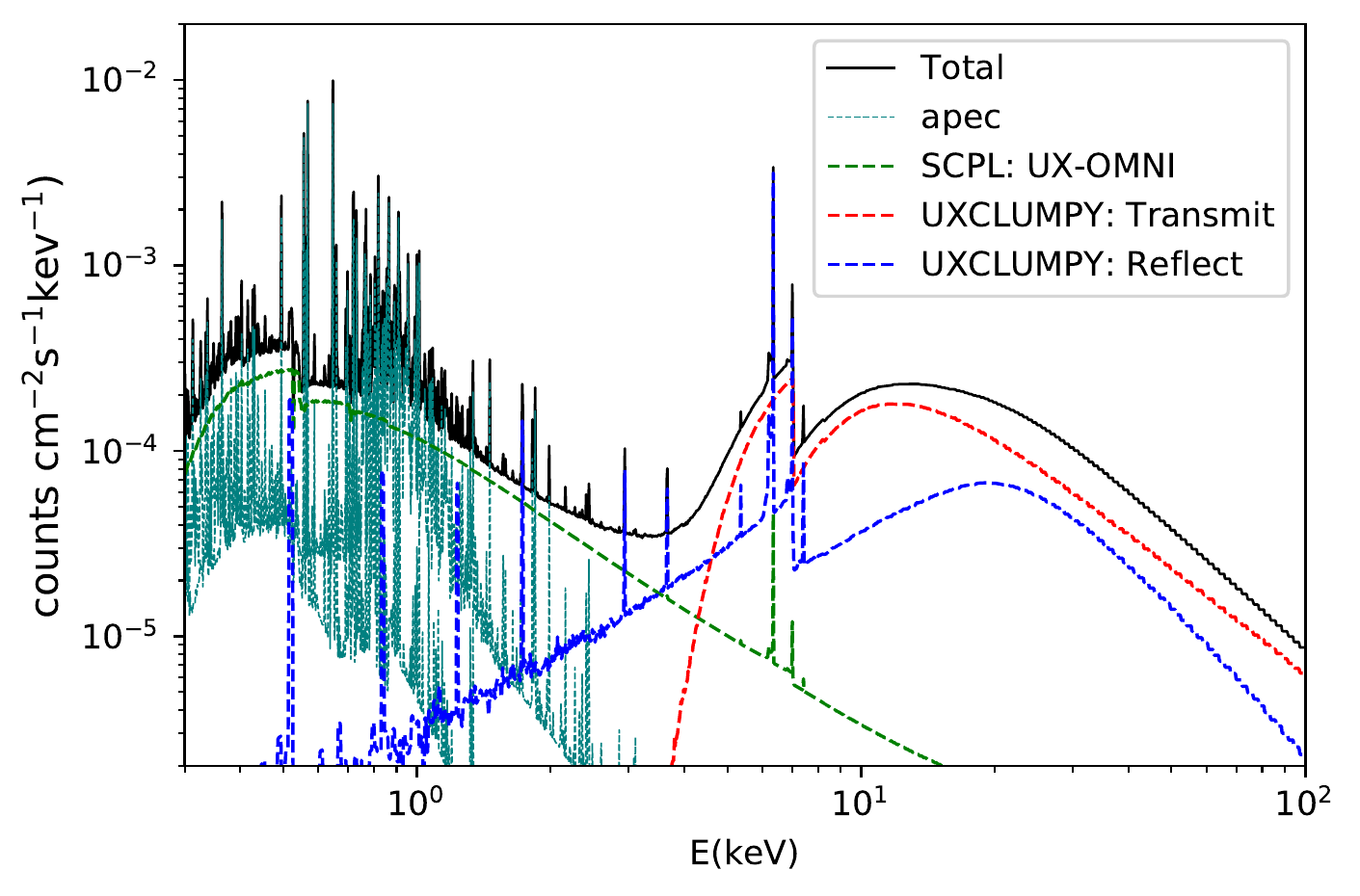}
	\caption{The plot illustrates the model components and how they contribute to the total spectrum. The plot is generated under UXCLUMPY with $N_{\rm H,los} = 100$, which implies a medium Compton-thick (MCT) regime, and it is clear that the zeroth-order continuum component has a higher flux contribution in the $E>5$~keV band. \label{fig-generic_model}}
\end{figure}

\begin{figure*}
\includegraphics[scale=0.66]{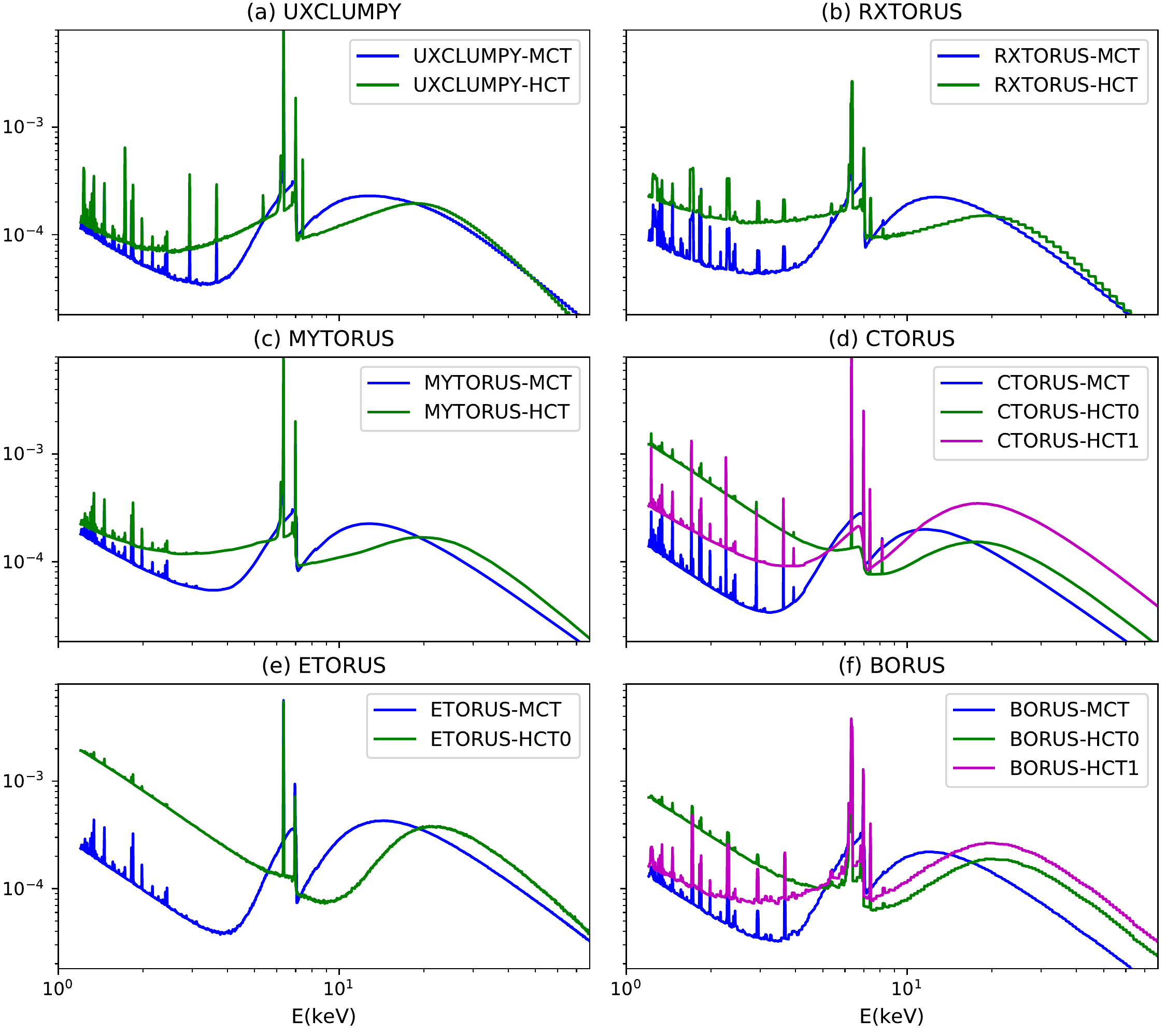} 
\caption{We plot overplot the spectrum of the input models for both the MCT and the HCT regime. HCT0 refers to the spectra where the CRH is comparatively weaker compared to the HCT1. For HCT0, the quantity $r = F_{\rm 2-10,torus}/F_{\rm 2-10,total} < 70\%$ whereas, for HCT1 it is $<70\%$. In our cases CTORUS and BORUS, $r_{\rm HCT1} \simeq 90 \%$ for both and $r_{\rm  HCT0} \simeq $ 65\% and 45\% for BORUS and CTORUS respectively.  \label{spectral_layouts}}
\end{figure*}

We simulated the data while implementing the same basic model components across all the torus models. For simplicity we choose one representative model, typical for X-ray obscured Seyferts in the nearby Universe. The model contains the following components:
\begin{itemize}

\item[(i)] Zeroth-order continuum:  The intrisic spectrum of the corona is assumed to be a simple power law or a cutoff power law (ICPL hereafter). In case of an obscured AGN, the ICPL from the AGN corona is attenuated by the column of the AGN torus, by the process of photoelectric absorption. This absorbed ICPL is referred to as the zeroth-order continuum \citep{murphy09} ($I_{\rm tor,t}(E,N_{\rm H,tor},\Gamma,\theta_{\rm i}, R , x_1,x_2,..)$, where $x_1,x_2,..$ are morphological parameters of the torus). The parameters that directly affect the zeroth-order continuum involve the parameters of the ICPL viz. $\Gamma$, $E_{\rm cut}$ and the $N_{\rm H,los}$ column of the torus. \textit{Throughout the paper $N_{\rm H,los}$ and $N_{\rm H,eq}$ are expressed in units of $10^{22}$ cm$^{-2}$ unless stated otherwise.} For the models MYTORUS, RXTORUS, CTORUS and UXCLUMPY, the zeroth-order continuum is published as FITS tables, whereas models like BORUS and ETORUS do not have a published FITS table for the same and hence one uses \texttt{ZTBABS $\times$ CABS\footnote{\href{https://heasarc.gsfc.nasa.gov/xanadu/xspec/manual/node234.html}{https://heasarc.gsfc.nasa.gov/xanadu/xspec/manual/node234.html}} $\times$ ZPOWERLAW} to simulate it.

\item[(ii)] Scattered/reflected continuum: In addition to photoelectric absorption, the incident X-ray photons undergoes Compton scattering (once or multiple times) in the torus material. This modifies the ICPL to give rise to the scattered continuum, $I_{\rm tor,R}(E,N_{\rm H,tor},\Gamma,\theta_{\rm i}, R , x_1,x_2,..)$, with distinct features like the CRH, fluorescent emission lines and a softer tail ($E<6$~keV) which are discussed below. For all the models the scattered component are published FITS tables. The scattered continuum contains the information about torus morphology. One aspect of the application of a torus model is the value relative normalization (denoted as $T/R$ hereafter) of the zeroth-order continuum with respect to the scattered continuum. The user has the freedom to either set $T/R=1$ or let $T/R$ vary. In a real scenario the scattered component is delayed with respect to the zeroth-order continuum due to light travel time. Thus, in such a case if the ICPL exhibits significant variability  $T/R \neq 1$ (see online malual\footnote{MYTORUS online manual: \href{http://mytorus.com/mytorus-instructions.html}{http://mytorus.com/mytorus-instructions.html}}) and should be kept free in the analysis.

\item[(iii)] Emission lines: The spectra also contains several fluorescent emission lines, of which the most notable is the iron emission line complex. The lines are accompanied by their corresponding Compton shoulders, which is the resultant of higher order down-scatterings. The treatment of lines are different in different models, e.g. MYTORUS provides only two emission lines, whereas UXCLUMPY, CTORUS, RXTORUS, BORUS provides numerous emission lines. ETORUS does not provide any lines, so the user needs to add an external model e.g. \texttt{zgauss} to simulate them.

\item[(iv)] Scattered power law/warm mirror reflection:  Many heavily absorbed AGN show a significant excess emission component in the soft band. This component might be the scattered component from diffuse gas which cannot be obscured by the torus \citep[e.g.][]{bianchi06,brightman14,buchner14,buchner19} or the scattered component from the volume filling interclump medium of a clumpy torus \citep{buchner19}. For simulation of data using MYTORUS, RXTORUS, BORUS, ETORUS and CTORUS the scattered power law or warm mirror (both referred as SCPL hereafter) was assumed to be a simple or cutoff power law. UXCLUMPY allows two setups for data simulation and fitting. In one setup, the user can use a simple \texttt{zcutoffpl} for the SCPL. We call this setup UX-ZCPL hereafter in the paper. In another setup, where the SCPL is the scattered emission from the interclump medium of a clumpy torus where scatterings of several orders are considered \citep{buchner19}, the user can use the published UXCLUMPY model component, \texttt{uxclumpy-cutoff-omni}. As a result, this component contributes its own weak reflection hump and Fe K$\alpha$ line. We call this setup UX-OMNI hereafter in this paper. The warm mirror spectrum thus emits its own mild CRH and a weak iron line. The ratio of the normalization ($C_{\rm sc-pl}$) of the SCPL with respect to the scattered component ranges from $10^{-3}$ to $10^{-2}$ times that of the scattered component for all simulations. In this work, we refer to the ratio of the scattered power law normalization to the torus normalization as $C_{\rm sc-pl}$.
   
\item[(v)] Soft X-ray thermal emission:  We can expect that X-ray spectra will almost always contain some potential contamination from host galaxy structures such as star-forming regions and point sources such as ULXs or XRBs, or cluster gas if one is studying an AGN in a cluster or X-ray-emitting shocks for radio-loud objects. To encorporate such soft contamination by significant emission lines caused by photonization \citep{kinkhabwala02}, we introduce two components of ionized gas emission. Correct modelling of the component would involve use of e.g. \textsc{photemis}\footnote{https://heasarc.gsfc.nasa.gov/xstar/docs/html/node106.html} from the package \textsc{xtardb}. This is however computationally prohibitive for the simulations we produce. Instead, we use collisionally ionized plasmas (\textsc{apec};(AtomDB version 3.0.9 \footnote{http://atomdb.org/} ). The different shape has no impact on how torus models are fit.

\item[(vi)] Galactic absorption: We use \texttt{TBABS} \citep{wilms2000} to simulate the Galactic absorption. We assume a value of $N_{\rm H,Gal}$ of $10^{21}$ cm$^{-2}$ for all the simulated data. 

\end{itemize}
The generic model of Seyfert-2 (see figure \ref{fig-generic_model}) implemented for all the available torus geometries and morphologies is: 
\begin{eqnarray}
    f(E)=
    && e^{-\tau(N_{\rm H,Gal},E)}[f_{\rm apec}(E,T_1)+f_{\rm apec}(E,T_2)+I_{\rm sc-pl}(E,\Gamma) \nonumber \\ 
    && + I_{\rm tor,T}(E,N_{\rm H,tor},\Gamma,\theta_{\rm i}, T ,x_1,x_2,..) \nonumber \\
    && + I_{\rm tor,R}(E,N_{\rm H,tor},\Gamma,\theta_{\rm i}, R , x_1,x_2,..) \nonumber \\
    && + I_{\rm tor,lines}(E,N_{\rm H,tor},\Gamma,\theta_{\rm i}, R , x_1,x_2,..)].
    \label{eq:1}
\end{eqnarray}
The regime of Compton-thick AGNs practically starts from $N_{\rm H,los} \simeq 100$ where the scattered continuum contributes a significant amount of flux to the spectra in addition to the zeroth-order continuum. If we move to higher values of the LOS column density, the zeroth-order continuum drops drastically and the scattered continuum starts dominating. At regime where $N_{\rm H,los} < 200$ the zeroth-order continuum is dominant in the $E>4$ keV band compared to the scattered continuum. In the regime of $N_{\rm H,los} > 200$ the scattered continuum starts overwhelming the zeroth-order continuum. So in context of this work, we phrase the regime with $N_{\rm H,los} <=200$ as medium Compton-thick (MCT) and $N_{\rm H,los}>200$ as heavy Compton-thick (HCT) regime. In the heavy Compton-thick regime we analyse two different classes of spectra for two models (CTORUS and BORUS). The classes are referred to as HCT0 and HCT1. The scattered power law (SCPL) is comparatively stronger in the HCT0 class compared to the HCT1 class in the 2--10~keV energy range (see Fig.~\ref{spectral_layouts}d and f). Its contribution is weaker with respect to the total flux ($F_{\rm 2-10,torus}/F_{\rm 2-10,total} \leq 70\%$) in the HCT0 case and the same ratio being stronger ($> 70\%$) in HCT1 case for each of two models, CTORUS and BORUS.

\subsection{Data Simulation}\label{sub:data-sim}
Data were simulated with the \textsc{fakeit} command in \textsc{xspec} for \textit{XMM--Newton} EPIC-pn \citep{jansen01} and \textit{NuSTAR} focus plane modules (FPM) A and B \citep{harrison13} based on the instrument responses. For our primary analysis, we assume joint simultaneous observations with both missions, simulating good exposure times after screening of 100 ks for \textit{XMM--Newton} ($t_{\rm XMM}$) and 50~ks per \textit{NuSTAR} FPM module ($t_{\rm NuSTAR}$); later, we explore the effects of using only one mission on the model fits. In our primary analysis, the 2--10~keV flux ($f_{2-10}$) of observed and absorbed flux for all models/instruments was $\sim 0.5$~mCrb. However, if specific cases require us to analyse a spectrum where 2--10~keV flux is significantly different from 0.5~mCrab i.e. $|f-0.5| \geq 0.09$ we specify the values of flux explicitly. For our primary analysis, we simulated data for 100~ks ($t_{\rm XMM}$) and 50~ks ($t_{\rm NuSTAR}$) on the \textit{XMM--Newton} EPIC-pn and \textit{NuSTAR} instruments respectively, when the analysis concerned the behaviour of the parameters. In a separate study (the results of which are presented in Section \ref{IM-bulk}), we intend to understand the distribution of statistical properties (e.g. the 90\% confidence region) of the parameters posteriors. To study the distribution of posterior properties we simulate 100 different spectral data sets under a select model for independent fitting using BXA.

For the purpose of studying the dependence of the Bayesian evidence ($Z$) on intrinsic flux of the object (see Section \ref{sec-6} below), we simulated 10 spectra with the 2--10~keV flux set to be $f=0.83/n$~mCrb, where $n$ runs from 1 to 10 with the exposure fixed at $t_{\rm XMM} = 100$~ks and $t_{\rm NuSTAR}=50$~ks.

\subsection{Fitting Methods}\label{sec-2.4}
We carry out Bayesian analysis on the simulated data sets for a given torus model to calculate the parameter posterior distribution. In this paper, we will frequently use common terminology used in simple (e.g. least squares) data-model fitting, but it should be clear to the reader, that despite using the terms that may also apply to simple data-model fitting, the process is Bayesian analysis unless explicitly mentioned otherwise. Specifically, we use nested sampling using \textsc{multinest} \citep{skilling04,feroz09} implemented via BXA and PyMultinest packages \citep{buchner14} for X-spec version 12.10.1f \citep{xspec96} for posterior calculation. The issues related to convergence e.g. getting trapped in a local maximum of the likelihood, in standard Goodman--Weare Markov Chain Monte Carlo (GW-MCMC) calculations are not present in nested sampling algorithms. Another convergence problem we have with MCMC is assumption on the length of the chain, which might not be enough for convergence and might require multiple burn-ins and several sequential runs. Nested sampling algorithms, including \textsc{multinest}, attempt to map out all of the most probable regions of parameter sub-space: it maintains a set of parameter vectors of fixed length, and removes the least-likely point, replacing it with a point with a higher likelihood, and thus shrinking the volume of parameter space in each calculation. The convergence of a \textsc{multinest} run is automatic and does not require any initial assumption like the chain length or burn-in. \textsc{multinest} calculates a parameter called evidence $Z$ which is the probability of data ($D$) given a hypothesis ($H$), $P(D|H)$, mathematically expressed as:
\begin{equation}
Z = \int \mathcal{L}(D|\Theta,H) \pi(\Theta|H) d^N \Theta
\label{eq:5}
\end{equation}
From a probalilistic point of view the evidence($Z$) is the probalility of data given the hypothesis, $P(D|H)$. There are two contributors to the final value of evidence ($Z$) or $P(D|H)$ namely, the likelihood function ($\mathcal{L}$) and the final posterior ($\pi$) (see equation \ref{eq:5}). The likelihood probability here quantifies the deviation of the data from the model and the posterior distribution $\pi(\Theta|H)$ quantifies the volume of the parameter space which signifies the effective implementation of Occam's razor criterion. The evidence value is a combined effect of these data-model deviations and Occam's razor. We use \textsc{multinest} version 3.10 with default arguments (400 live points, sampling efficiency of 0.8) set in BXA version 3.31. For all model fits we use uniform or log-uniform as our initial prior parameter distribution. We use \texttt{cstat} \citep{cash79} in XSPEC as the likelihood function for all model fits. The \texttt{cstat} likelihood is mathematically expressed as $\mathcal{L}(\Theta,D|H) = \prod_{i} \frac{(f_i(\Theta) t_{\rm exp})^{D_i}}{D_i!}e^{-f_i(\Theta) t_{\rm exp}}$, where $t_{\rm exp}$ is the exposure time for a given instrument, $f_i(\Theta)$ is the theoretical count rate for a model $H$ folded with instrument response and $D_{i}$ the count rate at the i-th channel or energy bin. 

We consider the best-fitting value of the parameters to be the median or 0.5th quantile of the posterior distribution and the lower and upper bound of errors are the 0.05th and 0.95th quantile of the distribution, unless stated otherwise. Our prescription of best fit value and bound on the error works best when the the posterior distribution is a monomodal distribution. Interpretation of the results for the posteriors which show multiple maxima i.e. having multiple solutions, depend on the specific situation concerning the model and the data set. Posterior distributions which are strongly non-gaussian (i.e. has a minima, are uniform or those that converge towards the edge of the prior range or any other `irregularities') will be termed in general as \textit{irregular} posteriors hereafter. Interpretation of \textit{irregular} posteriors will be contextual. Our prescription of best-fitting value and the errors might will be less reliable in such cases and a separate treatment might be required. 
To quantitatively express the goodness of the recovery of the parameters, we define some simple mathematical parameters to quantify constraints, discrepancy, and the quantitative estimate on the accuracy of the returned values of the parameters.
\begin{eqnarray}
R_{90} = q_{0.95} - q_{0.05}\\
\delta = \frac{0.5 R_{90}}{q_{0.5}} \\
\Delta q = |q_{0.5}-q_{\rm input}| \\
r_{\rm q} = \frac{2 \Delta q}{R_{90}}
\end{eqnarray}
Here $q_{r}$ is the $r$-th quantile of the posterior\footnote[1]{ The 2D posterior figures are produced using the package \texttt{corner} \citep{corner16}. All other plots are made using the package \texttt{matplotlib} \citep{matplotlib07} } distribution and thus $R_{90}$ quantifies the spread of the posteriors in the 90\% confidence region. $\delta$ is the proportional error about the median, thus the lesser the value of $\delta$ the lesser the spread of the 90\% confidence region about the median. $\Delta q$ quantifies the deviation of the median value of the posterior distribution from the input. $r_{\rm }$ is the ratio of the deviation ($\Delta $) to the 90\% confidence range ($R_{90}$). $r = 0$ indicates no discrepancy, whereas $r \sim 1$ indicates that the discrepancy is almost equal to the average statistical error. If $r > 1$ then the discrepancy is more than the average statistical error. In this work we take 1.5 to be the critical value for $r$, thus we term a parameter recovered if $0 \leqslant r \leqslant 1.5$.  

\section{RESULTS: INTRAMODEL FITS}\label{IM-fits}
In this section we present the results for the intramodel fits (IM-fits hereafter) for joint \textit{XMM Newton} EPIC-pn and \textit{NuSTAR} FPM-A and B data. We grouped all simulated spectra to 30 cts bin$^{-1}$ for our primary analysis.

\subsection{Medium Compton-thick regime (MCT)}
\begin{figure*}
\gridline{\fig{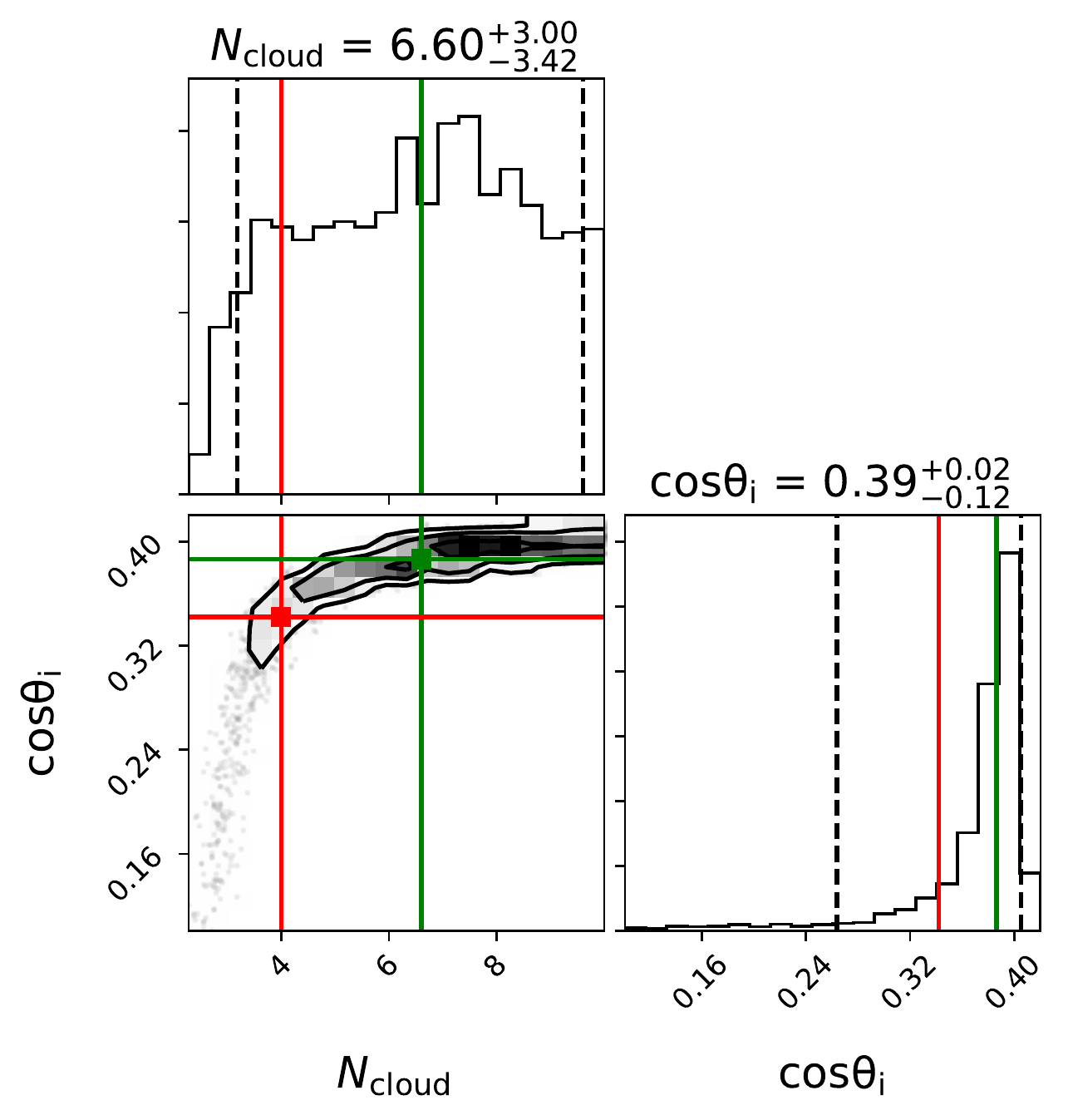}{0.32\textwidth}{(a)}
          \fig{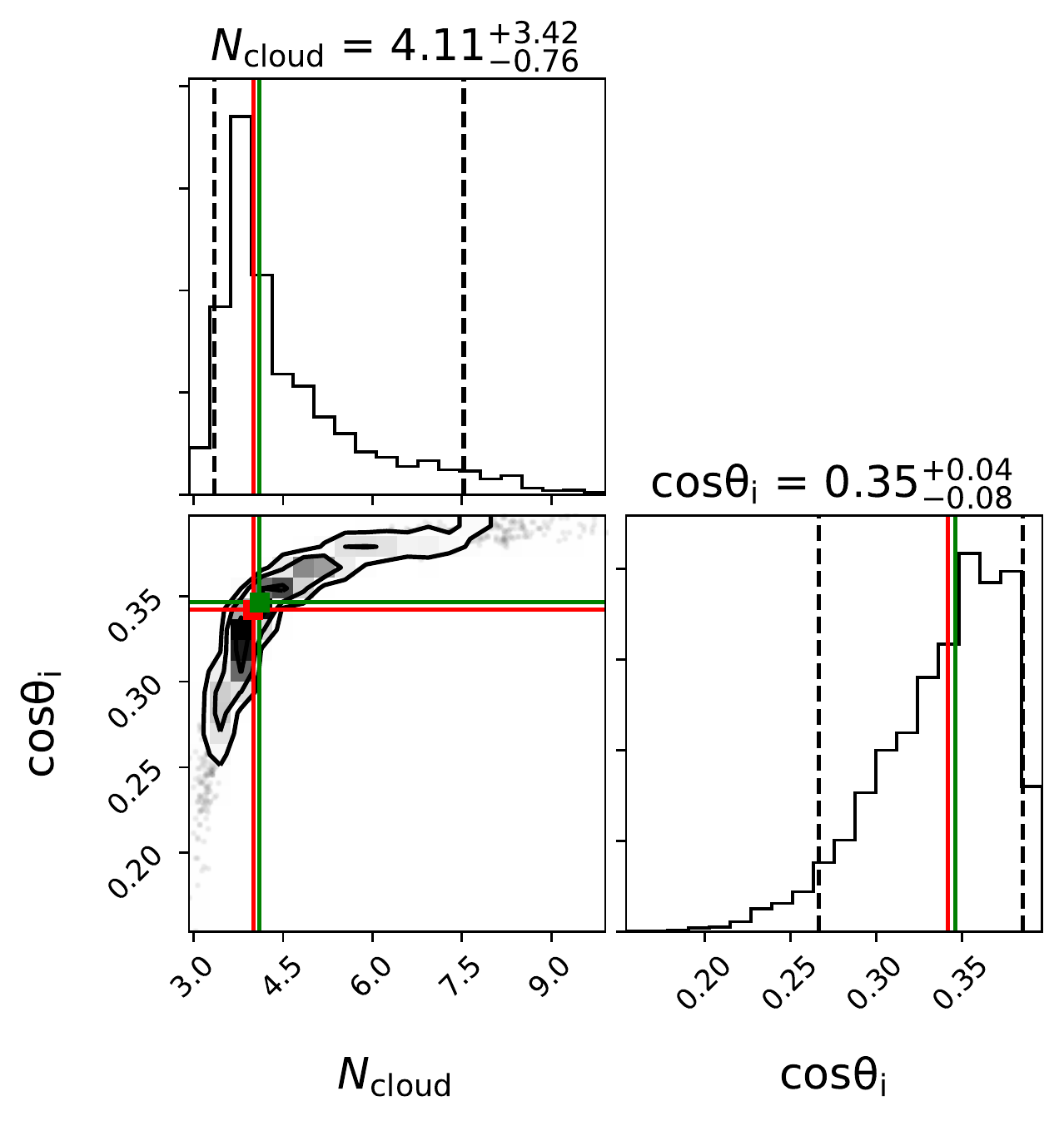}{0.32\textwidth}{(b)}
          \fig{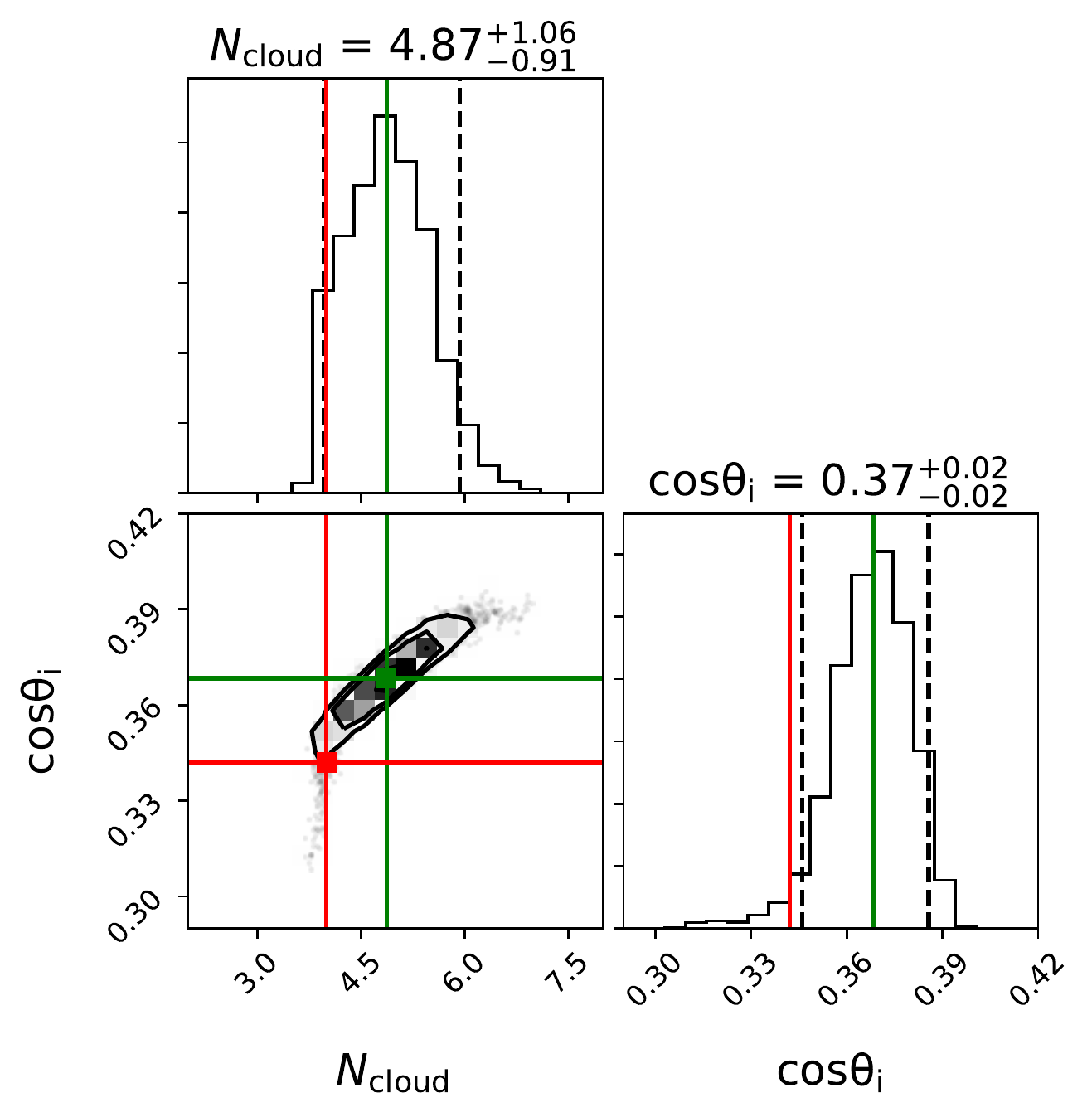}{0.32\textwidth}{(c)}
          }
\caption{Contours for CTORUS IM analysis in the (a)MCT: input $N_{\rm H,los} = 100$, (b)HCT0: input $N_{\rm H} = 500$, (b)HCT1: input $N_{\rm H,los} = 500$. The lines or cross-hairs coloured red denote the input value and those coloured green denote the median value calculated from the posterior distribution. The influence on $N_{\rm cloud}$ posterior on $\theta_{\rm i}$ due to its poor localization is noticeable in (a). But constraints improve considerably in (b). \label{ncl-mct-hct}}
\end{figure*}

In this subsection we discuss the results of intramodel fits to the simulated data in the MCT regime. Some models adopt LOS column density ($N_{\rm H,los}$) like BORUS, UXCLMPY and CTORUS. Other models adopt column density at equator ($N_{\rm H,eq}$) like MYTORUS, RXTORUS, and ETORUS. During simulation of the data, for the models which adopt $N_{\rm H,los}$ as a parameter, we have set $N_{\rm H,los}=100$. For the models that use $N_{\rm H,eq}$ as a parameter, we have set $N_{\rm H,eq} = 150$ where the implied value of $N_{\rm H,los}$ will be lower than $150$, for $\theta_{\rm i}$ situations where the line of sight interferes with the torus dust structure (e.g. $\theta_{\rm i} > \theta_{\rm o}$).
\begin{itemize}

\begin{figure*}
\includegraphics[scale=0.30]{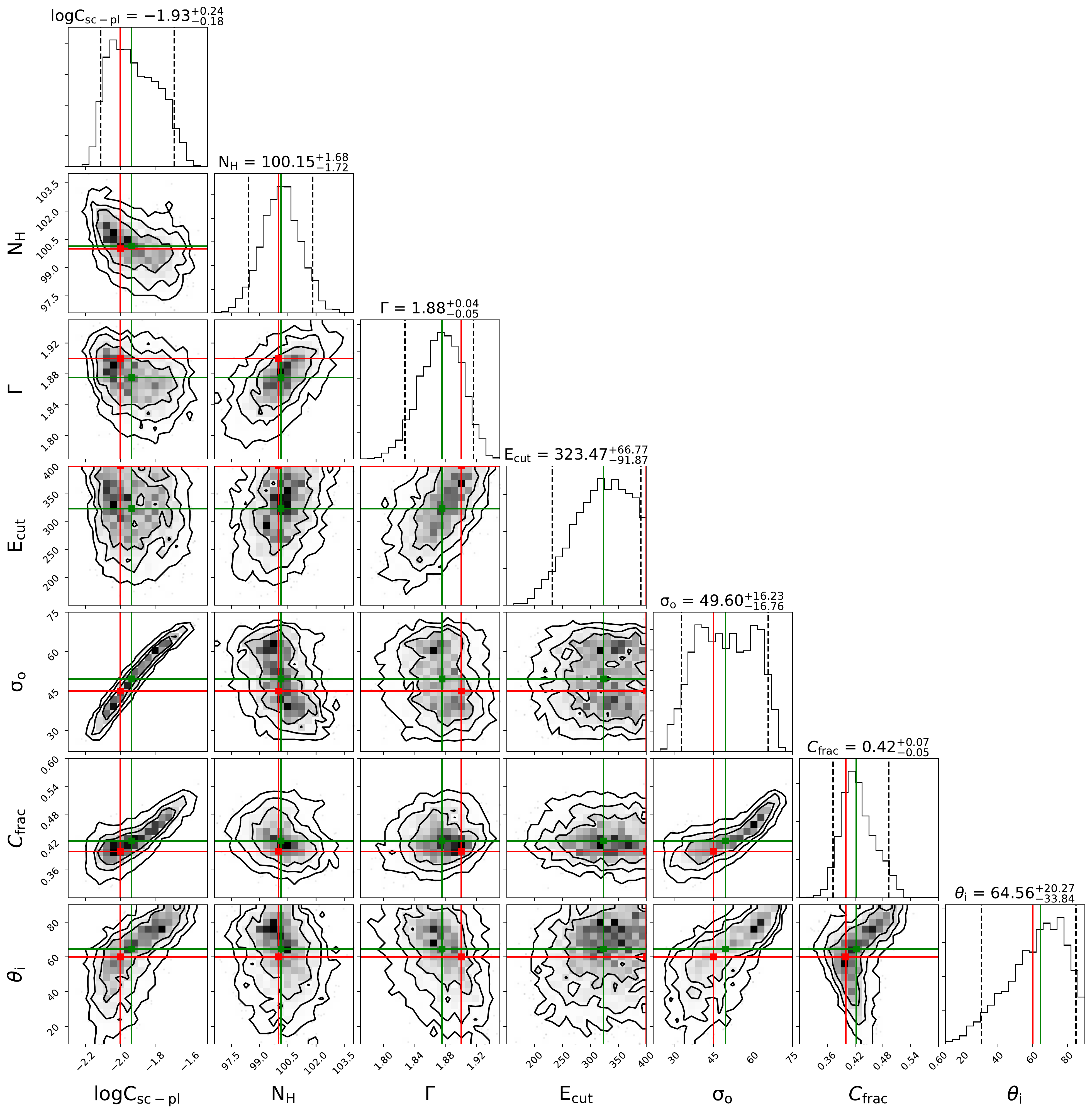}
\caption{Contours for UXCLUMPY IM analysis in the MCT regime, with $N_{\rm H,los} = 100$ as input. The degeneracy or correlation between the parameters (log$C_{\rm sc-pl}$--$\sigma_{\rm o}$--$C_{\rm frac}$) with extremely skewed contours can be noticed. The lines or cross-hairs coloured red denote the input value and those coloured green denote the median value calculated from the posterior distribution. \label{uxcl-mct}}
\end{figure*}

\item[(i)] \textbf{Column density ($N_{\rm H}$):} Both LOS column delsity ($N_{\rm H,los}$) in the models BORUS, CTORUS and UXCLUMPY and equatorial ($N_{\rm H,eq}$)  column density in MYTORUS, RXTORUS and ETORUS are recovered with very tight constraints for all models. Over all the tested cases and models, $N_{\rm H,los}$ is recovered, with maximum value of $\delta$=0.032 and the average values of $\delta$=0.023. In this regime the zeroth-order continuum contributes to the total observed flux more significantly than the CRH ($\left[\frac{F_{\rm T}}{F_{\rm R}} \right]_{\rm 3-100 keV} \sim$ 1.8 -- 5.0). The Fe K absorption edge and the rollover at $E<5$~keV are the distinct features of a zeroth-order continuum determining $N_{\rm H,los}$. The geometry of the system does not play a very significant role in determining the $N_{\rm H,los}$ in the MCT regime, both in terms of tightness of the posteriors and parameter recovery. This holds in the cases where the spectrum is dominated by the zeroth-order continuum. 

\item[(ii)] \textbf{Photon index ($\Gamma$):} We have simulated the data with SCPL-$\Gamma$ (scattered power law) and ICPL-$\Gamma$ (coronal power law) set to the same value ($\Gamma=1.9$) for all input models. When fit with the SCPL and ICPL $\Gamma$ tied, $\Gamma$ was recovered for all models with values of $R_{90}$,$\delta$ and $r_{\rm q}$ never exceeding 0.12, 0.03 and 0.92 respectively. Despite the degeneracy or correlation between $\Gamma$ and $E_{\rm cut}$ for (BORUS and UXCLUMPY), the constraints on $\Gamma$ are tight. We also test the case where the values of SCPL- and ICPL-$\Gamma$ were not tied for three models. The trends were model dependent in this case. For models CTORUS and MYTORUS (no $E_{\rm cut}$) constraints on ICPL-$\Gamma$ were unchanged. However, the behaviour of SCPL-$\Gamma$ worsened in both the cases as posteriors widened ($R_{90} = 0.27$ for CTORUS, $R_{90}=0.11$ for MYTORUS). In UXCLUMPY for data simulated in the UX-OMNI setup, $R_{90}$ value of ICPL-$\Gamma$ worsened for both the UX-OMNI and UX-ZCPL setup due absence of leverage from the SCPL and a simultaneous varying $E_{\rm cut}$, in the uncoupled configuration. The trends in SCPL-$\Gamma$ remained unchanged in the UX-OMNI setup. In the UX-ZCPL setup, the recovered value of SCPL-$\Gamma$ was flatter ($1.83 \pm 0.05$) and was just barely recovered ($R_{90}=0.1$, $r_q = 1.4$). For this case, the SCPL-$\Gamma$  lowered to account for the excess flux (noticeable in the $2 <E< 5 $~keV band) in the data (simulated in the UX-OMNI setup). Thus it can be concluded that the behaviour of SCPL-$\Gamma$ can affect constraints on ICPL-$\Gamma$ when they are tied, more so when they have dissimilar values.

\item[(iii)] \textbf{High-Energy Cutoff ($E_{\rm cut}$):} For most of the cases, we simulated the data sets with $E_{\rm cut}$ set to the highest possible limit allowed by the models, to have a minimum influence of $E_{\rm cut}$ on the other features (e.g. CRH) in the $E< 78$~keV band. When we keep $E_{\rm cut}$ free, the posteriors of $E_{\rm cut}$ are broad and the constraints are wide in both BORUS and UXCLUMPY. We also simulated two data sets in BORUS and UXCLUMPY with $E_{\rm cut,input}$ for the simulated data set to a value of 200~keV. In this case, the constraints on $E_{\rm cut}$ are recovered with regular monomodal posteriors. \\
\textbf{$\Gamma$--$E_{\rm cut}$ dependence:} For models which have $E_{\rm cut}$, $\Gamma$ shows a dependency on $E_{\rm cut}$ when it is kept free, this is clear from the skewness in $\Gamma$-distribution when $E_{\rm cut}$ is a variable parameter in the fits. For BORUS when SCPL-ICPL are uncoupled, posteriors of $\Gamma$ in the case of $E_{\rm cut}$ frozen fits are more asymmetric and have higher values of $R_{90}$ and $\delta$. For UXCLUMPY, the same situation holds, but the availability and dependence on SCPL down to 0.3 keV in \textit{XMM--Newton}, diminishes the effect of the $E_{\rm cut}$ in the fit where SCPL and ICPL are coupled. The overall trends suggest that $E_{\rm cut}$ might influence constraints on ICPL-$\Gamma$. 

\item[(iv)] \textbf{Angle of Inclination ($\theta_{\rm i}$):} The angle of inclination ($\theta_{\rm i}$) to the torus axis is determined the flux and the shape of the scattered continuum reaching the observer. Thus, the models will use the CRH and/or the low energy($E<6.0$ keV) tail of the scattered continuum, to estimate $\theta_{\rm i}$. For UXCLUMPY, in the UX-OMNI setup, in the \textit{XMM--Newton}+\textit{NuSTAR} joint fits, the value of $\delta$ for $\theta_{\rm i}$ is 0.41, indicating that it is not a well constrained parameter. For CTORUS, $\theta_{i}$ is skewed towards the higher angles. This is an artefact of the correlation of $\theta_{\rm i}$ with $N_{\rm cloud}$ (Fig.~\ref{ncl-mct-hct}a). The $N_{\rm cloud}$ posterior which covers the whole allowed range (see pt. vii) causes the $\theta_{\rm i}$ posteriors to get skewed to higher angles. In models based on bi-conical cutouts like BORUS and ETORUS $\theta_{\rm i}$ is constrained well if the torus opening angle ($\rm cos \theta_{\rm o}$ or equivalently $C_{\rm F,tor}$ of BORUS) constrained well. For cases when $\theta_{\rm o}$ shows trends of wider posteriors or bimodality $\theta_{\rm i}$ shows the same trend. Additionally bimodality in $\theta_{\rm i}$ was seen if $\theta_{\rm o}$ shows bimodality. In doughnut-based models like MYTORUS and RXTORUS, $N_{\rm H,eq}$ is the input parameter and $N_{\rm H,los} = f(N_{\rm H,eq},\theta_{i},c/a)$. Because of the dependence $N_{\rm H,los}$ is a well constrained parameter in this regime and thus can help narrow down the favourable $\theta_{\rm i}$ region in the parameter space. Further localization and limiting the $N_{\rm H,eq}$ -- $c/a$ -- $\theta_{i}$ degeneracy is provided by features of the scattered continuum. We find that the $N_{\rm H,eq}$ -- $\theta_{\rm i}$ correlation is strong when $c/a$ is frozen. However, when $c/a$ is kept free (in RXTORUS), the $N_{\rm H,eq}$--$\theta_{i}$ correlation weakens and the $c/a$ -- $\theta_{i}$ correlation strengthens.

\item[(v)] \textbf{Relative normalization between the transmitted and the reflected components ($T/R$):} In many models, the normalization of the zeroth-order continuum and the scattered continuum can was kept equal i.e. $T/R = 1$ or kept free as discussed in Section \ref{sub:model-implement}. We simulated data for two cases, (A) $T/R$ $> 1$ for CTORUS ($T/R$ = 1.8) and MYTORUS ($T/R$ = 1.5) (B) $T/R$=1 for all models. We kept $T/R$ free for only MYTORUS, CTORUS and ETORUS. For all these models, in both case-A and case-B the parameters were recovered. For MYTORUS, CTORUS and ETORUS ($r$,$\delta$) were found to be [A:(0.77, 0.08),  B:(0.12, 0.06)], [A:($8 \times 10^{-3}$, 0.154), B:(0.09, 0.17)], [B:(0.9, 0.29)], respectively.

\item[(vi)] \textbf{Relative Normalization of the SCPL ($C_{\rm sc-pl}$):} For the models MYTORUS, RXTORUS, BORUS, ETORUS and CTORUS $C_{\rm sc-pl}$ are recovered with $\delta \sim $ 0.067--0.54. For UXCLUMPY, in the UX-OMNI, $C_{\rm sc-pl}$ posteriors were recovered with values of $\delta$ and $r$ equal to 0.54 and 0.28 respectively. $C_{\rm sc-pl}$ is strongly correlated with $\sigma_{\rm o}$, $\theta_{\rm i}$, and $C_{\rm frac}$ of the inner ring. In the UX-ZCPL setup the torus $C_{\rm sc-pl}$ is recovered with $\delta \simeq 0.1$, $C_{\rm sc-pl}$ is correlated with $\sigma_o$ and $\theta_{\rm i}$, but not with $C_{\rm frac}$.

\begin{figure*}
\gridline{\fig{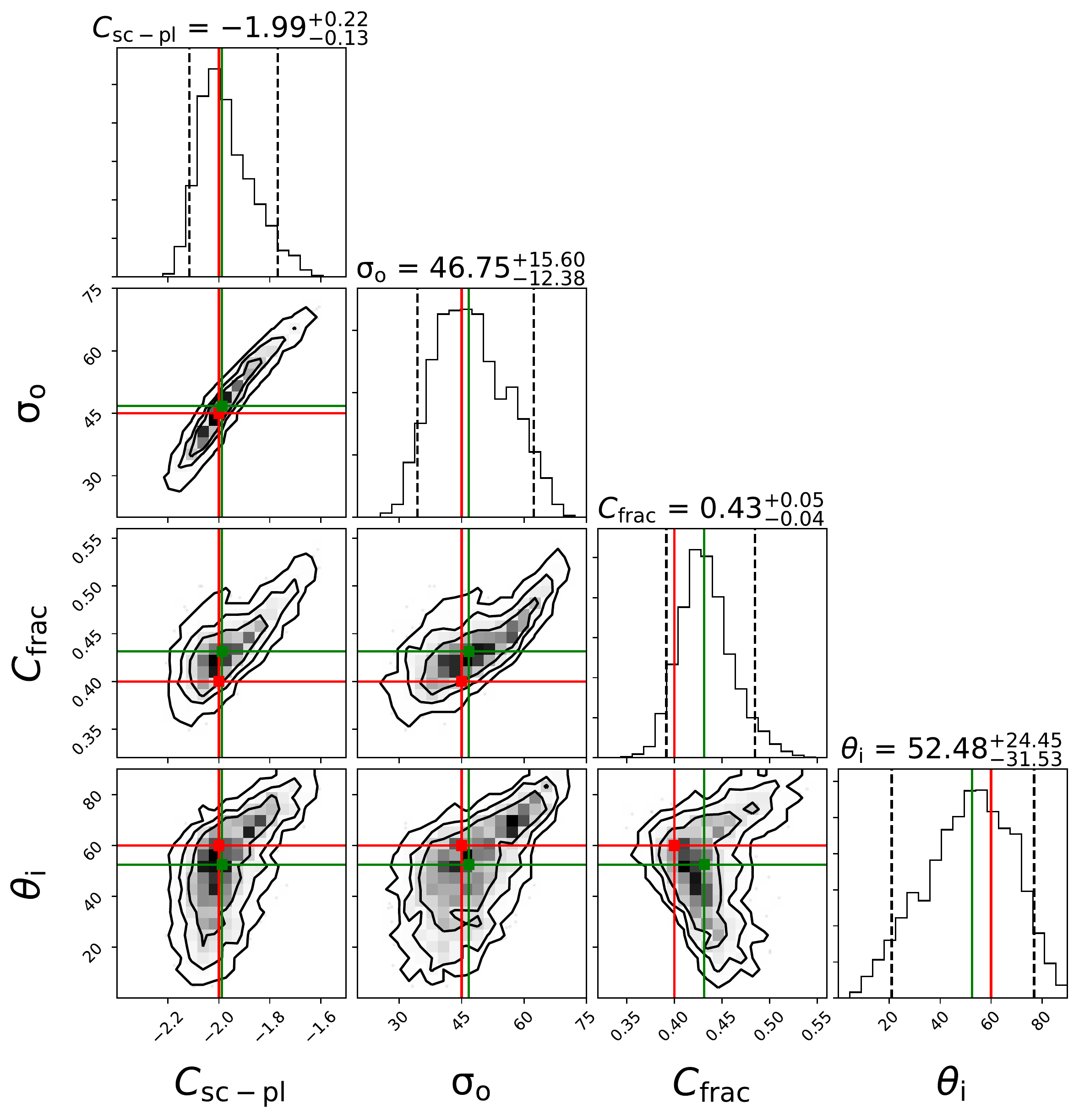}{0.38\textwidth}{(a)}
          \fig{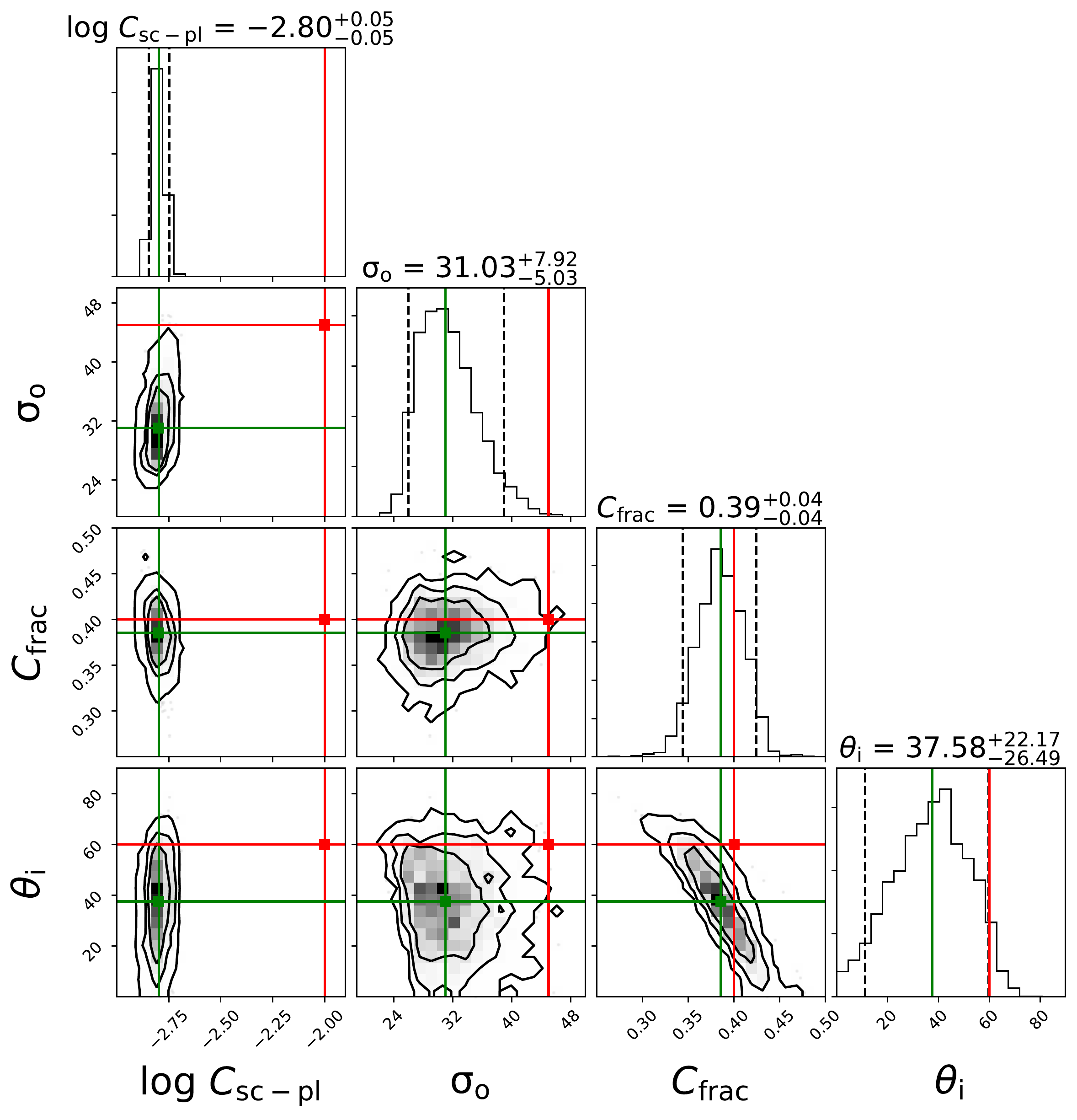}{0.38\textwidth}{(b)}
          }
\caption{Contour plots indicating the correlation of the parameters $C_{\rm sc-pl}$ (relative normalization of warm mirror or scattered power law), $\sigma_{\rm o}$, $C_{\rm frac}$ and $\theta_{\rm i}$ of UXCLUMPY, when data simulated under UX-OMNI is fit with the a)UX-OMNI  b)UX-ZCPL setup. The red and the green lines in the histograms and cross-hairs in the contour plots mark the input and the median values calculated from the distributions respectively. The nature of the correlation between the parameters changes depending on the SCPL model. In this case we keep the temperatures and normalizations of the apec components and the instrumental constants frozen at their input values. \label{zcpl-scpl}}
\end{figure*}

\item[(vii)] \textbf{Parameters of Torus morphology:}
	\begin{itemize}
	\item[(a)] \textbf{$c/a$ of RXTORUS:} The posteriors returned from the fits are regular monomodal distributions. The $c/a$-parameter shows strong correlation with the posteriors of $N_{\rm H,eq}$ and $\theta_{\rm i}$. $c/a$ in RXTORUS determines the opening angle of the doughnut, hence the $c/a$ -- $\theta_{\rm i}$ correlation. The constraints on $c/a$ are tight as $\delta \simeq 0.086$ and $r \simeq 0.024$.
	
	\item[(b)] \textbf{Opening angle ($\theta_{\rm o}$):} For ETORUS $\theta_{\rm o}$ is a direct input. However, in BORUS the opening angle is given by $\theta_{\rm o}= \cos^{-1} C_{\rm F,tor}$ for a contiguous torus. $\theta_{\rm o}$ shows different trends in ETORUS and BORUS. For ETORUS, $\theta_{\rm o}$ returned very broad posteriors with $\delta = 0.2$ when $T/R$ was frozen to the input value ($T/R=1$) and worsened to $0.35$ when $T/R$ was free, with strong $\theta_{\rm o}$ -- $T/R$ correlation. For BORUS ($T/R=1$ frozen in all cases) however the constraints are better with $\delta \simeq 0.02$. The parameter constraints are determined from the shape of the lower energy($E<6.0$~keV) tail of the reflected continuum and the shape of the CRH. 
	
	\item[(c)] \textbf{Tor-sigma ($\sigma_{\rm o}$) of UXCLUMPY:} When data simulated in the UX-OMNI setup and was fit it with the same, $\sigma_{\rm o}$ is recovered with $\delta \simeq 0.33$. The 90\% confidence region was spread over almost $\sim  65\%$ of the prior range for the given flux level (see Fig.~\ref{uxcl-mct}). The most distinct influence $\sigma_o$ and has is the soft band tail of the reflection component. $\sigma_o$ along with $C_{\rm frac}$ also affects the relative height of the CRH to the FeK$\alpha$ absorption edge. Additionally, the zeroth-order continuum is strongly present in the 7--25 keV band and dominates the reflection component. This invokes a strong $\sigma_{\rm o}$--$C_{\rm frac}$--$C_{\rm scpl}$ correlation and a large spread in the posteriors of $\sigma_{\rm o}$ (see Fig.~\ref{zcpl-scpl} or \ref{uxcl-mct}). When the UX-OMNI data set is fit with the UX-ZCPL setup, $\sigma_o$ was not recovered. While we get a lower value of $\delta \simeq 0.2$, the value of $r$ increases to 2.2, indicating a discrepancy. The SCPL excess was adjusted by decreasing $\sigma_o$, which increased the amount of the scattered component to make up for the comparatively higher contribution from the warm mirror. Fig.~\ref{zcpl-scpl} shows the comparison of the UXCLUMPY morphological parameters, for the UX-OMNI and UX-ZCPL cases. 
		
	\item[(d)] \textbf{Covering fraction ($C_{\rm frac}$) of inner ring of UXCLUMPY:} In the UX-OMNI setup, there exists a strong $C_{\rm frac}$--$\sigma_o$ degeneracy (see Fig.~\ref{uxcl-mct}). The posteriors were recovered with $\delta \simeq 0.14$. But strong degeneracy with $\sigma_{\rm o}$ can result in discrepancy with the input. The constraints on the $C_{\rm frac}$ are derived from the energy band of the scattered continuum between 7--25 keV. In the MCT regime, the dominance of the zeroth-order continuum over the scattered continuum in the 7--30 keV energy band results in poor constraints on $C_{\rm frac}$. In the UX-ZCPL setup, $C_{\rm frac}$ was recovered with $\delta \simeq 0.08$. From the contour plots (see Fig.~\ref{zcpl-scpl}) it can be seen that the nature of $\sigma_o$--$C_{\rm frac}$ correlation is different for the two setups. UX-OMNI shows a $\sigma_o$--$C_{\rm frac}$ Pearson's correlation coefficient of $0.75$ ($N_{\rm samples} \simeq 3718$), whereas for UX-ZCPL setup we get $\sigma_o$--$C_{\rm frac}$ correlation of $\sim 0.14$ ($N_{\rm samples} \simeq 2682$) with a confidence of greater than 99.99\%.
	
	\item[(e)] \textbf{$N_{\rm cloud}$ of CTORUS:} The input value of $N_{\rm cloud}$ of data simulated with CTORUS was set to 4. We simulated the data sets for $T/R = 1.0$ and 1.8 respectively (see point no. iv). The values of $R_{90}$ and $\delta$ for both the spectra with $T/R=1.8$ and $T/R = 1.0$ are similar ($R_{90} \sim$ 5 and $\delta \sim $ 0.6). The constraints on $N_{\rm cloud}$ are determined from two features of the reflected continuum: the $E<6$ keV tail and the CRH. The reflected continuum has lower flux by a factor of $\sim 1/5$ ($T/R = 1.8$) to $\sim 1/3$ ($T/R=1$) than the zeroth-order continuum in the 3--100~keV band in this regime; constraints from CRH become overwhelmed, leading to flattened posteriors of $N_{\rm cloud}$. Additionally, the soft tail ($E<6$~keV) of the scattered component is weaker compared to other models, leading the SCPL to be more dominant in this band, weakening the constraints on $N_{\rm cloud}$.
	
	\item[(f)] \textbf{Iron abundance ($A_{\rm Fe}$) of BORUS:} The zeroth-order continuum is simulated using the cutoff power law absorbed through \texttt{zTBABS}$\times$\texttt{CABS} and abundance is fixed at 1. Thus, technically there is no way to consistently simulate a generic spectrum with any value of $A_{\rm Fe}$ by linking the zeroth-order continuum-$A_{\rm Fe}$ to the scattered component-$A_{\rm Fe}$. Thus, to make the spectrum consistent, for the data simulated under BORUS: $A_{\rm Fe} = 1.0$ for the scattered component. $A_{\rm Fe}$ was recovered ($r_q \simeq 0.2$) with $\delta \simeq 0.05$ by measuring the features e.g. depth of the Fe K absorption edge, the height of the FeK$\alpha$ line of only the reflected component. 
	  
	\end{itemize}
\item[7.] \textbf{\textsc{apec} components:} \textsc{apec} parameters were genereally recovered when the lowest energy bin of the data is lower than the \textsc{apec} temperature included in the spectrum. Thus with this knowledge we kept the \textsc{apec} temperatures and normalizations frozen at the input values, except for a medium Compton-thick fit of UXCLUMPY where all the \textsc{apec} parameters are kept free.

\end{itemize}

\begin{figure*}
\includegraphics[scale=0.30]{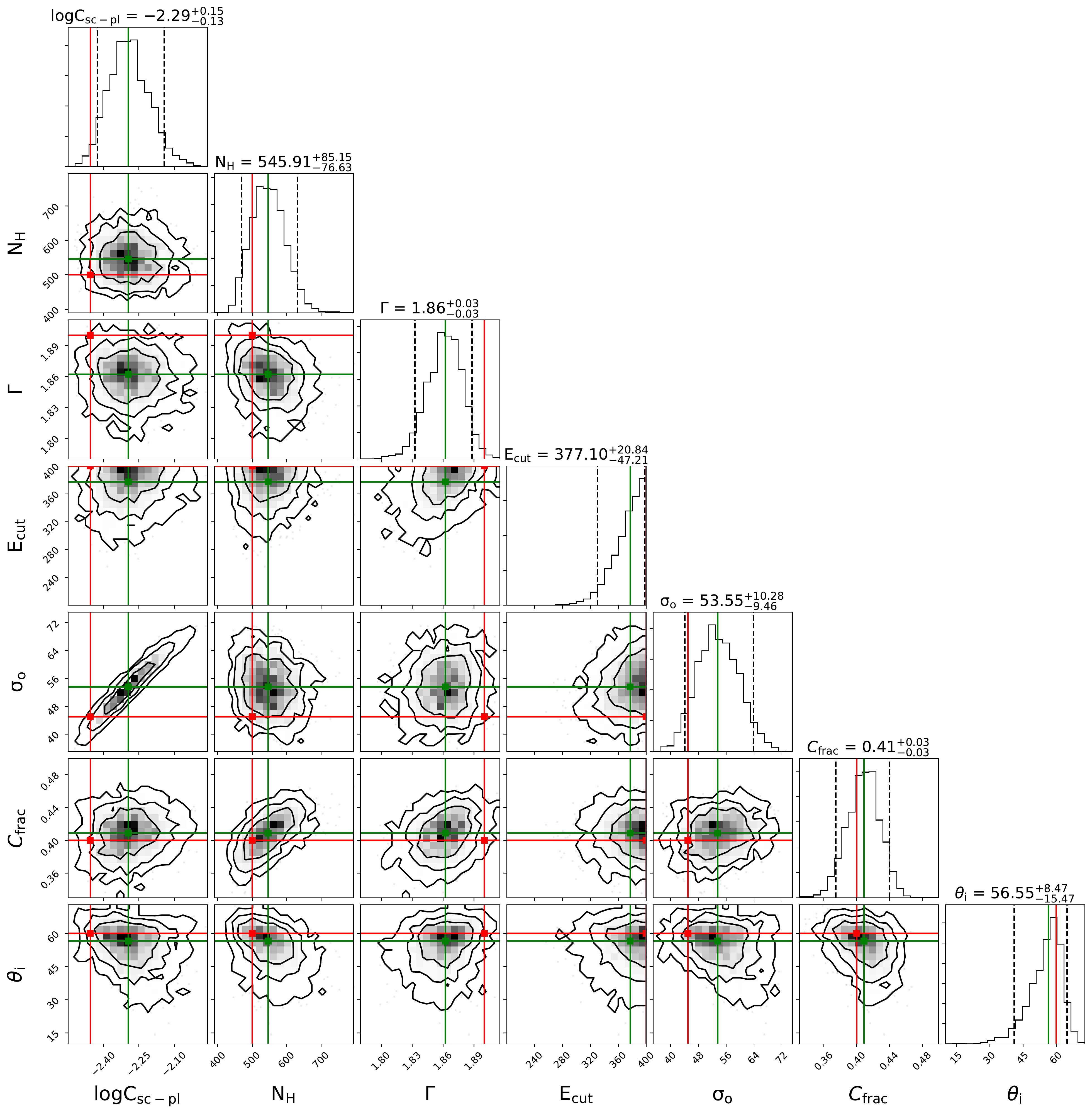}
\caption{Contour plot for UXCLUMPY IM analysis in the HCT regime, with $N_{\rm H,los} = 500$ as the input. A noticeable difference exists in the parameter posteriors in terms of (1) better posterior localization/constraints on parameters of morphology and (2) lesser level of parameter degeneracy exhibited by the nature of the 2D contours compared to the MCT regime (Figure \ref{uxcl-mct}). The lines or cross-hairs coloured red denote the input value and those coloured green denote the median value calculated from the posterior distribution.  \label{uxcl-hct}}
\end{figure*}

\begin{figure*}
\includegraphics[width=12cm,height=6cm]{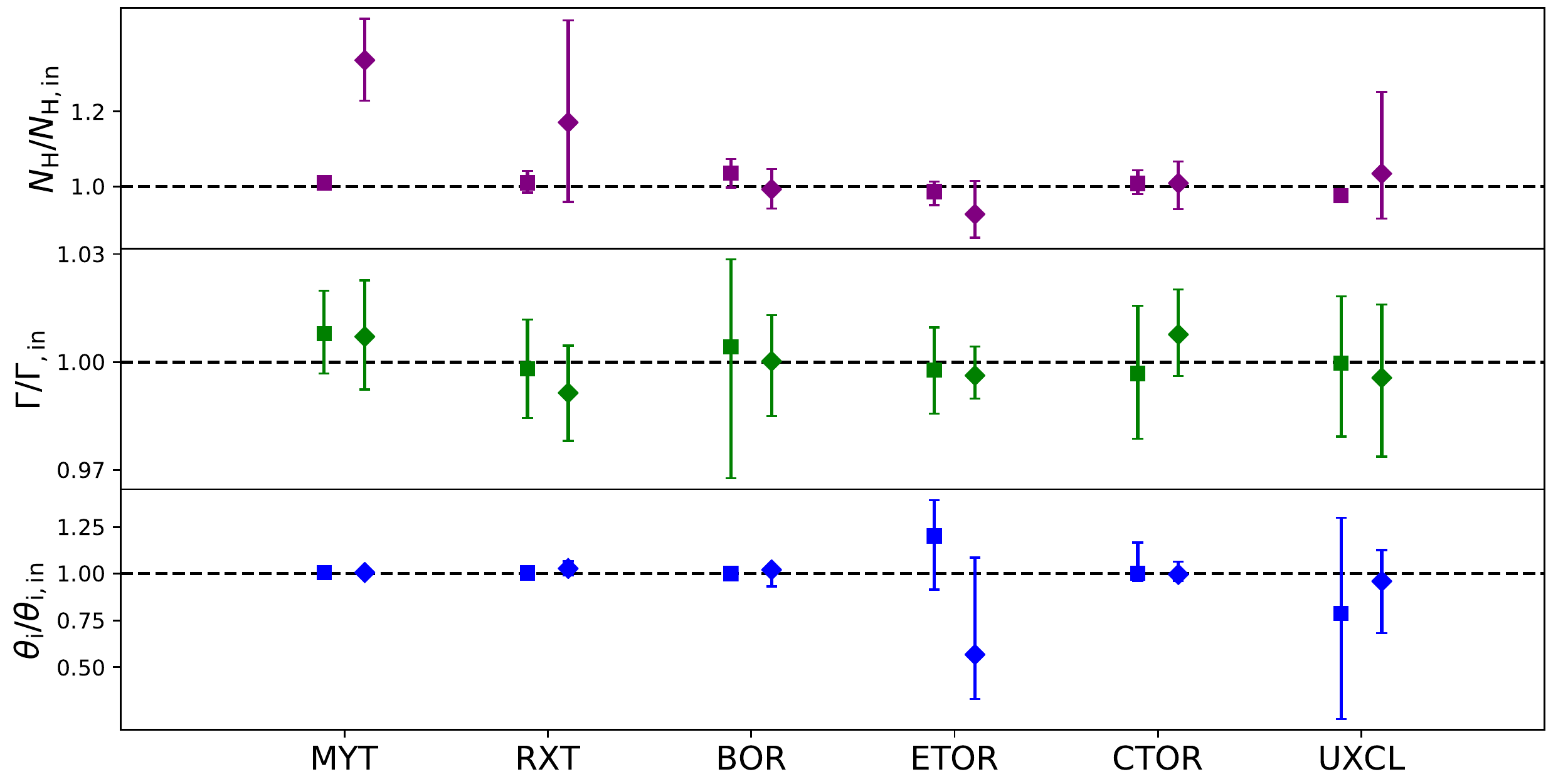} 
\caption{We plot the posterior medians and 90\% confidence ranges for the parameters that are common across all the models. The diamond markers are for the HCT regime and the square markers for MCT. Noticeable difference in the trends of parameters across regimes for different models can be seen e.g. $N_{\rm H,los}$ is well constrained in the MCT regime for all models. However, the extent of worsening of the constraints on $N_{\rm H,los}$ in the HCT regime is dfferent for different models. \label{commonpar}}
\end{figure*}

\subsection{Heavy Compton-thick regime}
\begin{figure*}
\includegraphics[scale=0.55]{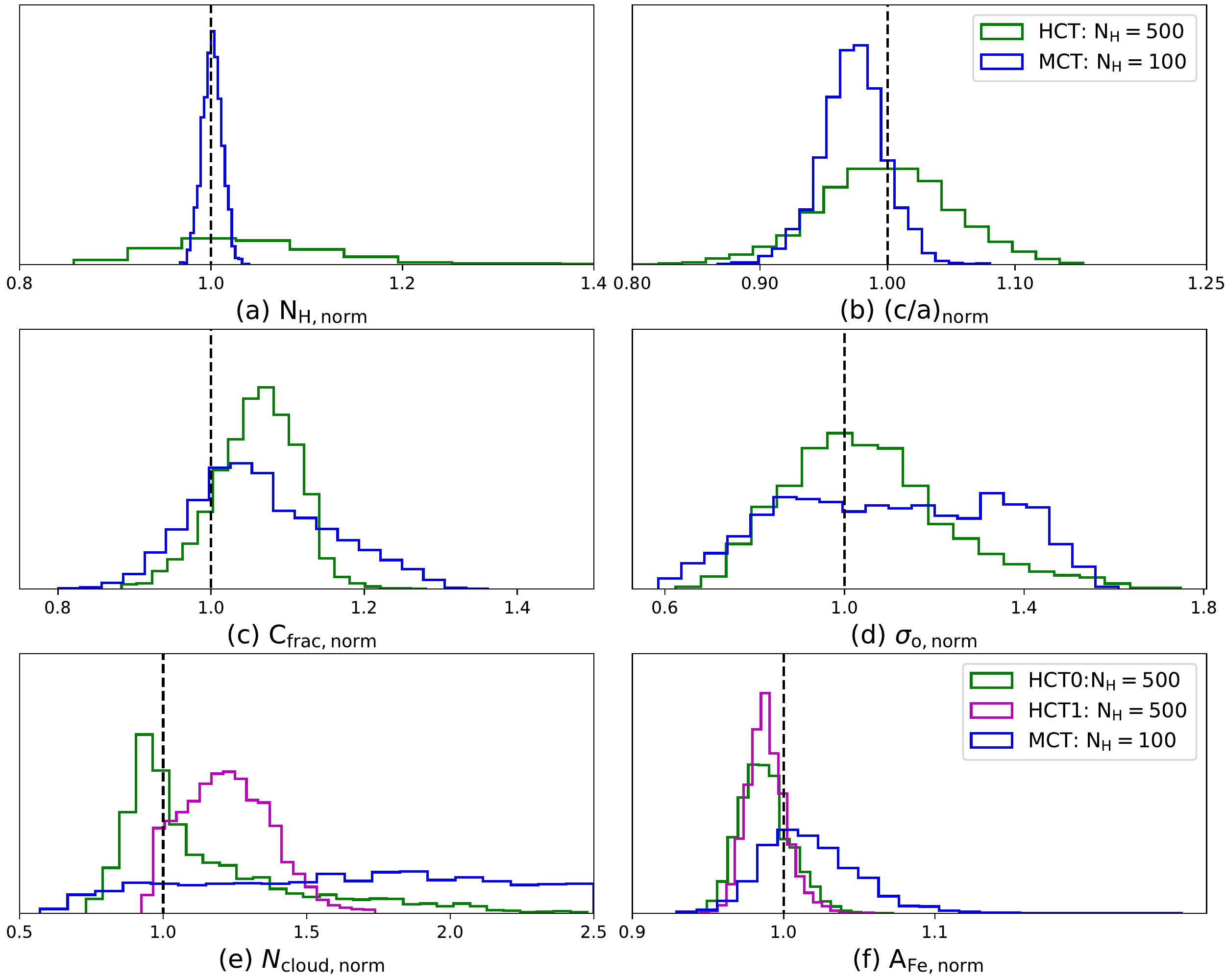}                   
\caption{Overplot of the distributions of $ X_{\rm norm} = X_{\rm posterior}/X_{\rm input}$ for MCT and HCT regime, where $X_{\rm posterior}$ represents the point corresponding to the posterior distribution and $X_{\rm input}$ is the input value used for data simulation. Noticeable differences exist in the distribution of the selected parameters in the MCT and HCT regimes. e.g. subplot (a) shows that the probability distribution of $N_{\rm H}$(note: in linear space) is more localized in the MCT regime. Posterior distribution of other parameters get more localized in the HCT regime by virtue of higher flux contribution from the scattered continuum, as shown in subplots (b)--(f), \label{mct-hct}}
\end{figure*}

The photon index $\Gamma$ did not show significant changes in the constraints. However several other parameters showed a significant change in terms of recovery, degeneracy, and constraints in the HCT regime compared to the MCT regime. We summarize the list of values of $\delta$ and $r$ for all parameters and models for the MCT and HCT regime in Table \ref{tab:2}. We discuss the results of fits for data simulated in the HCT regime, with $N_{\rm H,los} \simeq 500$ and 2--10~keV flux values similar to that of the MCT for all models. The most significant difference of this regime is the zeroth-order continuum gets overwhelmed by the scattered continuum ($\left[ \frac{F_{\rm T}}{F_{\rm R}}\right]_{\rm 3-100 keV} \simeq 0.1$). We discuss the results from the HCT regime in this subsection.
\begin{itemize}
\item[(i)] \textbf{Column density ($N_{\rm H}$) :} The constraints on $N_{\rm H,los}$ worsen (top panel of Figure \ref{commonpar} and Figure \ref{mct-hct}a) in this regime because of the decrement of $\left[ \frac{F_{\rm T}}{F_{\rm R}}\right]_{\rm 3-100 keV}$ by an order of magnitude. $\delta$ ranges between 0.03 to 0.21 (average $\delta \simeq 0.09$, table \ref{tab:2}) in the HCT-regime for IM fits of all models. For estimating the constraints on $N_{\rm H,los}$ in the HCT regime (in the absence of the zeroth-order continuum) the the features  of the scattered continuum e.g. the relative height of the CRH to the Fe K-edge, the $E<5$~keV tail and the shape of the Compton hump play the most significant role.

\item[(ii)] \textbf{Angle of inclination ($\theta_{\rm i}$) :} In all cases (except ETORUS) the constraints in $\theta_{\rm i}$ either improved or remained same as the MCT in the HCT regime. This improvement can be attributed to the improved prominence of the scattered continuum resulting in better constraints in the morphological parameters which affects $\theta_{\rm i}$ posteriors. For example in CTORUS, posteriors of $N_{\rm cloud}$ in the HCT1 case improved, thus improving $\delta$ for $\theta_{\rm i}$ to 0.05 (Fig.~\ref{ncl-mct-hct}). For UXCLUMPY the parameter posteriors of $\sigma_{\circ}$ and $C_{\rm frac}$ improved, leading to an improvement in the value of $\delta$ for $\theta_{\rm i}$. Although for the case of RXTORUS and MYTORUS, the constraints do not improve but remained almost the same as that of the MCT regime. However, ETORUS showed worse constraints on $\theta_{\rm i}$, when $T/R$ is free but better when $T/R$ if frozen (see next bullet point on $\theta_{\rm o}$). In the HCT regime, as an average generic trend posterior on $\theta_{\rm i}$ are better constrained, with an average value of $\delta$ lower than the MCT regime for the respective models (Table \ref{tab:2}).

\item[(iii)] \textbf{Parameters of morphology:}
\begin{itemize}
\item[(a)] \textbf{$c/a$ of RXTORUS:} The constraints on $c/a$ improved (Fig.~\ref{mct-hct}b) in this regime. The parameter was recovered, with $\delta \simeq 0.04$ which is nearly half of that calculated in the MCT regime. The value of $r_q$ on the other hand became worse (increased to 0.63 from 0.02 in the MCT regime), (see Table \ref{tab:2}) compared to the MCT regime. One reason for this change is the decrease in the spread of 90\% confidence region ($R_{90}$) in the HCT regime, thus increasing the ratio ($\Delta q/R_{90}$).    

\item[(b)] \textbf{Opening angle ($\theta_{\rm o}$) or $\cos \theta_{\rm o}$ :} For ETORUS the values of $\delta$ improved mildly to 0.14 (from 0.2 in the MCT) when $T/R$ is kept frozen at input in the HCT regime. When $T/R$ is kept free, the constraints worsen ($\delta \simeq 0.8$) compared to MCT and are spread over the whole prior range. The soft tail ($E<6$ keV) of the scattered continuum of ETORUS is considerably lower compared to the other models in the HCT regime. For BORUS we observe bimodality, with two disjoint peaks in the posteriors. The major peak corresponds to the input and the minor peak is significantly discrepant from the input. We can evaluate the quantiles independently for each peaks. When compared to the MCT regime the values of $R_{90}$ and $\delta$ are similar. $R_{90}$=1.51 and 2.12 for the major and minor peak respectively. The ratio of the number of samples corresponding to major and minor is $N_{\rm M}/N_{\rm m} \simeq 17$. It may be possible to reject the minor peak as it corresponds to lower probalility. The bimodality in $\theta_{\rm o}$ induces bimodality in $\theta_{\rm i}$ and $C_{\rm sc-pl}$.

\item[(c)] \textbf{Tor-sigma ($\sigma_{\rm o}$) of UXCLUMPY:} Because improved prominance of the scattered-continuum in the HCT regime, the influence of the effects of $\sigma_{\rm \circ}$ on 6--40 keV becomes evident unlike in MCT. In the HCT when the $\sigma_{\rm o}$ decreases, it increases the level of the soft band ($E<6$ keV) tail, it increase the warm mirror component if the spectra are simulated in the UX-OMNI setup and it approaches the Compton-thick ring and that introduces a mild depression in the 7--30 keV band similar to that of the Compton-thick inner ring. When $\sigma_{\circ}$ approaches 90$^{\circ}$ the system approaches an almost $4 \pi$ covered system, diminishing the soft band significantly. These features can potentially establish stronger constraints as observed (comparison of Fig.~\ref{uxcl-hct} with \ref{uxcl-mct} and posterior overplot in Fig.~\ref{mct-hct}(d)) in the HCT compared to MCT as the scattered continuum is the dominant part of the spectrum.

\begin{table*}
\begin{tabular}{|l|l|l|l|l|l|l|l}
\hline
\hline

 Parameter                      &       & $\delta$     &             &     &  $r_{q}$ &     &\\
\hline
                                & MCT    &  &  HCT     & MCT   &    & HCT  &\\
\hline  
 $N_{\rm H,los}$                &  0.02  &  & 0.09     & 0.44  &    & 0.51 &\\
 $\Gamma$                       &  0.018 &  & 0.014    & 0.35  &    & 0.46 &\\
 $\theta_{\rm i}$               &  0.18  &  & 0.13     & 0.60  &    & 0.49 &\\
 cos$\theta_{\rm o}$ of BORUS   &  0.035 &  & 0.021    & 0.84  &    & 0.23 &\\
 $\theta_{\rm o}$ of ETORUS     &  0.31  &  & 0.47     & 0.5   &    & 0.67 &\\
 $\sigma_{\rm o}$ of UXCLUMPY   &  0.32  &  & 0.2      & 0.37  &    & 0.37 &\\
 $C_{\rm frac}$ of UXCLUMPY ring&  0.13  &  & 0.08     & 0.51  &    & 0.84 &\\
 $N_{\rm cloud}$ of CTORUS      &  0.64  &  & 0.26     & 0.47  &    & 0.8  &\\
 $A_{\rm Fe}$ of BORUS          &  0.059 &  & 0.026    & 0.41  &    & 0.49 &\\
 $c/a$ of RXTORUS               &  0.086 &  & 0.043    & 0.024 &    & 0.63 &\\
      
\hline
\hline
\end{tabular}
\caption{In this table we compare the average values of $\delta$ and $r_{\rm q}$ for selected parameters in the MCT and HCT regimes. The averages are taken over all analysed sub cases of all intramodel fits of the models that have the given parameter. For most parameters of morphology, we get a better constraints except $\theta_{\rm o}$ of ETORUS which worsens and cos$\theta_{\rm o}$ of BORUS which does not show any significant improvement in the HCT regime.}
\label{tab:2}
\end{table*}

\item[(d)] \textbf{Covering fraction ($C_{\rm frac}$) of inner ring of UXCLUMPY:} The constraints improve (Fig.~\ref{mct-hct}-c) because of improved prominance of the features of the scattered continuum. This improved the posteriors of $\sigma_{\rm \circ}$ and $C_{\rm sc-pl}$, leading to elimination of the heavy correlation seen otherwise in the MCT regime. We found that $\delta$ for $C_{\rm frac}$ improves from 0.14 in the MCT regime to 0.08 in the HCT regime.

\item[$\blacktriangleright$] \textbf{$N_{\rm cloud}$ of CTORUS:} For the HCT regime we test two cases:
\begin{itemize}
\item[(1)] The SCPL is stronger and contributes much more flux (HCT0) than the scattered continuum from the torus (green curve in Fig.~\ref{spectral_layouts}d) in the 2--50 keV band $\left[ f_{\rm torus}/f_{\rm total} \right]_{2-10} = 0.45$. This case represents objects with higher diffused/scattered emission. For this case we get $\delta = 0.5$ and $R_{90} \simeq 4.18$. 
\item[(2)] The SCPL is weaker than the soft tail of the scattered continuum from the torus ( HCT1, magenta curve in Fig.~\ref{spectral_layouts}d): $\left[f_{\rm torus}/f_{\rm total} \right]_{2-10} \simeq 0.90$. In this case the CRH is much stronger and represents cases where the diffused or scattered emission is weak but the torus-continuum is strong. For this case we get $\delta = 0.2$ and $R_{90} \simeq 1.97$. This is because of the presence of the strong soft-tail from the torus scattered continuum.  
\end{itemize}
For both cases, the posteriors of $N_{\rm cloud}$ (Fig.~\ref{ncl-mct-hct}b and \ref{mct-hct}e) improve in the HCT regime, as the CRH and $E<4$ keV tail become dominant over the zeroth-order continuum.

\item[(e)] \textbf{Iron abundance ($A_{\rm Fe}$) of BORUS:} The constraints of BORUS improved in the HCT regime (Fig.~\ref{mct-hct}f), $\delta \simeq 0.02$. The problems introduced because of the inconsistency of the iron-abundance in the zeroth-order continuum and the scattered continuum would be minimal or absent in the HCT regime, because of the low value of $\left[ \frac{F_{\rm T}}{F_{\rm R}}\right]_{\rm 3-100 keV}$, which almost eliminates the zeroth-order continuum, thus enabling us to calculate $A_{\rm Fe}$ from the features of the dominant scattered continuum only. We found that $\delta$ improves to 0.02 in the HCT-regime from 0.07 in the MCT regime (see Table~\ref{tab:2}).

\end{itemize}
\end{itemize}

\begin{figure*}
\includegraphics[scale=0.60]{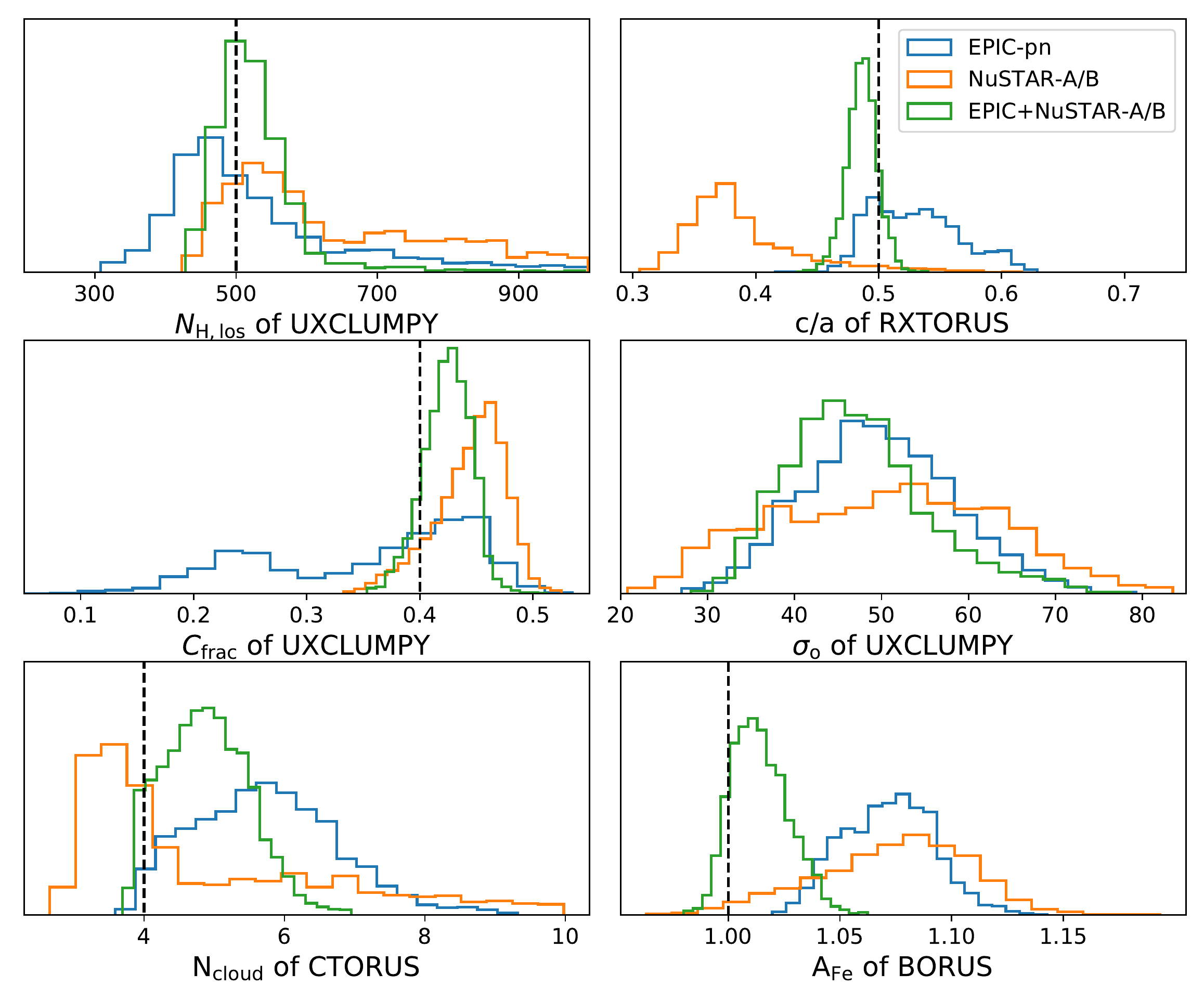}                
\caption{We compare the 1-D posterior distributions calculated from \textit{XMM--Newton}, \textit{NuSTAR} and joint \textit{XMM--Newton+NuSTAR} fits in this figure. We notice significant dependence in the nature of the posterior distributions on the instrument and their combinations. These posteriors prove that for decent constraints on some parameters e.g. $C_{\rm frac}$ of the inner ring of UXCLUMPY, \textit{NuSTAR} data is a requirement, whereas for $N_{\rm cloud}$ in the HCT regime joint data from \textit{XMM--Newton} and \textit{NuSTAR} is required. \label{instruments}}
\end{figure*}

\begin{table*}
\begin{tabular}{lllllllll}
\hline
\hline
Parameter        & \multicolumn{3}{c}{$R_{90}$} & \multicolumn{3}{c}{$\delta$} & \multicolumn{2}{c}{$r_{q}$} \\
\hline
                 & MCT        & HCT      &  \%-overlap  & MCT         & HCT   & per cent-overlap     & MCT        & HCT       \\
\hline
$\Gamma$         & $0.080^{+0.013}_{-0.015}$ & $0.081^{+0.018}_{-0.017}$ & $\sim$100\% &$0.021^{+0.004}_{-0.004}$  &$0.022^{+0.005}_{-0.005}$ & $\sim$100\% & $0.40_{-0.38}^{+0.85}$  & $0.63_{-0.50}^{+0.86}$ \\
$N_{\rm H,los}$  & $3.24^{+0.29}_{-0.31}$    & $143^{+251}_{-57}$        & $\sim$0\% & $0.016^{+0.0016}_{-0.0017}$ &$0.138^{+0.211}_{-0.048}$ & $\sim$0\%   & $0.56_{-0.52}^{+0.69}$ & $0.32_{-0.28}^{+0.76}$ \\
$\sigma_{\rm o}$ & $26^{+8}_{-6}$            & $20^{+5}_{-4}$            & 43\%      & $0.27^{+0.09}_{-0.08}$      &$0.22^{+0.03}_{-0.03}$    & 38\%        & $0.38_{-0.34}^{+0.91}$ &  $0.49_{-0.43}^{+0.82}$ \\
$C_{\rm frac}$  \\
of UXCLUMPY ring & $0.094^{+0.039}_{-0.016}$ & $0.07^{+0.008}_{-0.013}$  & 10\%      & $0.11^{+0.04}_{-0.02}$ & $0.09^{+0.01}_{-0.02}$        & 32\%        & $0.53_{-0.48}^{+0.66}$  & $0.47^{-0.44}_{+0.73}$ \\
$\theta_{\rm i}$ & $52^{+13}_{-14}$          & $29^{+9}_{-7}$            & 6\%       & $0.47^{+0.60}_{-0.19}$ & $0.28^{+0.2}_{-0.1}$          &53\%         & $0.41_{-0.40}^{+0.78}$ &  $0.53_{-0.45}^{+0.90}$  \\
\hline
\hline
\end{tabular}
\caption{In this table we show the distribution of the parameters characterizing posteriors distributions, calculated from the bulk fitting of 100 spectrum. The `overlap percentage' (abbreviated as per cent-overlap) is the fraction of the values of $R_{90}$ or $\delta$ from the wider distribution that overlaps with values lesser than the 0.95th quantile of the narrower distribution. The narrower which for the $\Gamma$ and $N_{\rm H,los}$ is MCT regime and for $\sigma_{\rm o}$, $C_{\rm frac}$ and $\theta_{\rm i}$ is the HCT regime. These values give an idea about how much the statistical errors ($R_{90}$, $\delta$ and $r_{q}$) might vary due to random process associated with the data generation.}
\label{tab-bulk-fits}
\end{table*}

\subsection{Bulk simulations}\label{IM-bulk}
The properties of the posterior distributions e.g. $R_{90}$, $\delta$ and $r_q$ are affected by statistical processes involved in data generation. This implies that for a given theoretical spectral model, the data sets generated in each case will be different and Bayesian analysis performed on these data sets will return posteriors of parameters with median values, $R_{90}$, $\delta$ and $r_q$ different from each other. However these distributions will follow a characteristic distribution for each parameter. To verify this we perform bulk simulations of spectra for the UXCLUMPY in the MCT and HCT regime. We limit ourselves to simulating 100 spectrum for each case, due to the limits of computational power and time. We perform intramodel fits on each of these spectra and calculate posteriors in each case.

We calculate the distribution of certain properties of the posteriors (e.g. $R_{90}$) of different regimes. We find several properties of the distribution which are consistent with our observations from the single fits. $N_{\rm H,los}$ remains well constrained in the MCT regime compared to the HCT regime. The distributions of $R_{90}$  and $\delta$ determined from the bulk fitting of the spectra of the torus-morphological parameters are localized to the lower values for the HCT regime compared to those of the MCT regime which is spread towards comparatively higher values. However, the distribution of $R_{90}$ from the two regimes show overlap with each other as shown in table \ref{tab-bulk-fits}. Thus the values of $R_{90}$ and $\delta$ estimated in the single spectral fit in the previous sections are parts of these distributions of respective parameters. That is for a given parameter the probability that the value of $R_{90}$ (or $\delta$) calculated from a single fit will remain enclosed in the respective values and the errorbar on the parameter quoted in Table \ref{tab-bulk-fits} is 90\%. The implies that $\delta$ values quoted in table \ref{tab:2} which reflect the statistical errors, should not be considered as absolute, as they can be affected by the statistical process involved in the data generation, implying a chance of getting slightly different values of $R_{90}$ and $\delta$ for data sets simulated for the same model under same conditions. The extent of variation in the $R_{90}$ and $\delta$ depends on the regime and the parameter of interest. Additionally table~\ref{tab:3} shows that a decrease in $R_{90}$ or $\delta$ values results in a decrease in the uncertainty on the same. In the context of intramodel fits, this means, as a parameter becomes better constrained, the statistical error on that parameter becomes less influenced by statistical processes involved in data simulation or generation.

Finally, as a test of validation of the errors for individual fits, we compare the average value of $R_{90}$ ($\langle{R_{90}\rangle} = \Sigma_i (q_{\rm 0.95} - q_{\rm 0.05})$) over all the individual fits (given a parameter) and compare it with the extent of 90\% confidence region of the distribution of median values of the posteriors($R_{\rm 90,med}$). We find that values of $\langle{R_{90}\rangle}$  and values    of $R_{\rm 90,med}$ are very similar for both regimes. For example, values of $\langle{R_{90}\rangle}$ and $R_{\rm 90,med}$ for $\Gamma$ are $\sim$0.81 and 0.60 for MCT regime and 0.81 and 0.64 for HCT regime respectively. The values of $\langle{R_{90}\rangle}$ and $R_{\rm 90,med}$ for $\sigma_{\rm o}$ are, respectively, 26.14 and 20.55 for the MCT regime, and 20.04 and 18.68 for the HCT regime. These similarities indicate the the properties fo the uncertainties obtained from individual fits are valid in that they are representative of the distributions of medians.

\subsection{Instrument dependence of model parameters}\label{ins}
In this section, we seek to understand which parameter constraints are leveraged by which instrument, and hence understand the required energy band coverage for measurement of a particular parameter. We fitted either \textit{XMM--Newton} data only, or \textit{NuSTAR} data only, and compare the results with those obtained above from joint fits. This was done for both MCT and HCT scenarios. We summarize our primary findings as follows:

\begin{itemize}
\item[(i)] Column density ($N_{\rm H,los}$) : In the MCT regime $N_{\rm H,los}$ parameter is more reliably determined with the \textit{XMM--Newton} data which facilitates the better constraints compared to \textit{NuSTAR} because of better energy resolution, thus facilitating better measurement of the Fe K absorption edge and the rollover at $E<5$~keV. In the HCT regime the scattered component is dominant and the strong attenuation of the transmitted component does not allow the measurement of a rollover or edge of the transmitted component. Both the instruments individually do not provide reliable measurement in the HCT regime. However joint fits in all cases are more reliable in terms of values returned and in terms of low statistical errors, which makes joint observation preferable when proper measurement of $N_{\rm H,los}$ is concerned.

\item[(ii)] $\sigma_o$ of UXCLUMPY: $\sigma_{o}$ of the cloud distribution is not a well constrained parameter in any case, with values of $\delta$ as large as $\sim 0.30$. The posteriors are wider (Fig.~\ref{instruments}d) for \textit{NuSTAR} compared to \textit{XMM--Newton} only, as the $E<4$~keV tail of the scattered component is absent in \textit{NuSTAR} only data.

\item[(iii)] Covering fraction ($C_{\rm frac}$) of inner ring of UXCLUMPY: The constraints are generally drawn from the shape of the CRH, that is, 7$< E  \lesssim 30$ keV band. We found that fitting only \textit{XMM--Newton} data returns very wide irregular posterior, whereas \textit{NuSTAR} only returns regular monomodal distribution, with stronger constraints. The constraints improved only by a small amount in the joint analysis. Thus, data from \textit{NuSTAR} is a compulsory requirement to constrain the covering fraction ($C_{\rm frac}$) of such a ring if present. 

\item[(iv)] $c/a$ of RXTORUS: The values of $\delta$ are similar for both the instruments. But the median value of \textit{NuSTAR} fit is discrepant compared to the input and posteriors are wider whereas the \textit{XMM--Newton} fit is not. On the other hand, the constraints on the lower limit, are stronger in the \textit{XMM--Newton} fit, whereas  constraints on the higher limit are stronger for \textit{NuSTAR} fit. The constraints on the joint fits, thus depend strongly on data from both the \textit{XMM--Newton} and \textit{NuSTAR} instruments. The median value of \textit{NuSTAR} posteriors differ from the input by nearly 24\%, due to unavailability of the E<4 keV tail of the scattered component.

\item[(v)] $N_{\rm cloud}$ of CTORUS: \textit{XMM--Newton}-only fits provides tighter constraints compared to \textit{NuSTAR} as shown in the figure (e.g.$R_{\rm 90,Nu}=6.48$ and $R_{\rm 90,XMM}=4.35$ for HCT). The joint posteriors improve significantly (e.g. $R_{\rm 90,joint}=1.97$ for HCT) under the influence of \textit{XMM--Newton}. Thus, for a more localized posterior on $N_{\rm cloud}$, joint fits with \textit{XMM--Newton} data is a requirement.

\item[(vi)] $A_{\rm Fe}$ of BORUS: Both the \textit{XMM--Newton} and \textit{NuSTAR} parameters have values quite different from the joint fits and the input. The posteriors of \textit{NuSTAR} are wider. We would require both \textit{NuSTAR} and \textit{XMM} data to constrain the posteriors better accuracy. 
\end{itemize}

Our analysis of the data from the instruments separately, also shows that simultaneous coverage of 0.3$<E<$ 78~keV band improves the posteriors significantly, compared to separate $E \sim$10~keV through \textit{XMM--Newton} or $E >$4~keV only through \textit{NuSTAR} coverages. We jointly show the 1D posterior plots of \textit{XMM--Newton} only, \textit{NuSTAR} only and \textit{XMM--Newton} plus \textit{NuSTAR} joint fits for a few parameters in Fig.~\ref{instruments}.

\section{RESULTS: CROSS-MODEL FITS}\label{CM-fits}
In this section, we discuss the results of cross-model fits (CM-fits hereafter). We use the notation $M_j \rightarrow D_i$ to denote that the model $j$ is fit to data simulated under model $i$. The indices i and j are the abbreviations for the model of which the reflection component is featured in the data simulations or fits. The abbreviations for the torus models are summarized in Table~\ref{tab:1}. For example, when the fitting model is UXCLUMPY, the notation would be $M_{\rm UXCL} \rightarrow D_{\rm MYT}$. For other cases, the abbreviations for model combinations are defined locally. The CM fits will identify and estimate the level of degeneracy between the models, show which parameters adjust themselves to fit a given data set when the assumed morphology is wrong.  We do not test all possible model combinations; instead, we concentrate on only the following cases:  
\begin{itemize}
\item $M_{\rm MYT} \rightarrow D_{\rm j}$,  $\rm j =$ RXT, UXCL : spectral change due to difference only in radiative process and how well a simple smooth doughnut can represent complex arrangements
\item $M_{\rm RXT} \rightarrow D_{\rm j}$,  $\rm j =$ MYT, UXCL : similar to MYT but for the variable doughnut parameter ($c/a$)
\item $M_{\rm BOR} \rightarrow D_{\rm j}$,  $\rm j =$ CTOR, UXCL  : what happens when a model with a simplified contiguous cut-out geometry is applied to clumpy morphologies
\item $M_{\rm CTOR} \rightarrow D_{\rm j}$, $\rm j =$ MYT, UXCL, BOR : how a clumpy torus model with an estimate on the number of clouds in the radial direction acts when applied to data simulated under other contiguous and clumpy models
\item $M_{\rm UXCL} \rightarrow D_{\rm j}$, $\rm j =$ MYT, RXT, BOR , CTOR : same as CTOR but UXCLUMPY has a more complex cloud distribution
\end{itemize}
Different models treat and make predictions for the shape/intensities of the emission complex of Fe K$\alpha$ and other fluorescent emission lines due to different scattering geometries and different physical processes in their radiative transfer codes e.g. MYTORUS contains Fe-lines with Compton shoulders only and other models contain a wider range of species. To account for such discrepancy between models, one can perform cross-model fitting ignoring certain narrow energy ranges thus focusing more on constraints from the continuum (see section \ref{CM-hct}). We can expect that there may be narrow ranges of data-model residuals corresponding to differently-modelled Fe K$\alpha$ and other line complexes. Future work on model comparison can be done when models offer a more consistent treatment of narrow features like fluorescent lines and Compton shoulders.

\subsection{Medium Compton-thick regime}\label{CM-MCT}
\begin{figure*}
\includegraphics[scale=0.56]{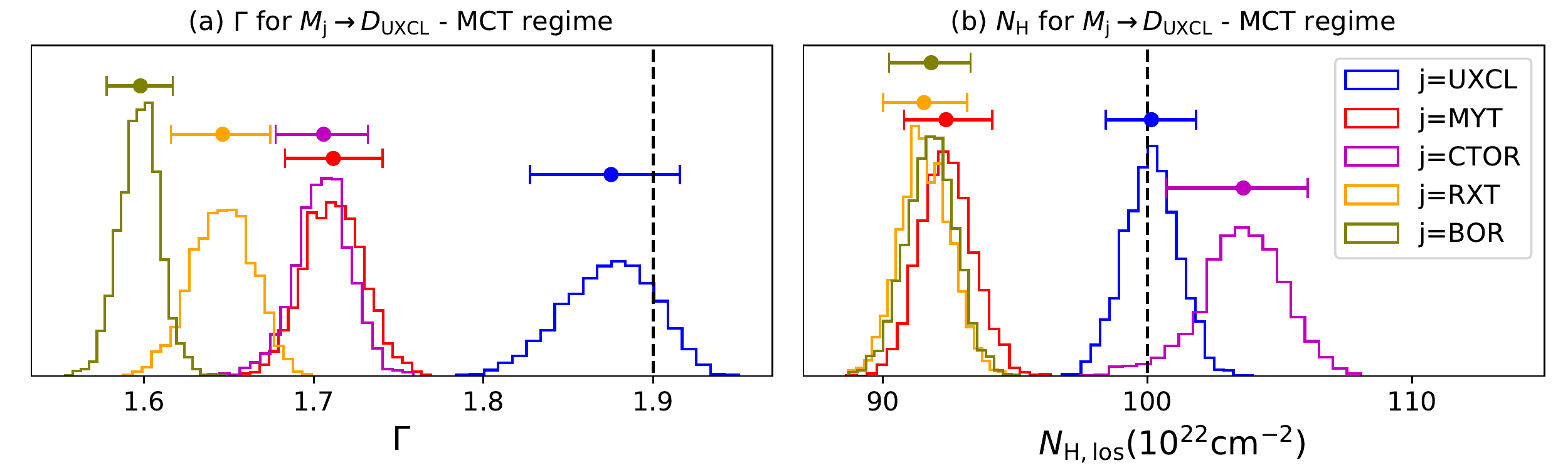}
\caption{ Cross- and intramodel fits for the case $M_{\rm j} \rightarrow D_{\rm UXCL}$ in the MCT regime, j = MYT, RXT, CTOR, BOR models. The discrepancies in $\Gamma$ are typically rather large, with a flattening by 0.2-0.3, and potentially impacting conclusions about the coronal power law. In contrast, discrepancies in $N_{\rm H,los}$ are much smaller, at only $\sim 10 \%$ which does not significantly impact the spectral fit. \label{cm_uxcl_mct}}
\end{figure*}

\begin{itemize}
\item[(i)]  \textbf{Column density ($N_{\rm H,los}$):} Across all models tested in the MCT regime, LOS column density ($N_{\rm H,los}$) is the only parameter that was practically recovered. The dominance of the zeroth-order continuum in this regime is the reason for the well-determined constraint and values consistent with the input for $N_{\rm H,los}$ irrespective of the model. The maximum discrepancy was observed for $M_{\rm CTOR} \rightarrow M_{\rm BOR}$, where the percentage discrepancy corresponds to 31.8\% and $r_{\rm q}$ corresponds to 8.2. With regard to specific trends we find that the clumpy torus models returned higher values of $N_{\rm H,los}$ when fit to contiguous torus and the converse also holds true. The values of $r_{\rm q}$ correspond to the variation of $N_{\rm H,los}$ by 30\%. In order to check the discrepancy statistically we also perform bulk CM-fits of the case $M_{\rm MYT} \rightarrow D_{\rm UXCL}$ to 100 spectra simulated under UXCLUMPY. The minimum and maximum discrepancy in $N_{\rm H,los}$ in the bulk fits were found to be 1.7\% and 8.6\% respectively. This discrepancy in the values of $N_{\rm H,los}$ will not create profound physical inconsistencies and will not significantly change the scientific implications in terms of the spectral shape and the components present. Thus, we conclude that $N_{\rm H,los}$ returned from the fits were mostly consistent with the input in the MCT regime.
 
\item[(ii)] \textbf{Photon index ($\Gamma$):} Both the CRH and zeroth-order continuum determine the value of $\Gamma$. The values of $r_{\rm q}$ that we observed for $\Gamma$ in the CM-fits range from 0.67 to 15 with $\Gamma_{\rm in} = 1.90$ for all tested cases. The best value was for the case $M_{\rm MYT} \rightarrow D_{\rm RXT}$ ($\Gamma_{\rm out} = 1.91$) and the worst value was obtained for $M_{\rm BOR} \rightarrow D_{\rm UXCL}$. In the MCT regime, for a specific case $M_{\rm j} \rightarrow D_{\rm UXCL}$, where j = MYT, RXT, BOR, CTOR, the fits systematically returned flatter values of $\Gamma$, $1.60 \lesssim \Gamma_{\rm out} \lesssim 1.75$ (see Table~\ref{tab:3}). In the converse case viz. $M_{\rm UXCL} \rightarrow D_{\rm j}$, where j = MYT, RXT, BOR, CTOR, the estimated values of $\Gamma$ were in the range $2.02 \lesssim \Gamma_{\rm out} \lesssim 2.17$. The high discrepancy (quantified by $r_q$) in values of $\Gamma$ is a direct implication of low statistical errors, caused by the intrinsic difference in the CRH shape for different models. Therefore, we can conclude that the retrieved value of $\Gamma$ is thus strongly model-dependent (see figure \ref{cm_uxcl_mct}).

\item[(iii)] \textbf{Angle of inclination ($\theta_{\rm i}$):} The values of $\theta_{\rm i}$ returned from the CM fits are dependent on the torus morphology. However there might be cases where interpretation of the values can be contextual and is dependent on morphology. We observed specific trends for certain models in the CM-fits. We list these in the following bullet points:\\
\textbf{(A)} In CTORUS the zeroth-order continuum is independent of $\theta_{\rm i}$. The information about the only parameters of morphology ($N_{\rm cloud}$) and geometry ($\theta_{\rm i}$) is in the scattered component. In CTORUS the obscuring clouds are located in a hypothetical biconically cutout thick shell between 60$^{\circ}$ and 90$^{\circ}$ \citep{liu14} with respect to the symmetry axis. In the $M_{\rm CTOR} \rightarrow D_{\rm UXCL}$ fit when we assume a prior $[0^{\circ},90^{\circ}]$, we get an inconsistent value of $\theta_{\rm i} < 60 ^{\circ}$ according to the geometry proposed in \cite{liu14}. However, a possible misinterpretation is the existence of a stray clump(s) or obscurer at $\theta = \theta_{\rm i} < 60^{\rm \circ}$ which is outside the hypothetical shell enclosure of CTORUS. This stray clump or obscurer can be thought to be contributing to $N_{\rm H,los}$. This is just a result of spectral or model difference and hence an artefact of fitting (also see the similar discussion for $\cos \theta_{\rm o}$ for BORUS). When we limit the prior range to $[60^{\circ},90^{\circ}]$, the posterior monotonically increases towards the $60^{\circ}$ edge.\\
\textbf{(B)} For the CM-fits, $M_{\rm UXCL} \rightarrow D_{\rm j}$ where j = CTOR, MYT, BOR, $\theta_{\rm i}$ assumes an edge-on configuration with monotonically increasing posteriors consistent with $90^{\circ}$. The only exception is $M_{\rm UXCL} \rightarrow D_{\rm RXT}$, where the median value calculated from the posterior is $\simeq 66^{\circ}$ for an input $\theta_{\rm i} = 70^{\circ}$; we suspect that the excess contribution in the softer bands from Rayleigh scattering might be the result of the lower values of $\theta_{\rm i}$. The input value of $\theta_{\rm i}$ and the value returned from fit $M_{\rm UXCL} \rightarrow D_{\rm RXT}$ value might agree by chance.\\
We discuss the case of $\theta_{\rm i}$ for the $M_{\rm RXT} \rightarrow D_{\rm MYT}$ case along with $c/a$ in the fifth bullet point.

\item[(iv)] \textbf{Relative normalization between the transmit and the reflect component ($T/R$) :} We first discuss the results from the case: $M_{\rm j} \rightarrow  D_{\rm UXCL}$ ($\rm j$ = RXT, MYT, CTOR, BOR). We find that $T/R \neq 1$ (input). For the case $M_{\rm BOR} \rightarrow D_{\rm  UXCL}$, we get an exceptionally high value of $T/R \simeq 6.38$. Physically, the zeroth-order continuum across all models has the same shape for given a $N_{\rm H,los}$ ($A_{\rm Fe} = 1$), whereas the scattered continuum is very different for different models. The tendency of $T/R > 1$ in CM-fits may be an attempt to get a better fit. Physically if the value of $T/R$ is inconsistent with 1, it would imply variability of the coronal power law, however as demonstrated here this effect might be an artefact of model difference. 

\begin{table*}
\begin{tabular}{lc|lllllll}
\hline
\hline

           & Input           &                  &                 & Fitting model &             &               &                 \\
\hline       
Parameters & Input for $D_{\rm UXCL}$  & $M_{\rm UXCL}$ & $M_{\rm MYT}$  &$M_{\rm CTOR}$  &$M_{\rm RXT}$  &$M_{\rm BOR}$(coupled)  \\               
\hline
$C_{\rm sc-pl}$ & -2   &$-1.93^{+0.24}_{-0.18}$  & $-2.28^{+0.04}_{-0.04}$&$-2.19^{+0.05}_{-0.09}$&$-2.24^{+0.05}_{-0.05}$        &$-2.49^{+0.03}_{-0.04}$\\
$N_{\rm H,eq}$  & -    & -                    & $211^{+40}_{-34}$      & -                     & $218^{+20}_{-25}$                 &-                     \\
$N_{\rm H,los}$ & 100  &$100^{+1.7}_{-1.7}$   & 92                     &$104.5^{+0.24}_{-0.29}$& $92$                             &$91.83^{+1.5}_{-1.6}$  \\
$\Gamma$        & 1.9  &$1.88^{+0.04}_{-0.05}$& $1.72^{+0.03}_{-0.03}$ &$1.71^{+0.03}_{-0.03}$ &$1.8^{+0.03}_{-0.03}$             &$1.60^{+0.02}_{-0.02}$   \\
$E_{\rm cut}$   & 400  &$323^{+67}_{92}$      & -                      & -                     &-                                 &500(frozen)             \\
$\sigma_{\rm o}$& 45   &$49^{+16}_{-17}$      & -                      & -                     &-                                 &-                      \\
$C_{\rm frac}$  & 0.4  &$0.42^{+0.07}_{-0.05}$& -                      & -                     &-                                 &-                      \\
$N_{\rm cloud}$ & -    &-                     & -                      &$3.36^{+3.95}_{-1.2}$  &-                                 &-                      \\
$T/R$        &1(frozen)&1(frozen)             & $1.90^{+0.10}_{-0.10}$ &$3.03^{+0.53}_{-0.44}$ &$2.42^{+0.16}_{-0.15}$            &$6.38^{+0.73}_{-0.73}$ \\
$c/a$        &-                               & -                      & -                     & - & $0.88^{+0.03}_{+0.02}$        & -             \\
$C_{\rm Frac,tor}$ or $\rm cos \theta_{\rm o}$ & -  &-                     & -                     & -                  &-               & $0.85^{+0.06}_{-0.12}$ \\
$\theta_{\rm i}$ or $\rm cos \theta_{\rm i}$(for BOR and CTOR) & 60.0 &$64.56^{+20}_{-33}$ & $63.25^{+1.43}_{-0.92}$ &$0.53^{+0.29}_{-0.04}$ &$36.85^{+3.48}_{-2.62}$  &$0.88^{+0.05}_{-0.14}$ \\
$A_{\rm Fe}$ (for BOR) & - & - & - & - & - &  $0.96^{+0.06}_{-0.09}$  \\
$L_{\rm 7-50}/L_{\rm in,7-50}$ of coronal power law & 1.0 &$1.04\pm 0.04$ & $0.87 \pm 0.02$ & $1.08 \pm 0.04$ & $0.89 \pm 0.02$ & $0.96 \pm 0.03$\\
\hline
$\chi^2/$dof    &-     &1.004                 &1.14                  & 1.09                   &1.11                    &1.34                    \\
\hline
\hline
For evidence estimates: $1.2 \leq E \leq 78$ keV\\
\hline
$\rm logZ$ &-     &-592.96                &-626.01              &-654.60                  &-625.09                  &-800.12                  \\
$\rm log BF = log Z_{j}-log Z_{UXCL}$ & -      & 0                  &  -33.06                 &  -61.64 &   -32.13  & -207.16\\
\hline
\hline 
\end{tabular}
\caption{ We summarize the input parameters that are used for data simulation under UXCLUMPY and the median values and errors calculated from the parameter posteriors in this table, for the cases $M_{\rm j} \rightarrow D_{\rm UXCL}$, where j = UXCL, MYT, RXT, CTOR and BOR in the MCT regime. For parameter estimation the full energy range allowed by a model has been taken into consideration and the $\chi^2$/dof value is also calculated over the entire energy range. The values of evidence and Bayes Factor are quoted for the case when all the energy range of the data used for analysis is 1.2-78.0 keV for all models.}
\label{tab:3}
\end{table*}

\begin{figure*}
\includegraphics[scale=0.34]{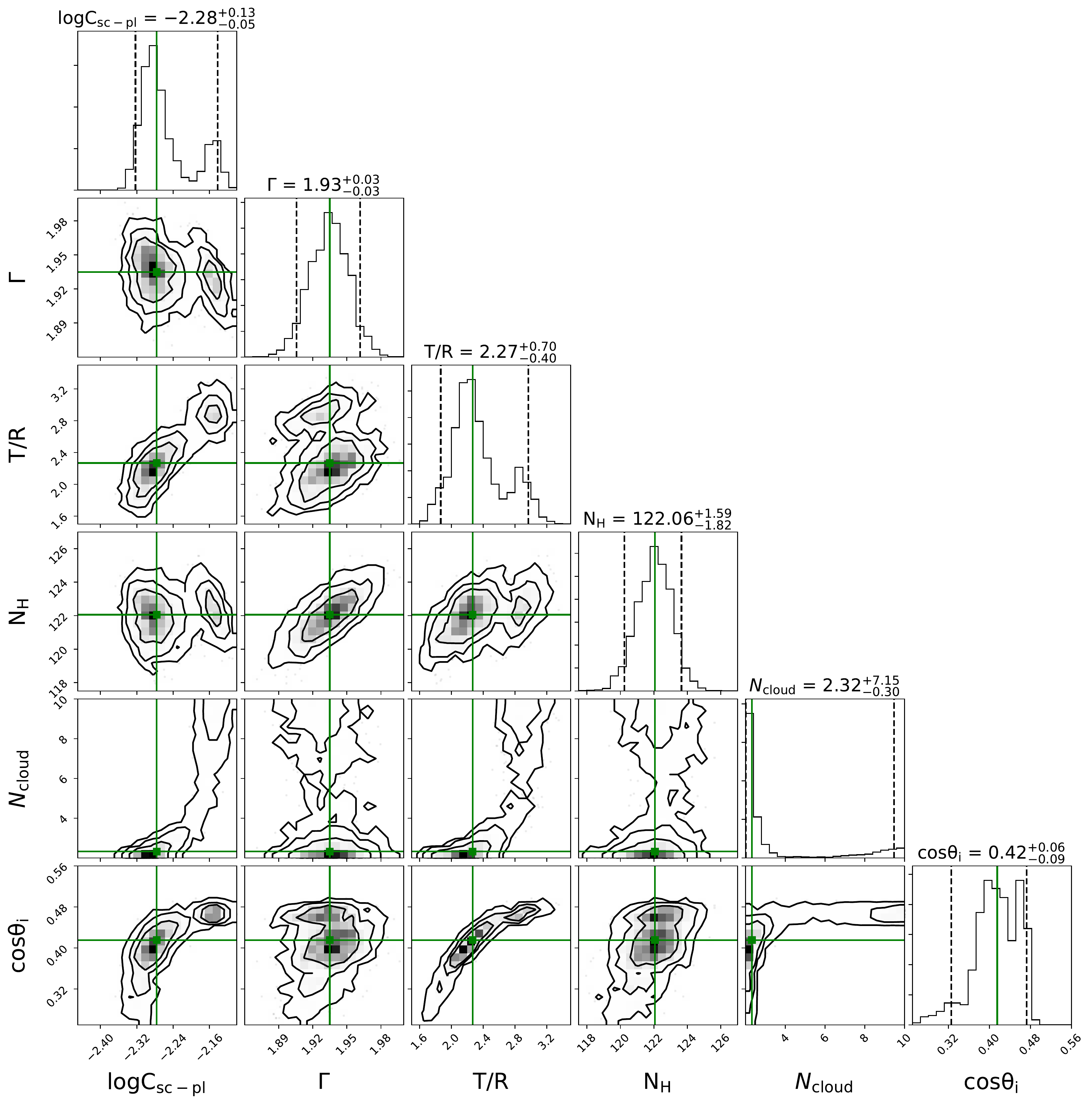}
\caption{Contour plot results for the analysis $M_{\rm CTOR} \rightarrow D_{\rm MYT}$ in MCT regime. The posterior of $N_{\rm cloud}$ is irregular as there exists no maxima in the allowed prior range, but a minima, with ridges at the prior edges. This `bifurcation' in the PDF of $N_{\rm cloud}$ gives rise to bimodality in the other parameters. \label{myt_ct_mct}}
\end{figure*}

\begin{figure*}
\gridline{\fig{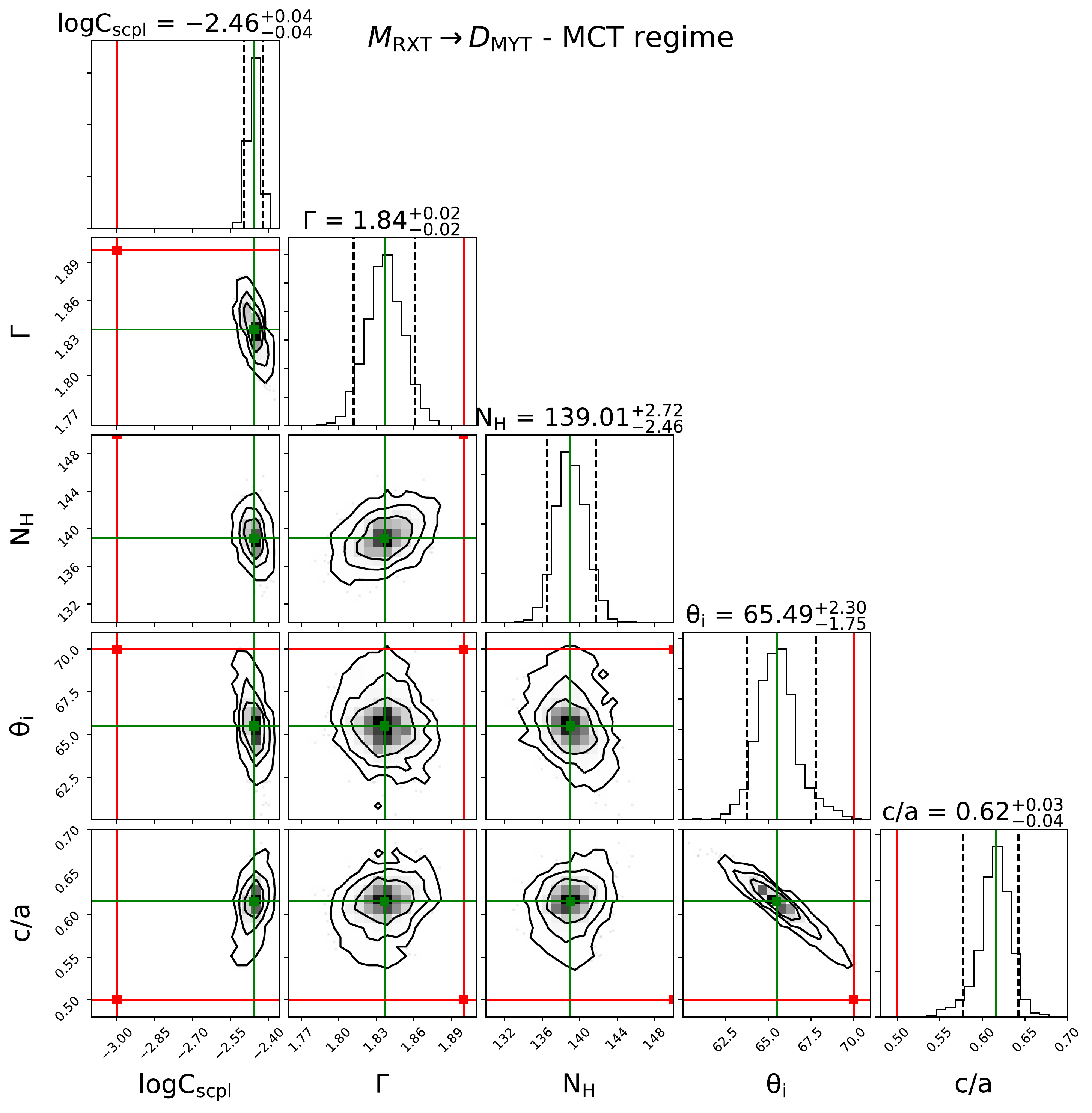}{0.52\textwidth}{(a)}
          \fig{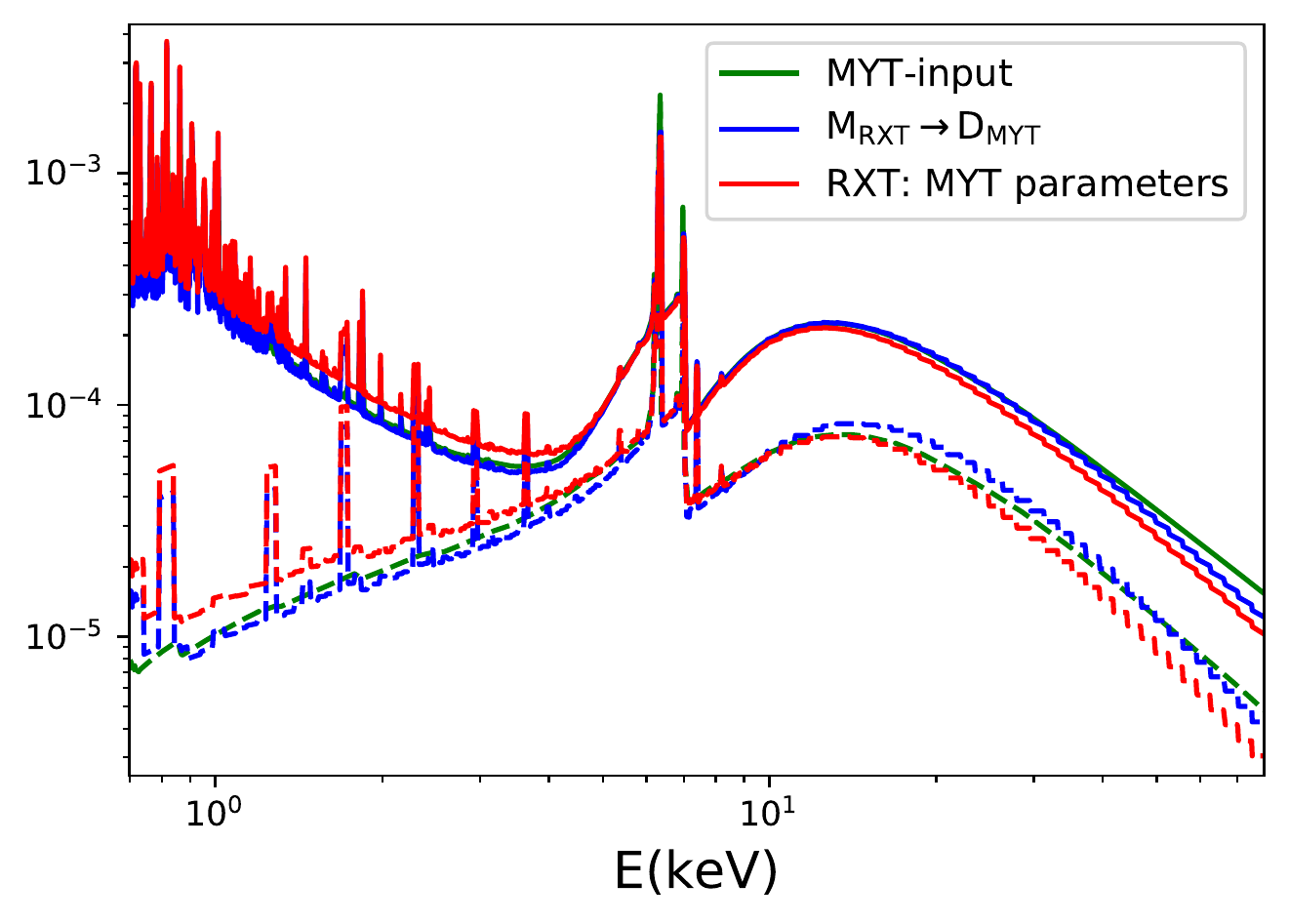}{0.4\textwidth}{(b)}
          }
\caption{(a) Contour plot results for the analysis $M_{\rm RXT} \rightarrow D_{\rm MYT}$ in MCT regime. The noticeable discrepancy can be seen in the parameter $c/a$ which in turn influences the constraints on log$C_{\rm SCPL}$, $\theta_{\rm i}$ posteriors. (b) spectral overplot of the input model (green), the best fit spectrum from $M_{\rm RXT} \rightarrow D_{\rm MYT}$ (blue) and the RXTORUS model under MYTORUS input parameters (red). The dashed spectra are the scattered continua of the respective model spectra. The scattered components of RXTORUS and MYTORUS spectra under same parameters are different because of the difference in radiative process assumptions. $\chi^2$/dof = 1.11 for the best-fit parameters. \label{rxt_myt_mct}}
\end{figure*}

\item[(v)] \textbf{Parameters of Torus Morphology:} 
\begin{itemize}
\item[(a)] \textbf{Opening angle ($\rm cos \theta_{\rm o}$ or $C_{\rm frac,tor}$) :} For $M_{\rm BOR} \rightarrow D_{\rm CTOR}$ we get $\cos \theta_{o} \simeq 0.4 $ ($\theta_{\rm o} \simeq 67^{\circ}$) with corresponding $\theta_{i} \simeq 69.6^{\circ}$. However in CTORUS the obscuring clumps are distributed in the region $\theta \geq 60^{\circ}$. This can be due to the difference in the total covering fraction of between the two torus. For the case $M_{\rm BOR} \rightarrow D_{\rm UXCL}$ in the coupled configuration ($N_{\rm H,los} = N_{\rm H,eq}$ in this case, \citep{balokovic18} ), we get a case where  $\cos \theta_{\rm o}$ < $\cos \theta_{\rm i}$ which implies the line of sight does not intersect the main gas distribution of the torus, but returns a non-zero line of sight absorption ($N_{\rm H,los} \simeq 92$). This dichotomy can be mis-interpreted as a different matter distribution (e.g. a stray clump, same $N_{\rm H,los}$ as the main gas distribution) lying along the line of sight. In the uncoupled configuration of $M_{\rm BOR} \rightarrow D_{\rm UXCL}$, the $\cos \theta_{\rm o}$ > $\cos \theta_{\rm i}$, but the $N_{\rm H,los} \simeq 97$ and $N_{\rm H,eq} \simeq 1300$ heavily discrepant and becomes inconsistent with respect to the defined morphology. However, a physical interpretation for this case can be made, if the torus is clumpy, in which case $\rm \cos \theta$ loses its meaning and hence can be interpreted as covering fraction, $C_{\rm frac,tor}$.

\item[(b)] \textbf{$c/a$ of RXTORUS:} For $M_{\rm RXT} \rightarrow D_{\rm MYT}$,  $c/a = 0.62 \pm 0.03$ (Fig.~\ref{rxt_myt_mct}) which is discrepant with the fixed assumed value of 0.5 (effective $\theta_{\rm o} = 60^{\circ}$) in MYTORUS despite the fact that both are doughnut models. Given a set of parameters the scattered continuum of MYTORUS and RXTORUS have different shapes because of the radiative difference. The scattered continuum has more photon counts due to the presence of Rayleigh scattering in the $E<6$~keV band and lesser photon counts at the CRH because of having $E_{\rm cut}$ fixed at 200~keV (Fig.~\ref{rxt_myt_mct}). The parameter $c/a$ thus adjusts itself to a higher value decreasing the effective $\theta_{\rm o}$ to $51.7^{\circ}$. $\Gamma$ and $\theta_{i}$ were also reduced to $1.83$ and $65^{\rm \circ}$ respectively (Fig.~\ref{rxt_myt_mct}). For the $M_{\rm RXT} \rightarrow D_{\rm UXCL}$ case $c/a \simeq 0.88$, which can be considered consistent with the extended and random cloud distribution. This also led to $\theta_{\rm i}$ for this case to take a comparatively low value, $\sim 37^{\rm \circ}$.
  
\item[(c)] \textbf{$\sigma_{\rm o}$ of UXCLUMPY:} The value of $\sigma_{\rm o}$ follows different trends for data sets simulated with different models. For all the cases of $M_{\rm UXCL} \rightarrow D_{\rm j}$, j = MYT, RXT, BOR and CTOR, values of $\sigma_{\rm o}$ were systematically found to be $\leq 30^{\rm \circ}$ (Fig.~\ref{sig-ctk-mct-cm}) which means most of the clouds aggregate closer to the equator. For all the fits $\delta < 0.1$, which is lower compared to the IM fits. In order to confirm that this trend is due to the effect of systematics (model difference) we performed an additional spectrum of MYTORUS spectrum for an exposure of $t_{\rm XMM} = 1.0 $ Ms and $t_{\rm NuSTAR} = 0.5$ Ms and fit with UXCLUMPY ($M_{\rm UXCL} \rightarrow D_{\rm MYT}$). We find that the median values and the errorbars (the diamond and circular markers in \ref{sig-ctk-mct-cm}) are consistent with each other. This indicates that the trends for the high exposure simulations are similar to that of the typical exposure case of $M_{\rm UXCL} \rightarrow D_{\rm MYT}$. This proves that the trends are systematic. We simulate data under UXCLUMPY using the best-fit parameters obtained from $M_{\rm UXCL} \rightarrow D_{\rm MYT}$ and perform an IM fit. This is partially similar to the `surrogate' method \citep{press02}. However, our analysis differs from \cite{press02}, as we use a single `surrogate' data set and our analysis of it is purely Bayesian. Consequently, the objective is comparison of the properties of the IM-posteriors obtained from the `surrogate' data set with that of the CM-posteriors obtained from the real data set. We find that when the $M_{\rm UXCL} \rightarrow D_{\rm MYT}$ CM-fit returned $\delta \simeq 0.07$ for $\sigma_{\rm o}$, the IM fit for the `surrogate' data set returned are wider with (Table \ref{original_surrogate}) $\delta \simeq 0.46$ for $\sigma_{\rm o}$, which is $\sim 6$ times higher.

\item[(d)] \textbf{Covering fraction ($C_{\rm frac}$) of the inner ring of UXCLUMPY:} The $C_{\rm frac}$ of the inner ring in the MCT regime returned posteriors with median values of $C_{\rm frac}$ < 0.1 ($M_{\rm UXCL} \rightarrow D_{\rm j}$) or \textit{irregular} posterior converging at the lower-bound of the prior range implying $C_{\rm frac} \sim 0$ (Fig.~\ref{sig-ctk-mct-cm}) (all the other cases). We compare the posterior of $C_{\rm frac}$ from the CM-fit $M_{\rm UXCL} \rightarrow D_{\rm MYT}$ and the posterior obtained from IM-fit of the `surrogate' data set. The posterior of $C_{\rm frac}$ in the CM-fit is irregular as it converges towards the lower-edge (see Fig.~\ref{sig-ctk-mct-cm}). Similar to the $\sigma_{o}$ the high exposure spectral fit ($M_{\rm UXCL} \rightarrow D_{\rm MYT}$), returns trends consistent with the typical exposure spectrum, implying a systematic trend. The median is $\sim 0.01$. However, the $C_{\rm frac}$-posterior obtained from the same UXCLUMPY IM-fit of the `surrogate' data set does not show irregularity. We find that $C_{\rm frac, median} \sim 0.19$ with $\delta \simeq 1$ and $R_{\rm 90} \simeq 0.1$ range is significantly large compared to that from $M_{\rm UXCL} \rightarrow D_{\rm MYT}$.

\item[(e)] \textbf{$N_{\rm cloud}$ of CTORUS} The posteriors of the $N_{\rm cloud}$ for all the CM-fits, are \textit{irregular} in the sense that the probability function has a minimum or is a monotonically increasing or decreasing function in the allowed prior range. The best values are either consistent with 2 or 10 or both (e.g. Fig.~\ref{myt_ct_mct}) all cases. In a real scenario, this can be an indication that CTORUS is a \textit{wrong} model to apply in this case.
\end{itemize}
\end{itemize}

\begin{figure*}
\includegraphics[scale=0.56]{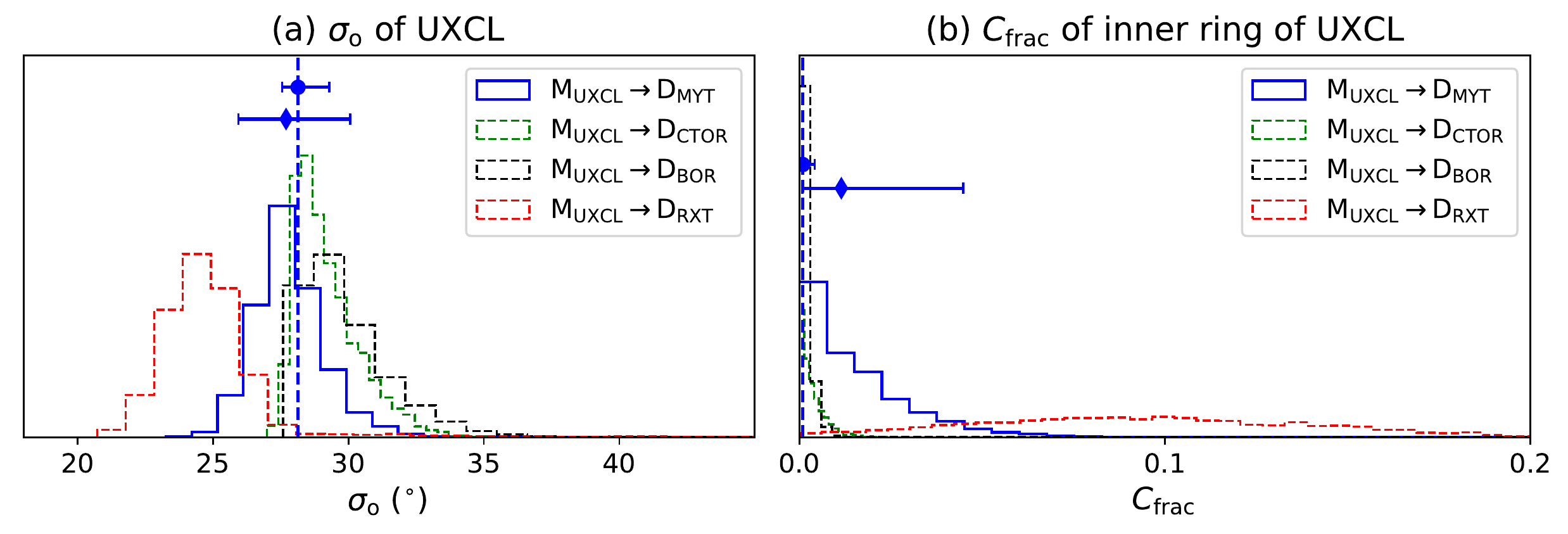}
\caption{ We show the behaviour of morphological parameters of UXCLUMPY for $M_{\rm UXCL} \rightarrow D_{\rm j}$-fits in the MCT regime(a)$\sigma_{\rm o}$ and (b)$C_{\rm frac}$ of the inner ring. $\sigma_{\rm o}$ is always $\leq 30^{\rm \circ}$ and $C_{\rm frac}$ in most cases have irregular posteriors consistent with 0 and a regular posterior with median value consistent $\simeq$ 0.1. The blue markers with errorbars represent the median values and uncertainty on $\sigma_o$ and $C_{\rm frac}$ for typical of exposure (diamond marker: $t_{\rm XMM} = 100ks$,$t_{\rm NuSTAR}=50ks$) and a deeper observation (circle marker: $t_{\rm XMM} = 1.0$Ms,$t_{\rm NuSTAR}=0.5$Ms) for $M_{\rm UXCL} \rightarrow D_{\rm MYT}$. The dashed vertical lines mark the median values obtained from the spectrum with high exposure. The median values and the errors obtained from the typical and deeper observations show that trends are systematic and occuring due to model difference. See discussion in Section \ref{CM-MCT}.\label{sig-ctk-mct-cm} }
\end{figure*}

\begin{figure*}
\includegraphics[scale=0.56]{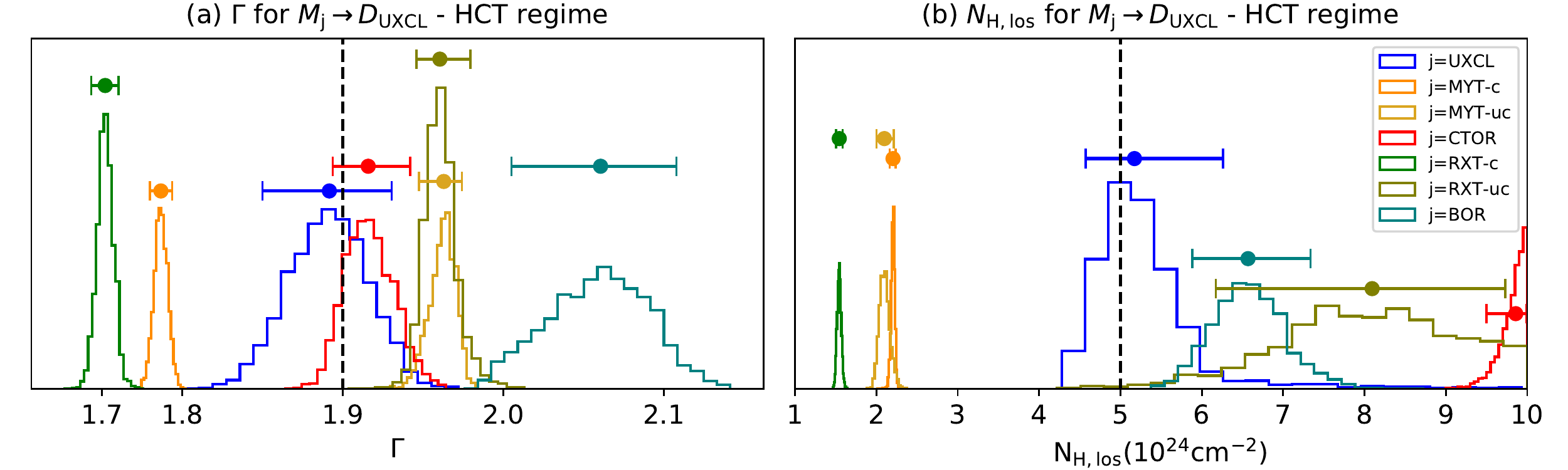}
\caption{ Cross- and intramodel fits for the case $M_{\rm j} \rightarrow D_{\rm UXCL}$, j = MYT, RXT, CTOR, BOR models. The labels: -c and -uc refers to coupled and uncoupled configuration respectively. The discrepancies in $\Gamma$ are due to both flattening and steepening of $\Gamma$. The discrepancies in $N_{\rm H,los}$ are also large amounting to a maximum of $\sim 100 \%$. \label{cm_uxcl_hct}}
\end{figure*}

\begin{figure*}
\gridline{\fig{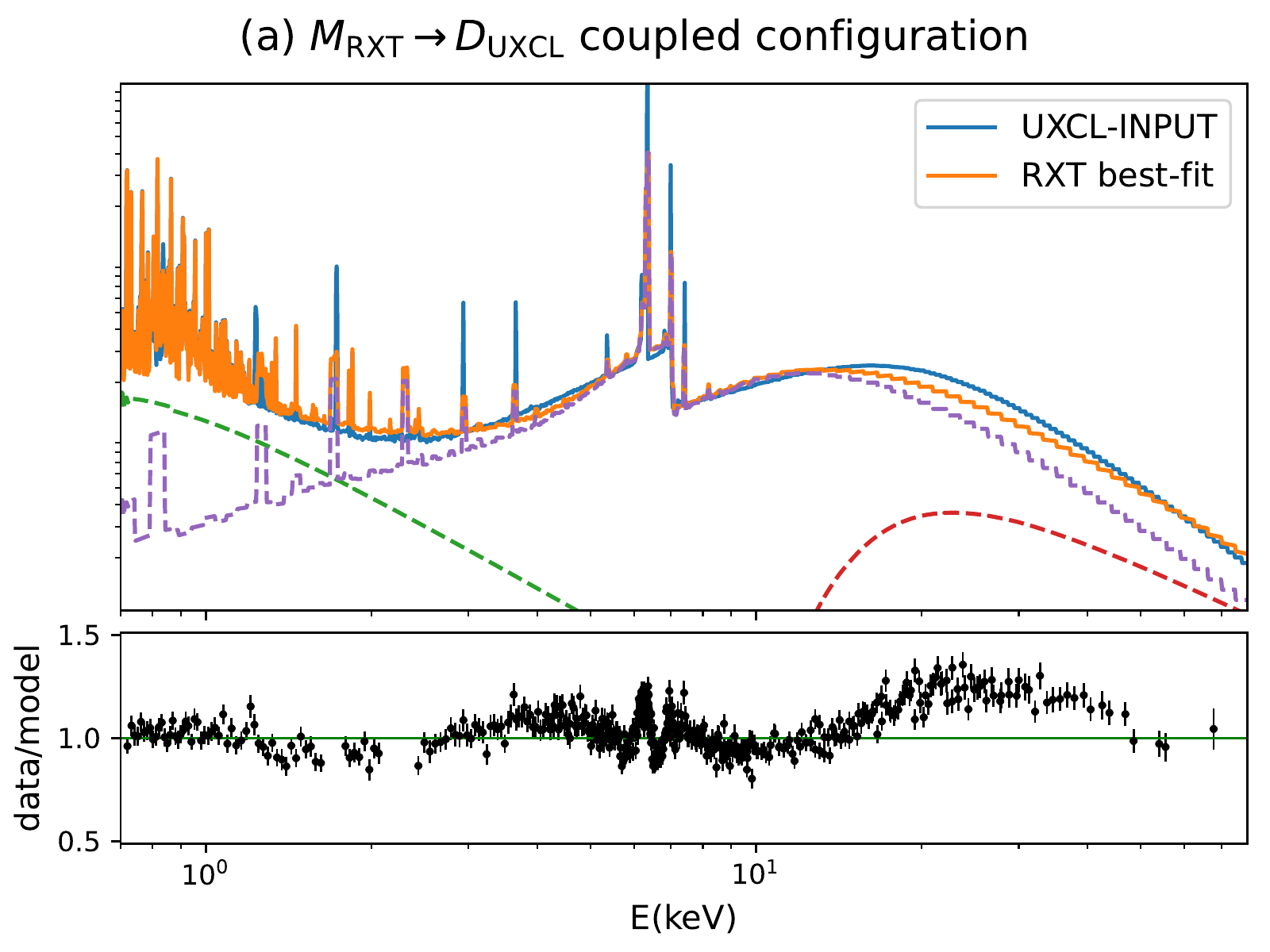}{0.46\textwidth}{}
          \fig{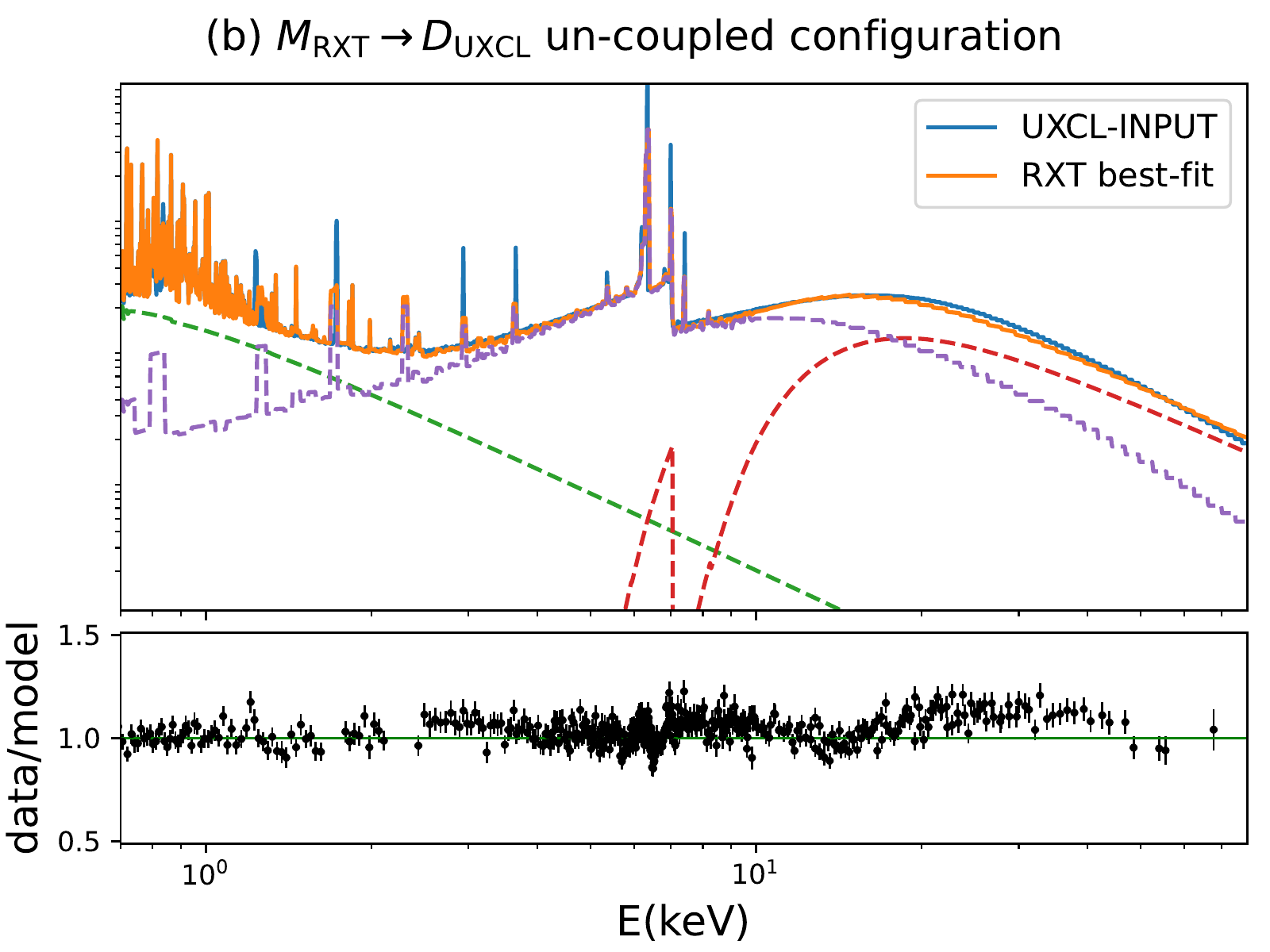}{0.46\textwidth}{}
          }
	\caption{ $M_{\rm RXT} \rightarrow D_{\rm UXCL}$ fit for two cases: Z-S coupled with $\chi^2$/dof $\simeq$ 1.4 and log BF $\simeq$ -364.99; Z-S un-coupled: with $\chi^2$/dof $\simeq$  1.16 and log BF $\simeq$ -148.56. The Bayes Factors are calculated using the log $\rm Z_{\rm UXCL}$ value obtained from the UXCLUMPY intramodel fit. Most of the discrepancy in the spectra happens in the continuum. \label{rxt_uxcl_hct}}
\end{figure*}

\begin{table*}
\begin{tabular}{ccccccc}
\hline
\hline
Model parameters of $M_{\rm UXCL}$ & \multicolumn{2}{c}{MYT-MCT regime}      & \multicolumn{2}{c}{CTOR-HCT regime}      & \multicolumn{2}{c}{BOR-HCT regime}\\
\hline
\multicolumn{7}{c}{$R_{90}$ values for CM-fit and surrogate data set} \\
\hline
    & $D_{\rm MYT}$ & $D_{\rm UXCL}$(surrogate) & $D_{\rm CTOR}$ & $D_{\rm UXCL}$(surrogate) &  $D_{\rm BOR}$ & $D_{\rm UXCL}$(surrogate) \\
\hline
$\sigma_{\rm o}$(in $^{\rm \circ}$)& 4.13 & 28.95  &   1.42      & 2.50       & 1.57  & 2.68    \\
$C_{\rm frac}$ of inner ring       & 0.04 & 0.26   &   0.004     & 0.027      & 0.008 & 0.034    \\
$\theta_{\rm i}$                   & 3.63 & 9.47   &   0.15      & 1.31       & 0.26  & 1.08    \\
\hline
\hline                     
\end{tabular}
\caption{Parameter comparison for the cross-model fits ($M_{\rm UXCL} \rightarrow D_{\rm j}$, j = MYT-MCT, CTOR-HCT, BOR-HCT) and surrogate data sets simulated under the fitting model (UXCLUMPY) for the best-fit parameters. $\theta_{\rm i}$ and the parameters of morphology exhibit lower value $R_{90}$ compared to the intramodel-surrogate fit for UXCLUMPY.}
\label{original_surrogate}
\end{table*}

\subsection{Heavy Compton-thick regime}\label{CM-hct}
In the HCT regime we do not get good fits (compared to the MCT regime) for a lot of cases when models are applied in their consistent form. So in many cases we had to apply the fits with the zeroth-order continuum and the scattered continuum uncoupled and or ignoring the data where there is an emission line (one can alternatively add gaussian lines), to get a good fit. We ignored the energy ranges connected with difference in the emission lines in the following specific cases: $M_{\rm MYT} \rightarrow D_{\rm UXCL}$(1.16-1.31 keV, 1.68-1.78 keV, 3.5-3.61 keV), $M_{\rm BOR} \rightarrow D_{\rm UXCL}$ (2.15-3.26 keV, 1.64-1.75keV), $M_{\rm RXT} \rightarrow D_{\rm UXCL}$ (2.15-3.26 keV, 1.64-1.75keV), $M_{\rm CTOR} \rightarrow D_{\rm UXCL}$ (2.15-3.26 keV, 1.64-1.75keV) and $M_{\rm MYT} \rightarrow D_{\rm RXT}$ (1.6-1.9 keV, 2.2-2.4 keV).

\begin{itemize}
\item[(i)] \textbf{Column density ($N_{\rm H,los}$) :} The values of $N_{\rm H,los}$ were not consistently retrieved (e.g. Fig.~\ref{cm_uxcl_hct}-b) for the CM-fits in the HCT regime. In the context of the input model, the strong extinction on the transmitted component worsens the recovery of $N_{\rm H,los}$ when wrong models are fit. For the HCT regime all models were simulated with $N_{\rm H,los} \simeq 500$ and $T/R=1$.  The dissimilarity in the shapes of the CRH lead to bad fits in many cases when $T/R=1$ (frozen) in the coupled configuration. So, $T/R$ is kept free in most of the cases. We fit the doughnut (MYTORUS,RXTORUS) and biconical cut-out model (BORUS) to UXCLUMPY, in both coupled ($N_{\rm H,los} = f(N_{\rm H,eq})$) and the uncoupled configuration (where $N_{\rm H,los} \neq f(N_{\rm H,eq})$). The doughnut fits were always significantly better in the uncoupled configuration but the $N_{\rm H,los}$ values are discrepant in all cases. The fits using BORUS seems practically uneffected in the uncoupled configuration. However, severe discrepancy and \textit{irregular} posteriors in $N_{\rm H,los}$ are seen many cases ($ 0.88 \leq r_q \leq 27.0$, for the cases with \textit{regular} $N_{\rm H,los}$ posteriors). There exist no specific trend methodology by which we can discern a correct value of $N_{\rm H,los}$. \textit{The fitting is based on the attempt to adjust $N_{\rm H,los}$, $\Gamma$ and $T/R$ such that it best mimics a CRH generated under a different morphology. This is unlike the MCT regime where the the rollover and the FeK-edge from the zeroth-order contunuum determines $N_{\rm H,los}$ and the effect of CRH on $N_{\rm H,los}$ relatively less.}

\item[(ii)] \textbf{Photon index ($\Gamma$):} $\Gamma$ is discrepant (e.g. Fig.~\ref{cm_uxcl_hct}-a) in most cases due to the varying shapes of the CRH. For all models the input value of $\Gamma$ was 1.9. From the fits we find $0.67 < r_q < 21$ holds for all tested model. The best value of $r_q$ was obtained for $r_q=0.67$ for $M_{\rm RXT} \rightarrow D_{\rm UXCL}$ (at $\Gamma_{\rm out} = 1.91$) and the worse is obtained for $M_{\rm CTOR} \rightarrow D_{\rm BOR}$ (at $\Gamma_{\rm out} = 1.69$). Thus in general it can be concluded that the recovered value of $\Gamma$ is strongly model dependent. 

\item[(iii)] \textbf{Relative normalization between the transmitted and the reflected component ($T/R$) :} When $T/R$ was kept free, we find irregular posteriors and or severe discrepancy with the input value  ($[T/R]_{\rm input}=1$ for all spectral simulations). One particular example case is $M_{\rm RXT} \rightarrow D_{\rm UXCL}$ where $[T/R]_{\rm out} = 4.92$ in the uncoupled configuration. This results in stronger zeroth-order continuum in the fit (originally much weaker in the $D_{\rm UXCL}$). The $\sim$~5-fold increase in $T/R$ effectively tries to fit the $E>10$~keV region of the data. (Fig.~\ref{rxt_uxcl_hct}). In other cases, $T/R$ is reduced to very low values, thus arbitrarily decreasing the zeroth-order continuum (which is otherwise attenuated by $N_{\rm H,los}$). This results relatively lower values of $N_{\rm H,los}$ (e.g. $M_{\rm MYT} \rightarrow D_{\rm UXCL}$: $[T/R]_{\rm out} \sim 0.5 $, $N_{\rm H,los} \simeq 200$). Several other cases where inconsistency in $T/R$ occur are: $M_{\rm BOR} \rightarrow D_{\rm UXCL}$, $M_{\rm CTOR} \rightarrow D_{\rm UXCL}$, $M_{\rm CTOR} \rightarrow D_{\rm BOR}$ etc. Physically, $T/R \neq 1$ would imply variability of the coronal power law; however as demonstrated here \textit{this can arise due to morphological difference resulting in varying shape of a CRH}. The large discrepancy of $T/R$ can result in heavy discrepancy in the intrinsic luminosity of the corona ($L_{\rm corona}$).

\item[(iv)] \textbf{Parameters of morphology:}
\begin{itemize}
\item[(a)] \textbf{$c/a$ of RXTORUS:} For the case of $M_{\rm RXT} \rightarrow D_{\rm UXCL}(C_{\rm frac} = 0)$ in the coupled configuration, $c/a \simeq 0.11$. The geometry was thus assumed to be more of an annulus than a doughnut but $\chi^2$/dof=1.6. For the uncoupled configuration $c/a$ returned values much higher, $\sim 0.71$. However, the interpretation of $c/a$ at face value is not possible. The geometrical interpretation of the results: the absorption happens through a Compton-thick absorber placed close to the axis of symmetry ($\theta_{\rm i} \simeq 3^{\rm \circ}$) which is independent of a separate $\sim$ medium Compton-thick ($N_{\rm H,eq} \simeq 86$, point 1) doughnut shaped gas distribution. For $M_{\rm RXT} \rightarrow D_{\rm UXCL}(C_{\rm frac} = 0)$, $c/a$ posteriors were bimodal, which resulted in bimodality in other posteriors. For both coupled and uncoupled configuration $M_{\rm RXT} \rightarrow D_{\rm UXCL}(C_{\rm frac} = 0.4)$ returned $c/a$>0.75 indicating an increased covering fraction of the torus. $M_{\rm RXT} \rightarrow D_{\rm MYT}$ also returned bimodal $c/a$-posterior.

\item[(b)] \textbf{cos$\theta_{\rm o}$ or $C_{\rm Frac, tor}$:} In the case of $M_{\rm BOR} \rightarrow D_{\rm CTOR}$ we get cos$\theta_{\rm o} \simeq 0.39$, which is consistent with cos$\theta_{\rm i}$. The implications are similar to that of the fit obtained from the MCT regime. In the case of $M_{\rm BOR} \rightarrow D_{\rm UXCL}$, the uncoupled configuration ($N_{\rm H, los} \neq N_{\rm H,eq}$) returns a bimodal $\rm cos \theta_{\rm o}$ or $C_{\rm Frac,tor}$ (one major and one minor peak in the posterior), which leads to bimodality in $\theta_{\rm i}$, $A_{\rm Fe}$ and $C_{\rm scpl}$. For this case we find $\cos \theta_{\rm i} < \cos \theta_{\rm o}$. Adding this up with the fact that $N_{\rm H,los} \neq N_{\rm H,eq}$ it is evident the otherwise `$\cos \theta_{\rm o}$'-parameter makes more sense when interpreted as covering fraction $C_{\rm frac,tor}$ of a clumpy torus as described in the fifth point in section \ref{CM-MCT}.

\item[(c)] \textbf{$\sigma_{\rm o}$ of UXCLUMPY:} $M_{\rm UXCL} \rightarrow D_{\rm CTOR}$ returns $\sigma_{\rm o} \simeq 14^{\rm \circ}$ with $\delta \simeq 0.1$ and $M_{\rm UXCL} \rightarrow D_{\rm BOR}$ returns $\sigma_{\rm o} \simeq 12^{\rm \circ}$ and $\delta \simeq 0.1$. In both the cases the joint analysis of \textit{XMM--Newton} and \textit{NuSTAR} returned fits for which $1.2 \leq$ $\chi^2$/dof $\leq 1.3$ (mainly due to discrepancy in the emission lines).  We simulate `surrogate' data sets under the best fit parameters of UXCLUMPY under both $M_{\rm UXCL} \rightarrow D_{\rm CTOR}$ and $M_{\rm UXCL} \rightarrow D_{\rm BOR}$ best fits parameters and carry out IM-fits. We find that the IM-fits in the surrogate data sets return wider posteriors (Table \ref{original_surrogate}). For $\sigma_{\rm o}$, $\delta_{\rm IM}/\delta_{\rm CM}$ is 1.75 for BORUS and 1.77 for CTORUS. Both the fits $M_{\rm UXCL} \rightarrow D_{\rm RXT}$ and $M_{\rm UXCL} \rightarrow D_{\rm RXT}$ returned very low values of $\sigma = 7^{\rm \circ}$.

\item[(d)] \textbf{$C_{\rm frac}$ of the inner ring of UXCLUMPY:} Unlike the MCT regime, values of $C_{\rm frac}$ returned from the $M_{\rm UXCL} \rightarrow D_{\rm CTOR}$ and $M_{\rm UXCL} \rightarrow D_{\rm BOR}$ were clearly non-zero and assumed significantly higher values ($C_{\rm frac} \simeq 0.3$ for both) with tighter constraints $\delta \sim 0.01$ for $M_{\rm UXCL} \rightarrow D_{\rm BOR}$ and $\delta \sim 0.001$ for $M_{\rm UXCL} \rightarrow D_{\rm CTOR}$. For the \textit{NuSTAR} only fits as described in the previous point, the values of $C_{\rm frac}$ and $\delta$ are similar to those of the joint fits for $M_{\rm UXCL} \rightarrow D_{\rm BOR}$. For the $M_{\rm UXCL} \rightarrow D_{\rm CTOR}$ fits the $C_{\rm frac}$ value decreased but the $\delta$ increased to 0.5. The `surrogate' data sets under IM fits returned broader posteriors (Table \ref{original_surrogate}). For $C_{\rm frac}$, $\delta_{\rm IM}/\delta_{\rm CM}$ is 4.04 for BORUS and 6.67 for CTORUS. For both the RXT and MYT fits the posteriors are similar in shape and the median values are similar.

\item[(e)] \textbf{$N_{\rm cloud}$ of CTORUS:} We carried out the CM fits $M_{\rm CTOR} \rightarrow D_{\rm j}$ where j = MYT, UXCL and BOR. For $M_{\rm CTOR} \rightarrow D_{\rm UXCL}$ and $D_{\rm MYT}$ we do not get a good fit ($\rm 1.2 \leq \chi^2/dof \leq 2.0$) mostly because of the difference in the emission line profiles between UXCLUMPY and CTORUS. We get an \textit{irregular} posterior for $N_{\rm cloud}$ converging towards the higher extreme edge of the prior range for both cases. For $M_{\rm CTOR} \rightarrow D_{\rm BOR}$ the $N_{\rm cloud}$ posteriors are extremely narrow.

\end{itemize}
\end{itemize}

\begin{table*}
\begin{tabular}{|l|c|c|c|c|c|c|c|c}
\hline
\hline
Parameters & $A_{\rm Fe}$  & log $\xi_{\rm i}$ & $N_{\rm H,los}$ (cm$^{-2}$)  & $\Gamma$ & $\sigma_{\rm o}$ ($^{\circ}$) & $C_{\rm frac}$ & $\theta_{\rm i} $ ($^{\circ}$) & $\chi^2$/dof\\
\hline
Input&&& \\
\hline
MCT &  1.0   & 1.0 & 100.0 &   1.9    &  45.0    &  0.4   &  60.0  & --     \\
HCT &  1.0   & 1.0 & 500.0 &   1.9    &  45.0    &  0.4   &  60.0  & --     \\
\hline
$M_{\rm UXCL} \rightarrow D_{\rm UXCL+RELXILL}$&&& \\
\hline
MCT &- &- & $102.0\pm1.24$    &$1.83\pm0.03$ &$32.0^{+8.1}_{-4.4}$ &$0.57^{+0.02}_{-0.04}$ &$46.6^{+5.6}_{-4.3}$ & 0.978  \\
HCT &- &- & $547^{+58}_{-57}$ & $1.72 \pm 0.02$ & $32.8^{8.9}_{-4.6}$& $0.47^{+0.01}_{-0.01}$&  $24.0^{9.7}_{-14}$ & 1.046 \\ 
\hline 
$M_{\rm UXCL+KD-XIL} \rightarrow D_{\rm UXCL+ RELXILL}$&&& \\
\hline 
MCT &  $0.61^{+0.14}_{-0.09}$  &  $0.62^{+0.64}_{-0.54}$ & $97 \pm 2$ & $1.86\pm 0.03$  & $48.6^{+11.5}_{-14}$  & $0.34^{0.09}_{-0.1}$   & $77^{+6.5}_{-14}$ &  0.974 \\
HCT &  $3.36^{+5.6}_{-2.41}$ & $2.31^{+1.5}_{-2.0}$ & $516^{+106}_{55}$ & $1.9^{+0.04}_{-0.03}$ &  $51^{+12}_{-11}$ &  $0.44^{+0.06}_{-0.08}$   &  $50.6^{+11.4}_{-25}$ & 1.035 \\
\hline 
\hline 
\end{tabular}
\caption{Table for input and the fit parameters for data simulated under torus and relativistic disc. Two cases are shown in this table. In the first case, only the torus ($M_{\rm UXCL}$) is fit to the data where both the torus and disc components $D_{\rm UXCL+ RELXILL}$  are present. In another case, the fit was performed with both the relativistic disc and the torus.} 
\label{relxill_fit}
\end{table*}

\section{RESULTS: DETECTABILITY OF AN ADDITIONAL RELATIVISTIC DISC-REFLECTION COMPONENT} 
\begin{figure*}
\includegraphics[scale=0.61]{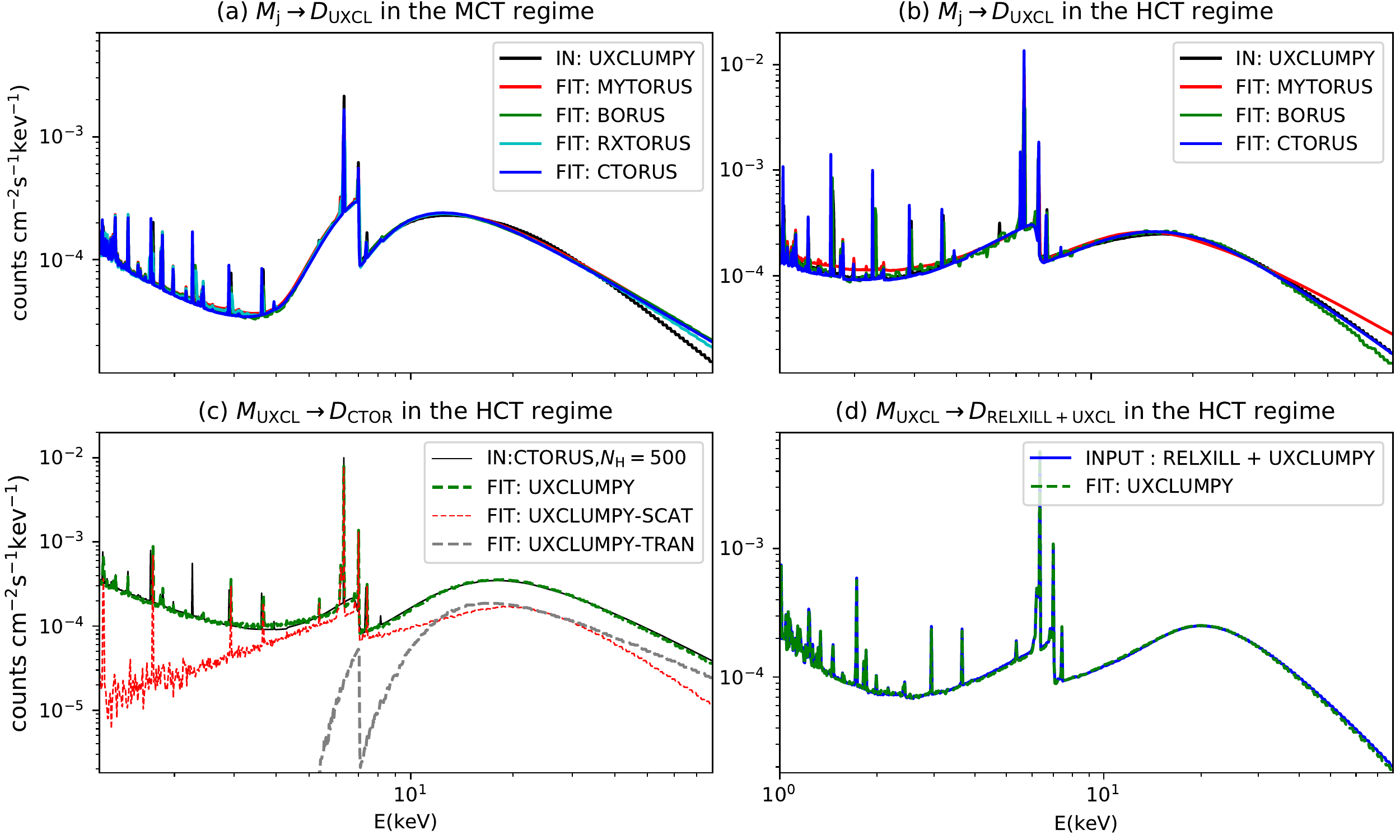} 
\caption{Cross model fits: overplot of UXCLUMPY input model and fitting models (a)in the MCT regime (b)for the HCT regime with the Compton-thick ring absent ($C_{\rm frac} = 0$) (c) Overplot of CTORUS input and $M_{\rm UXCL} \rightarrow D_{\rm CTOR}$ best-fitting model. The zeroth-order continuum adjusts itself to replicate the CRH of CTORUS (d)Overplot of input model comprised on a weak disc component(RELXILL) and torus component(UXCLUMPY) and fitting model with only a torus model(UXCLUMPY)\label{spec_overpl}}
\end{figure*}

In a Compton-thick AGN, the transmitted and reprocessed emission of the torus is the dominant part of the spectrum. However, there may be an additional component present in the spectrum originating due to relativistic reflection from the inner accretion disc. The question thus is, would it be possible to detect such a component from spectral model fits? To answer the question, we simulate a data set that has both the CRH-dominated torus component and a relativistic disc-reflection component. We use UXCLUMPY to simulate the torus and RELXILL \citep{relxill14} to simulate the disc-reflection component ($D_{\rm UXCL+RELXILL}$). Since the disc reflection component originates close to the ISCO, it should be heavily absorbed by the torus column. Thus the model expression in \texttt{xspec} terminology is:\\
\texttt{MODEL = TBABS$\times$(APEC(1)+APEC(2)+ UXCLUMPY-SCPL + ZTBABS$\times$CABS$\times$RELXILL + UXCLUMPY)} \\
The \texttt{ZTBABS} and \texttt{CABS} $N_{\rm H, los}$ was tied to UXCLUMPY as the obscurers are one or multiple torus clouds. The parameters of RELXILL like $\Gamma$ and $\theta_{\rm i}$ were tied to the UXCLUMPY as the torus and the disc are illuminated by the same coronal power law. In the flux range of 4 to 50~keV, the ratio of the flux of the disc absorbed through the torus column with respect to the total torus emission $F_{\rm 4-50,Disc}/(F_{\rm 4-50,trans} + F_{\rm 4-50,refl})_{\rm Torus}$ is 0.36 to 0.38, which makes the total torus emission (comprising of the torus CRH and the transmitted component) the dominant feature of the spectra. In the published UXCLUMPY model, the CRH from the torus is calculated assuming a power law. But when a disc reflection is present, the torus should reprocess both the power law and disc-reflection component. It was not possible for us to implement this in the current UXCLUMPY setup. This will put some limitations on the accuracy of our test but we do not expect it to significantly change the results. When our fitting model has only torus component (UXCLUMPY) and the data has both the disc \texttt{RELXILL} and torus component the notation would be $M_{\rm UXCL} \rightarrow D_{\rm UXCL+RELXILL}$ and in the converse case, the disk is modelled with \texttt{KDBLUR(XILLVER)}, the notation is $M_{\rm UXCL+KD-XIL} \rightarrow D_{\rm UXCL}$. During analysis, the prior of the relative normalization of the disc-component $C_{\rm KD-XILL}$ was set to be a log-uniform prior extending from $10^{-10}$ to 1. All other parameters of the disc that were kept free (ionization parameter log$\xi_{\rm i}$ and abundance $A_{\rm Fe}$) were given uniform priors extending accross the entire prior range. The methodology of analysis of the data is described in the following bullet points:
\begin{itemize}
\item[(i)] \textbf{TORUS COMPONENT ONLY:} We fit the data  with torus component only i.e. $M_{\rm UXCL} \rightarrow D_{\rm UXCL+RELXILL}$ to see if it is possible for only a torus modelled by UXCLUMPY to replicate the spectra. We keep the APEC components and $E_{\rm cut}$ frozen to their input values. We find that the spectral shape was replicated and there no significant model or data excess in the residuals for both MCT and HCT regime. The over-plot of the input-model and fit model for the HCT regime is shown in Fig.~\ref{spec_overpl}. At the median values of the parameter posteriors, we obtain $\chi^2/$dof = 1.046 for HCT and 0.978 for MCT. The parameters like $\Gamma$, $C_{\rm frac}$ of UXCLUMPY and $\theta_{\rm i}$ showed large discrepancies with the input. $\sigma_{\rm o}$ was consistent with the input. The fit returned a lower value of $\Gamma$ and a higher value of $C_{\rm frac}$ (see Table \ref{relxill_fit}).

\item[(ii)] \textbf{TORUS AND DISC COMPONENT:} We fit the simulated data with UXCLUMPY and an attenuated disc component modelled by \texttt{ZTBABS} $\times$ \texttt{CABS} $\times$  \texttt{kdblur}\footnote{\cite{kdblur91}}(\texttt{XILLVER}\footnote{\cite{xillver13}}). This produced a better fit compared to the previous case, with $\chi^2$/dof $\simeq$ 1.035 for HCT and 0.974 for MCT, with no significant residuals. The \texttt{XILLVER} parameters like $A_{\rm Fe}$ and $\chi_{\rm i}$ show very broad posteriors, which is attributable to the heavy extinction through the torus column.
\end{itemize}
Detecting an additional relativistic disc reflection component which is heavily attenuated in the torus absorbing material with certainty is impossible with simple  $\chi^2$/dof values. In all the fits with only the torus model (UXCLUMPY), the model replicates the data with almost indistinguishable values of $\chi^2$/dof. But the low Bayes factor (see $\S$\ref{sec-6} for definition), BF=$\rm Z_{torus}/Z_{torus+disc} \simeq 1.21 \times  10^{-19}$ for HCT and $9.25 \times 10^{-8}$ for MCT clearly hints at the existence of a relativistic disc reflection component.
We also test the possibility for false detection of a disc component in the HCT regime and carry out the $M_{\rm UXCL+KD-XIL} \rightarrow D_{\rm UXCL}$ fit. We found that the parameters of UXCLUMPY got recovered, but the parameters of XILLVER showed \textit{irregular} posteriors. Specifically, the tendency of the $C_{\rm KD-XILL}$ posterior-distribution to converge towards the lower limit ($\sim 10^{-8}$) indicated the absence of the disc reflection component. The log-evidence values for the $M_{\rm UXCL+KD-XIL} \rightarrow D_{\rm UXCL}$ and $M_{\rm UXCL} \rightarrow D_{\rm UXCL}$ are log$\rm Z_{torus+disc} \simeq -652.19$ and log$\rm Z_{torus} \simeq -653.67$ respectively. Thus, the bayes factor is BF=$\rm Z_{torus}/Z_{torus+disc} \simeq 0.033$, the value being greater than our threshold of 0.01 (Section \ref{sec-6}) for model distinction. Nevertheless, the results of $M_{\rm UXCL+KD-XIL} \rightarrow D_{\rm UXCL}$ is consistent with the absence of a relativistic disc refelction.
 
\begin{figure*}
\includegraphics[scale=0.61]{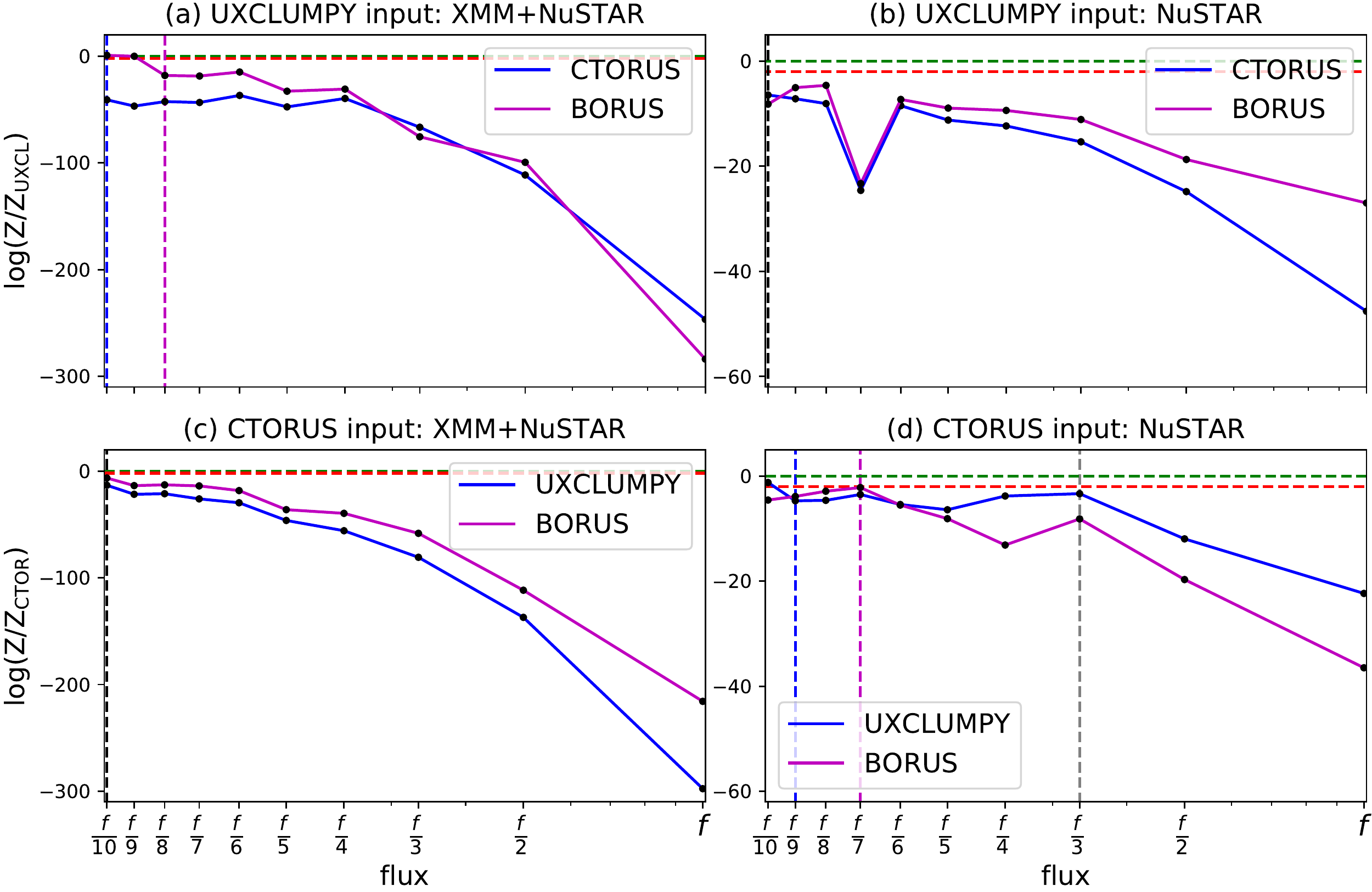}
\caption{log Bayes Factor vs flux level plot: Spectra simulated for \textit{XMM--Newton} and \textit{NuSTAR} exposures of 100 ks and 50 ks respectively with an intrinsic flux level of $f_n = 0.83/n$ mCrb. (a) and (c) demonstrate the variation of evidence values with flux   when both \textit{XMM--Newton} and \textit{NuSTAR} data are used. (b) and (d) demonstrate the same, but considering \textit{NuSTAR} only data. The blue and magenta dashed lines denote the flux level at which the relation $\rm log Z_{\rm input} - log Z_{\rm fit} \rm \geq 2$ holds for the wrong model. The black and green dashed lines marks the logBF=2 and logBF=0 level, respectively. 
\label{bf_vs_flux}}
\end{figure*}

\section{RESULTS: DEPENDENCE OF EVIDENCE ON FLUX LEVELS} \label{sec-6}
To study the effect of flux levels on model evidence, we simulate 10 spectra ($S_n$, $n$ runs from 1-10) for 10 different flux levels under UXCLUMPY and CTORUS for an intrinsic 2--10~keV flux of $0.83/n$ mCrb in the HCT regime. The exposures on all the simulated spectra are $100$ ks on \textit{XMM--Newton} and 50 ks on \textit{NuSTAR}. All spectra are grouped into 30 counts bin$^{-1}$. We run BXA sims on the simulated spectra for both: (A)joint \textit{XMM--Newton} and \text{NuSTAR} (B)\textit{NuSTAR} only cases. Our Bayesian fitting strategy can be summarized in the following representation: $M_{\rm j} \rightarrow D_{\rm UXCL}$ and $M_{\rm j}\rightarrow D_{\rm CTOR}$, where j = UXCL,CTOR,BOR. In our work, Bayes factor is defined as the ratio of the Bayesian evidence of a fitting model ($Z_{\rm fitting \: model}$) to that obtained from an intramodel fit ($Z_{\rm input}$): $\rm BF = Z_{\rm fitting \: model}/Z_{input}$. We plot the Bayes factor values to the corresponding flux levels (Fig.~\ref{bf_vs_flux}). For effective distinguishing of models, we set the following condition on Bayes factor (BF): $\rm BF < 10^{-2}$, which practically means that the flux value below which the \textit{$\rm log BF$ value rises above $-2$ for the first time} is considered as the lower limit on flux, below which effective distinguishing of \textit{correct} from the \textit{wrong} model is not possible. We expect that joint fits are more effective in distinguishing models. Our results confirm this from the fact that we can distinguish the models more effectively down to an intrinsic flux value of $\sim 0.1$ mCrb for joint \textit{XMM--Newton} plus \textit{NuSTAR} fits. For the \textit{NuSTAR} only fits, for which this lower threshold value was slightly higher than $\sim 0.1$ mCrb (Fig.~\ref{bf_vs_flux}b and d). It is also worth noticing that systematically lower values of logBF and local fluctuations in logBF (e.g. as seen in the \textit{NuSTAR}-only fits; Fig.~\ref{bf_vs_flux}) can push the threshold to higher limits (grey vertical line at flux level $f/3$ in Fig.~\ref{bf_vs_flux}d). These fluctuations which might have occurred during the data simulation or detection process are more likely to influence the evidence values at the lower flux regimes or lower bandwidth data, thus reducing the precision in evidence-based model distinction. Thus, it is safer to assume more conservative thresholds determined from synthetic data simulations, when applying Bayes factor for model distinction in case of real data . We discuss these aspects in detail in section \ref{discussions}.

\section{DISCUSSIONS}\label{discussions}
\subsection{Guidelines to the X-ray Community}\label{disc-1}
\subsubsection{Parameter Estimates}
We perform joint analysis of synthetic data simulated under \textit{XMM--Newton} and \textit{NuSTAR}. The baseline flux level for all the input models was $\sim 0.5$~mCrb in the 2--10~keV band. The exposure times on \textit{XMM--Newton} and \textit{NuSTAR} were 100~ks and 50~ks respectively. For \textit{XMM--Newton} EPIC-pn, given the flux and exposure time in the 2--10~keV band, the average photon-count is $\sim 8.2 \times 10^{4}$. From the results of our analysis of the synthetic data, we assert the following prescriptions for the parameter posteriors:
\begin{itemize}
\item[(i)] From the cross-model fits, it is quite clear that \textit{the photon index ($\Gamma$) cannot be determined with certainty when a wrong model is applied}. Almost all the posteriors are regular monomodal distributions with very small statistical errors, but the discrepancy (represented by the $r_{\rm q}$ values) between the input and fit values are extraordinarily high ($\Delta \Gamma \simeq 0.3$ and $r_q$ as high as 16). This discrepancy in $\Gamma$ for different models is also reported by \cite{buchner21}. Even multiple models returning similar values of $\Gamma$ should not be taken at face value (as discussed in $\S$\ref{CM-fits}). \textit{Our results indicate that precision in the value $\Gamma$ is no guarantee for its correctness.}  In such a scenario, where the systematic discrepancies originating due to model-difference cannot be corrected, we recommend that users apply caution in the interpretation of $\Gamma$ (e.g., when used to estimate intrinsic luminosity or accretion rate relative to Eddington), particularly if very low values of $\Gamma$ are obtained. Alternate, indirect methods to determine the value of $\Gamma$ may be preferred. For example, \citet{brightman13} noted an empirical relation between $\Gamma$ and FWHM of optical broad emission lines in X-ray selected, broad-line radio-quiet AGN. However, such a sample is biased to be X-ray-unobscured, and this relationship might not hold in X-ray obscured AGNs. \textit{The community would thus benefit from exploration into whether any empirical relation between $\Gamma$ and other measurable properties of Compton-thick AGN exists.}

\item[(ii)] Consistent values of LOS column density  ($N_{\rm H,los}$) were returned in the medium Compton thick (MCT) regime, because of the strong zeroth-order continuum. In the heavy Compton thick (HCT) regime, the values of $N_{\rm H,los}$ were recovered but were less constrained as seen from most of the intramodel fits (IM-fits). However, when \textit{wrong} models were applied (CM-fits) many cases returned discrepant values and irregular posteriors. The determination of the precise value of $N_{\rm H,los}$ starts to become difficult in the HCT regime, because the suppression of a zeroth-order continuum by the scattered continuum. However also in the HCT regime, in the CM-fits there are cases when the intensity of the zeroth-order continuum is modulated by the $T/R$-parameter (arbitrarily taking values > or < 1) to adjust the shape of the CRH (Fig.~\ref{rxt_uxcl_hct}). \textit{The large discrepancy of $T/R$ can result in an equally sized mis-estimation of the intrinsic accretion luminosity ($L_{\rm corona}$). ($L_{\rm X,corona}$), which can be severely discrepant} ($L_{\rm corona,fit}$ upto $\sim$ 10 times $L_{\rm corona,input}$ in our simulations).

\item[(iii)] The IM-fits returned comparatively narrower posteriors (Fig.~\ref{mct-hct}) in the HCT regime for most parameters of torus morphology. The constraints on the parameters of morphology are drawn from the $E<6$ keV tail, Fe K edge, and CRH of the scattered continuum which are dominant in the HCT regime. \textit{However, these comparative levels of improvements in the constraints in the HCT regime w.r.t. to MCT can be dependent on the exact spectral shape.} However, it can be inferred that under the \textit{correct} model assumption parameters of morphology on average can be better constrained in the HCT regime, as the HCT regime has lower `contamination' from the zeroth-order continuum. However, the behavior of the posteriors of the morphological parameters shows different trends for different models for intramodel fits and cross-model and are open to interpretation:
\begin{itemize}
\item[(a)] Continuous torus: In the continuous torus model there are only two morphological parameters: $c/a$ for doughnuts and $\theta_{\rm o}$ or $\cos \theta_{\rm o}$ or $C_{\rm frac, tor}$ in the biconical cutouts. The doughnut thickness parameter $c/a$ (of RXTORUS) shows \textit{regular} posteriors (section \ref{sec-2.4} for definition) in all cases. However, low values of $c/a$ (e.g. $c/a < 0.2$) were observed for the baseline (coupled) configuration, which indicates an annular geometry rather than a doughnut. \textit{Meanwhile for the uncoupled configuration one interpretation of the value of $c/a$ can indicate the reflector geometry generating the scattered component (see section \ref{CM-hct}).} Several other interpretations with different setups for an otherwise doughnut model are explained in \cite{yaqoob12}. $\cos \theta_{\rm o}$ or $C_{\rm frac,tor}$ of BORUS returns posteriors which are regular for most cases. However for some fits to \textit{wrong} models in the coupled configuration ($N_{\rm H,los} = N_{\rm H,eq}$) we get $\theta_{\rm o} > \theta_{\rm i}$. This suggests an unobscured configuration. However these cases returned a $N_{\rm H,los}$ consistent with a Compton thick obscured case. For our case, we know that the applied model is incorrect. However in real situation where the input spectrum is unknown and these class of solutions are encountered there may a create potential chance of misinterpretation of an otherwise wrong solution (see the point on $\cos \theta$ or $C_{\rm frac,tor}$ in the section \ref{CM-MCT}).

\item[(b)] Clumpy torus: In the clumpy torus models there are three morphological parameters: $\sigma_{\rm o}$ and $C_{\rm frac}$ of inner ring of UXCLUMPY and $N_{\rm cloud}$ of CTORUS. When UXCLUMPY is the wrong model (CM-fits) $\sigma_{\rm o}$ systematically showed narrower posteriors (lower values of $R_{90}$ and $\delta$) compared to the case when it is the correct model (IM-fits) at the same flux levels. The covering fraction ($C_{\rm frac}$) of the inner C-thick ring of UXCLUMPY returned \textit{irregular} posteriors (section \ref{sec-2.4} for definition) with $C_{\rm frac} \sim 0$ for majority of the tested cases in the MCT regime. However in the HCT regime the posteriors of $C_{\rm frac}$ were monomodal with comparatively higher values ($C_{\rm frac} > 0.1$). \textit{This could be interpreted as the tendency of the UXCLUMPY geometry to adjust the overall covering fraction of the clump distribution to match the shape of the data generated under another morphology.} $N_{\rm cloud}$ of CTORUS returned mostly \textit{irregular} posteriors when CTORUS is the \textit{wrong} fitting model, which had either a minimum or converged at one of the extreme edges of the priors. For the intramodel fits, however, the $N_{\rm cloud}$ posteriors are wide but mostly \textit{regular}.
\end{itemize}
Although it might not be a generic trend there might exist differences in the nature of the posteriors for a \textit{wrong} and \textit{correct} model. Thus in real data analysis a rigorous test of the behavior of the parameters of the fitting model becomes a necessity. \textit{This can be done by simulating synthetic data under the fitting model for the values of the best-fit parameters and then performing an intramodel fit. The posteriors properties (e.g. $\delta$, $R_{90}$) for the real data and the intramodel fit can then be compared.}

\item[(iv)]  Posteriors of $\theta_{\rm i}$ are generally regular-monomodal in case of application of the correct model (IM fits). However, in several cases the proportional errors are large. In general, the constraints on $\theta_{\rm i}$ are dependent on the constraints on other morphological parameters, which are better determined in the HCT regime with a dominant scattered continuum allowing better constraints. However, when the applied model is wrong (CM-fits), in several cases we got \textit{irregular} posteriors (converging at the edge) especially those involving the fitting of UXCLUMPY. In general, $\theta_{\rm i}$ remains a parameter dependent on morphological assumptions and hence \textit{should not always be accepted at face value}.

\item[(v)] As described before a significant fraction of cases where a \textit{wrong} model is applied (CM fits), returned \textit{irregular} posteriors  for some of the parameters [e.g. $\theta_{\rm i}$, $N_{\rm cloud}$, $C_{\rm frac}$ etc. (see Table \ref{tab:1}) for a summary of the parameters]. However, for a significant number of fitting models, the best value (typically the median) derived from the posteriors of the cross-model fits returned $\chi^2/dof < 1.2$, which is otherwise acceptable. Also in most of these cases, it is possible to improve the value of $\chi^2/dof$ significantly by adding emission lines or by decoupling the zeroth-order and the scattered continuum (see section \ref{CM-hct}), etc. This might return meaningful values of certain parameters (e.g. $N_{\rm H,los}$), but the morphological and radiative aspects remain open to (mis-)interpretation (e.g. point-3 of sections \ref{CM-MCT} and \ref{CM-hct}). \textit{The danger of misinterpretation is less in the case for \textsc{multinest} as in addition to $\chi^2/dof$ estimates we can apply Bayes factor to weigh up the model preferences and explore the nature of intramodel posteriors.}
\end{itemize}
We thus strongly recommend that the users perform their data simulations and use global Bayesian algorithms for data analysis. The methodology used here and the exercise undertaken in this work can be treated as an example. We advocate simulating data under selected models of choice to understand potential limitations both in cases of preparing observing proposals and when modeling already obtained data. We leave it to the readers to use the responses/ARFs of the selected instrument(s) as appropriate, and perform intramodel and/or cross-model fits as needed to understand the behavior of the posteriors for the given models, instrument, exposure time, etc.

\subsubsection{Limitations of using Bayes factor values}\label{bf_limitations}
From the results in $\S$6, it is clear that values of log-evidence (log$Z$) and log-Bayes factor (log BF) are strongly affected by the number of degrees of freedom, which is a strong function of the number of data bins, and availability (or lack thereof) of certain instruments and bandpasses. For a given flux/exposure time, maximizing the energy bandpass used by, for example, performing joint \textit{NuSTAR} and \textit{XMM--Newton} fits instead of fitting just one instrument maximizes the ability to distinguish between models based on Bayes Factor.  In the case of joint \textit{NuSTAR} and \textit{XMM--Newton} fits to the synthetic Compton-thick AGN spectra (simulated under the selected models; section \ref{sec-6}) , adopting a threshold log-Bayes factor of $\gtrsim 2$ was effective in model distinction for cases with 2 -- 10 keV fluxes brighter than $\sim 0.08$ mCrb (Fig.~\ref{bf_vs_flux}), for 50/100 ks exposures with \textit{NuSTAR/XMM--Newton} respectively. For the \textit{NuSTAR}-only fits the availability of fewer channels (data points) brings the logBF (Bayes factor) value closer to 0 (1), for a given intrinsic flux value. Additionally, as demonstrated in section \ref{sec-6}, `local' random fluctuations in values of logBF occur (Fig.~\ref{bf_vs_flux}), but whether these fluctuations will impact distinction of a \textit{wrong} model from a \textit{correct} model depends flux or availability of data in a certain band. For our examples (Section \ref{sec-6} and Fig.~\ref{bf_vs_flux}), for the joint fits, the values of logBF are more negative and lie comfortably below the selected threshold (up to a flux range of $f/9$ or 0.09 mCrb) despite random fluctuations. Meanwhile, for the \textit{NuSTAR}-only fits the values of logBF are closer to the threshold and thus run more into the risk of crossing into the logBF$>0$ zone under the influence of random fluctuations leading to a false inference on the morphology. The grey and vertical dashed line in Fig.~\ref{bf_vs_flux}d marks the flux value where logBF for $M_{\rm UXCL} \rightarrow D_{\rm CTOR}$ comes dangerously close to the logBF-threshold (but does not cross it) for a \textit{NuSTAR}-only fit. Thus, there exists a possibility that for a given threshold on logBF, the threshold on flux required for an effective model distinction can get pushed to a higher value in the absence of a certain band/instrument and or presence of random fluctuations in values of logBF. Hence, these threshold values are approximate estimates and should not be considered as absolute. It should also be noted that these results from our examples are based on the assumption that the soft band emission (e.g. emission modelled with APEC) is modelled perfectly, which might not be the case in real situations(see 7.2). This is because of the involvement of several complicated factors related to measurement, contamination from other sources in the field of view and physical modeling of the system, \textit{we advocate that the readers perform their simulations setting a reasonable limiting value of logBF, to estimate the appropriate threshold on intrinsic flux necessary for model distinction.}

\subsection{Limitations of our analysis}\label{limits}
\begin{itemize}
\item[(i)]The spectral shape can determine counts in a band and various input parameters used for data simulation. For example, different values of $N_{\rm H,los}$, different levels of contribution from the scattered power law, etc.\ can contribute to enhancement or suppression of torus features in the 2--10 keV band.  This might also affect the relative spectral counts collected by \textit{XMM--Newton} and \textit{NuSTAR} given an exposure time, thus resulting in variation in the constraints. Effects of variation in spectral shape on parameter constraints were shown for a few parameters like $N_{\rm cloud}$ (Fig.~\ref{ncl-mct-hct}). Comprehensive testing of all possible cases of spectral shapes for all models is out of the scope of this work. However, in the case of real data analysis, these spectral shapes should be taken into consideration.

\item[(ii)] In real data reduction, the given spectrum has unknown parameters and we model it as best we can. But there will always be some discrepancy between the modelled parameters and the real spectral parameters. This consequently results in statistical and systematic variations in the location estimates (e.g. median) of the parameter posteriors during the data generation process. A similar situation is encountered for simulated data. To undertake such a study one can simulate a very large number of spectra (of order $10^4$) \citep[e.g.][]{gonzalez_martin19} using \textsc{fakeit} in \textsc{xspec} \citep[e.g.][]{markowitz06} and then carry out Bayesian analysis in bulk. We performed such a study on a smaller scale in $\S$\ref{IM-bulk}, wherein we simulated and fit 100 spectra. Given the heavy computational requirements for the study, we limit ourselves to only 100 spectral fits. The distributions of the posterior properties can be studied better with a much larger number of simulated data sets.

\item[(iii)] The models used here differ in two aspects namely, the assumed morphology and the assumptions on radiative physics. The results of our simulations have been limited by the simultaneous occurrence of both radiative and morphological differences between the models. Thus, we could not effectively differentiate between the effects of the radiative and the morphological differences in our current study. To test solely the effect of the morphological difference of the torus, it is necessary to set the radiative properties across different morphologies to be constant, so that the various properties of the CRH are just the effect of torus morphology only. Therefore, only one radiative transfer code e.g. XARS should be used to simulate different spectra for different morphology which in turn should be used for model testing. Conversely, to test the variation that radiative physics, different radiative transfer codes like \texttt{XARS, REFLEX, GEANT}, etc. could be used to simulate the spectra for a fixed morphology.

\item[(iv)] In our analysis, we assumed only simplistic situations while modeling the emitted radiation from a Compton-thick obscured AGN. More specifically we did not take into account any additional emission from other X-ray bright sources which might be present in the host galaxy. In a real scenario, there might exist significant contamination from other X-ray bright sources like ULXs, X-ray binaries, etc.\ This has been demonstrated in \cite{arevalo14} for the case of the Circinus galaxy, where several X-ray bright sources contaminate different energy bands at different levels. In our work, we have demonstrated the energy/band dependence of certain parameters. Thus, contamination can lead to wrong estimates of the parameter. It is thus required that the contaminants are accounted for and are modelled correctly.

\item[(v)] In the future, it is expected that more physically self-consistent torus models and complex-torus models \cite[shown schematically in e.g.][]{esparza21} will be developed, incorporating additional morphologies, cloud distributions, and physical processes. In addition, the community can look forward to data from new X-ray missions, including \textit{XRISM} and \textit{ATHENA}: each of these missions will feature a calorimeter that can resolve subtle features in fluorescent emission line complexes such as the Compton shoulder \citep[e.g.][]{yaqoob10_1,tanimoto19}. Even presently, data from currently active missions other than \textit{XMM--Newton} and \textit{NuSTAR} can be used for analysis. Thus, performing additional tests similar to that performed in this work is a necessity, when one uses a model and data from an instrument not been tested here.
\end{itemize}

\section{SUMMARY}\label{summary}
In this project we test the reliability of the estimates of the model parameters for an X-ray obscured Compton-thick AGN. We simulate synthetic data under six models with the instrument functions of \textit{XMM--Newton} and \textit{NuSTAR} at a flux level of $\sim 0.5$~mCrb in the 2--10~keV band. The exposure times were 100~ks on \textit{XMM--Newton} and 50~ks on \textit{NuSTAR}. Using Bayesian analysis techniques we perform intramodel (IM) and cross-model fits on the synthetic data using these models. We investigate the level of degeneracy between the parameters, the ability to distinguish models and the different ways a wrong model fitting results in incorrect interpretation of the morphological and central engine parameters (e.g. photon index). We point out the challenges and limitations concerning the analysis of single epoch spectra.

In the intramodel fits, we find that the parameters are mostly recovered. However many morphological parameters showed very wide posteriors covering more than $30\%$ of the prior range in the medium Compton-thick regime ($N_{\rm H,los} < 2 \times 10^{24}$~cm$^{-2}$). Constraints on $N_{\rm H,los}$ worsen and the constraints on the morphological parameters improve in the heavy Compton-thick regime ($N_{\rm H,los} > 2 \times 10^{24}$~cm$^{-2}$). We also demonstrated the effect of a particular energy band (or instrument) which is necessary for determining reasonably good constraints on the parameters.

We also discuss a few ways in which a wrong model can fit a given data set and lead to false conclusions on the morphology along with other properties (e.g. $N_{\rm H,los}$/$N_{\rm H,eq}$) of the obscurer and or reflector. When a wrong model is fit to a particular data set, the photon index ($\Gamma$) of the coronal power law (a common parameter across all models) in most of the cases returned posteriors showing that $\Gamma$ is constrained with a proportional error of $<$ 3\% but with significant discrepancy with the input. However for wrong model fits in case of the parameters which are present in the fitting model but not in the model used for data simulation, a different method can be adopted. For a few cases involving fitting with UXCLUMPY viz. $M_{\rm UXCL} \rightarrow D_{\rm j}$ where j = MYTORUS, CTORUS and BORUS, we compare the posteriors obtained from the CM-fit to an that obtained from and IM-fit to a `surrogate' data set (simulated under fitting model with the median values obtained from the CM-fits). The analysis of the `surrogate' data set returned broader posteriors compared to the CM-fits for the morphological parameters $\sigma_o$ and $C_{\rm frac}$ of the inner ring (see model description). \textit{It is thus established that the precision in the value obtained from a fit is no guarantee for its correctness and does not necessarily indicate the application of a model consistent with the real morphology.} Our results from the simulations thus indicate the dangers of accepting fit results at face value.

The parameter spaces of Compton-thick AGN with limited data are complicated and sometimes multi-modal, and thus demand global parameter exploration algorithms. With the use of Bayes factor from \textsc{bxa/multinest} runs it is possible to rule out the least preferred models for a given data set, at least at the higher end the available data quality. However, the use of synthetic data (simulated under the models in use) to assess the possibility of random fluctuations in the values of Bayes factor, \textit{which can potentially lead to false model conclusions} (Section \ref{sec-6}, \ref{bf_limitations} and Fig.~\ref{bf_vs_flux}), is recommended given an instrument combination and or flux regime. We thus conclude that measuring the properties of Compton-thick obscurers or reflectors with single epoch data obtained jointly from \textit{XMM--Newton} and \textit{NuSTAR} bands is not straightforward, even without the complicated effects of contamination. However, a study with multi-epoch data can help us study variability (to detect variations in $N_{\rm H,los}$) and thus partial conclusion on the morphology can be drawn \cite[e.g.][]{markowitz14,balokovic18,laha2020,esparza21}. Thus time-variability analysis coupled with spectral analysis can yield better constraints on morphology.

The results of all the analysis carried out in this work thus suggest the importance of performing synthetic data simulation and analysis alongside real data using the instruments and the models of interest. This will help understand the potential limitations of the analysis, the models and the instruments. Hence, it is recommended that the users of these models perform similar simulations considering this work as example and noting corresponding potential limitations.

\section*{ACKNOWLEDGEMENTS}
The authors thank Prof. Dr. hab. Agata {R{\'o}{\.z}a{\'n}ska} for her valuable feedback on the scientific results and their broader significance in the X-ray astronomy. The authors also thank the anomymous referee for the insightful comments. This research has made use of data and software provided by the High Energy Astrophysics Science Archive Research Center (HEASARC), which is a service of the Astrophysics Science Division at NASA/GSFC. This work was financially supported by Polish National Science Center (NCN) grant number: 2018/31/G/ST9/03224. AM also acknowledges partial support from NCN grant number: 2016/23/B/ST9/03123. 

\section*{Data Availability}
The corner-plots of all the analysed cases can be found in \href{https://users.camk.edu.pl/tathagata/tsaha1996.html}{https://users.camk.edu.pl/tathagata/tsaha1996.html}. Any other data products will be made available on reasonable request.


\bibliographystyle{mnras}
\bibliography{version5}
\appendix


\bsp	
\label{lastpage}
\end{document}